\documentclass[twocolumn,showpacs,preprintnumbers,amsmath,amssymb]{revtex4}
\usepackage{graphicx}
\begin{document}
\title{Spin-orbital and spin Kondo effects in parallel coupled quantum dots}
\author{D. Krychowski and S. Lipi\'nski}
\affiliation{%
Institute of Molecular Physics, Polish Academy of Sciences\\M. Smoluchowskiego 17,
60-179 Pozna\'{n}, Poland
}%
\date{\today}
\begin{abstract}
Strong electron correlations and interference effects are discussed in parallel-coupled single-level and orbitally doubly degenerate  quantum dots. The finite-U mean-field slave boson approach is used to study many-body effects. The analysis is carried out in a wide range of parameter space including both atomic-like and molecular-like Kondo regimes and taking into account various perturbations, like interdot tunneling, interdot interaction, mixing of the electrode channels and exchange interaction. We also discuss the influence of singularities of electronic structure and the impact of polarization of electrodes. Special attention is paid to potential spintronic applications of these systems showing how current polarization can be controlled by adjusting interference conditions and correlations by gate voltage. Simple proposals of  double dot spin valve and bipolar electrically tunable spin filter are presented.
\end{abstract}

\maketitle

\section{Introduction}

Coherent transport in the presence of strong electron-electron interactions is one of the central issues in the field of nanoscale systems examined  both  experimentally ~\cite{Goldhaber, Kouwenhoven, Cronenwett, Glazman, vanderWiel} and theoretically ~\cite{Georges, Aguado, Aono, Lopez, Ding, Busser}. Perhaps the most interesting regime occurs when electrons in the dot acquire antiferromagnetic correlations with electrons in the leads, giving rise to the well-known Kondo effect ~\cite{Hewson, Liang, Park}. Another theme that has received considerable attention is the role of interference between different current paths. Interference conditions can be changed by magnetic field (Aharonov - Bohm oscillations ~\cite{Gerland, Hofstetter}) or modification of geometry. The example of the latter are  Fano or Dicke resonances ~\cite{Fano, Bulka, Guevara, Sztenkiel, Nahm, Trocha} present when background and resonance scattering processes occur simultaneously. The simplest system to study the interplay between interference and many-body correlations are coupled parallel quantum dots (PDQD) ~\cite{Chen}. The easy  control of  the  couplings in these systems allows examination of a broad variety of transport regimes. Most of the papers devoted to Kondo-assisted transport through double dots have focused on dots connected in series ~\cite{Georges, Aguado, Aono, Lopez, Izumida, Oosterkamp, Jeong, Craig, Tanaka, Lim, Ribeiro}, where interference does not appear, and much less attention has been  paid to parallel dots ~\cite{Lopez, Chen, Boese, Kawakami, Bonca, Okazaki, Wong}. Recently it has been also  realized  that PDQDs   are much more experimentally  suitable than the  dots in series  for studying spin entangled state composed of coherent Kondo resonances ~\cite{Chen}. Present paper is devoted to analysis of  transport in parallel coupled dots of different symmetries.  We discus how to manipulate the Kondo state by interference conditions in order to reach new device functionality particularly  in spintronic applications.  Our general discussion based on variants  of two-impurity Anderson model that take into account degeneracy and various perturbations is addressed to double dot devices formed in various materials, including GaAs two-dimensional electron gas ~\cite{Kouwenhoven, Cronenwett, Glazman}, semiconductor nanowires ~\cite{Pfund} and carbon nanotubes ~\cite{Sapmaz, Goss}. Carbon nanotube quantum dots (CNTQDs) are very attractive not only due to the potential applications, but also from cognitive point of view due to high degeneracy of the energy levels leading to  the appearance of exotic  many body effects of enhanced symmetry. Besides the spin, also other degrees of freedom e.g. orbital ~\cite{Sasaki}  or charge ~\cite{Wilhelm, Holleitner}  can trigger the Kondo correlations and moreover spin and orbital degeneracies can also occur simultaneously leading to highly symmetric Kondo state of SU(4) symmetry. In this case simultaneous screening of orbital and spin degrees of freedom is caused by tunneling processes causing spin, orbital pseudospin and spin-orbital fluctuations. Experimentally the spin -orbital  SU(4) Kondo  has been investigated in carbon nanotube quantum dots (CNTQDs) ~\cite{Herrero, Makarovski, Rasmussen, Wu},  in vertical semiconductor QDs ~\cite{Sasaki}, in capacitively coupled QDs ~\cite{Keller} and Si-fin-type field effect transistors ~\cite{Tettamanzi}. Orbital degeneracy commonly occurs also in various organic molecules, such as metal phtalocyanine, metal porphyrine etc. and thus similar many-body effects as these observed in CNTQDs should be also realized for these molecules in the range of weak coupling with electrodes ~\cite{Minamitani}. SU(4) symmetry is also believed to arise in graphene nanostructures due  electron's spin and valley degrees of freedom ~\cite{Guclu}. The problem of simultaneous screening of charge or orbital degrees of freedom and spin has been also discussed theoretically ~\cite{Simon, Borda, Choi, Loss, Busser2, Zhukov, Anders, Lipinski, Filippone}.  Due to the enhanced degeneracy Kondo temperature in these systems is typically at least one order of magnitude higher than for the common spin Kondo effect, what makes them interesting for practical applications.  SU(4) group is minimal  group allowing orbital-spin entanglement, which guarantees rotational invariance in spin and orbital spaces. The four-state entanglement is interesting for quantum computing, because each four-state bit is equivalent to two state bits, so the four-state bits double the storage density. It is interesting to note that it is possible to switch between higher and lower symmetries. Partial breaking of the degeneracy by magnetic field or by difference of gate voltages of the dots results in a crossover from SU(4) to SU(2) Kondo physics either in spin or orbital sectors giving  additional tool for manipulating  the transport regimes ~\cite{Lee, Krychowski, Aguado2, Chudnovskiy, Schmid}. In addition to tunneling also central to the design circuits for logic and quantum information processing based on double dots is an  examination of capacitive coupling and interdot exchange.  These topics have been  discussed both theoretically ~\cite{Aono, Lopez, Halperin, Pohjola, vonDelft, Galpin, Ramsak, Varma, Kotliar, Burkard, Lei, Oreg, Vojta, Michalek, Chung, Martinek, Zhang}  and experimentally ~\cite{Molenkamp, Blick, Wilhelm2, Chan, Hubel}, but only for SU(2) symmetry. Another theme that has received considerable attention is the role of polarization of electrodes in weakening of Kondo correlations. This problem has been widely discussed for single  quantum dots (see e.g. ~\cite{Sergueev, Bulka2, Martinek2, Choi2}),  but only a few of papers have been devoted to this issue for double dots ~\cite{Tanaka2, Trocha2, Weymann2} and all of them are restricted also only to SU(2) symmetry.

	In this paper we perform a comparative study of strongly correlated parallel coupled quantum dots of SU(4) and SU(2) symmetries examining the impact of interdot tunneling, interdot interaction, interdot exchange and mixing of electrode channels on the conductance. We also analyze the role of singularities of the electronic structure of electrodes and the effect of magnetic polarization. Discussing the problem of coupled SU(4) dots we concentrate mainly  on the case of separate electrodes. Particular emphasis in this work is put on displaying of the potential inherent in playing polarity of the electrodes  together with interference effects  and many body correlations in PDQDs. To highlight this issue we give few examples of spintronic proposals. We elucidate  transfer of  polarization of conductance by tunneling  from the dot coupled to magnetic electrodes to the dot attached to nonmagnetic electrodes. We discuss the spin valve properties of PDQDs and present the method of drastic modification of gate voltage characteristic of tunnel magnetoresistance. We show that the double dot system coupled to common pair of polarized electrodes can work as an efficient bipolar spin filter. We also consider the case when the dots are additionally coupled  by exchange and discuss  a competition between Kondo correlations and  interdot spin antiferromagnetic correlations. We indicate on the possibility to perform a swap operations between entangled bonding and antibonding Kondo resonances  (singlet - triplet) by the change of interdot tunneling or gate voltage.

\section{Model and slave boson mean-field  formulation}

We consider a system of two  single-level or two-level quantum dots  connected in parallel to electron reservoirs. Our predominant discussion concerns the case of  the separate leads attached to each of the dots as presented in  Fig. 1a, but we compare also some of the results with the case of common electrodes (Fig. 1b). The systems are modeled by the  generalized  two-impurity(single level/double level) Anderson Hamiltonian:
\begin{eqnarray}
&&{\mathcal{H}}=\mathcal{H}(1)+\mathcal{H}(2)+\mathcal{H}^{1-2}_{dir}+\mathcal{H}^{1-2}_{ind}\nonumber\\
&&+\mathcal{H}^{1-2}_{int}+\mathcal{H}^{1-2}_{exch}
\end{eqnarray}
where, $\mathcal{H}(i)$ denote Hamiltonians of quantum dots coupled to the leads, and $\mathcal{H}^{1-2}$ Hamiltonians describe interdot couplings: direct tunneling coupling $\mathcal{H}^{1-2}_{dir}$, indirect coupling via the channels of common electrodes $\mathcal{H}^{1-2}_{ind}$,  interdot interaction $\mathcal{H}^{1-2}_{int}$, and interdot exchange  $\mathcal{H}^{1-2}_{exch}$.  For brevity we give below the corresponding expressions for the two-orbital case, in the single orbital case the orbital index takes only value $l=1$  and obviously vanishes interorbital Coulomb interaction
\begin{eqnarray}
&&\mathcal{H}(i)=\sum_{l=1,2\sigma}E_{il\sigma}n_{il\sigma}+\mathcal{U}n_{il\uparrow}n_{il\downarrow}
+\mathcal{U}\sum_{\sigma\sigma'}n_{i1\sigma}n_{i2\sigma'}\nonumber\\
&&+V\sum_{kl\sigma\alpha=L,R}(c^{\dag}_{kil\alpha\sigma}d_{il\sigma}+h.c)
+\sum_{kil\sigma\alpha}\varepsilon_{kil\alpha\sigma}n_{kil\sigma}
\end{eqnarray}
where $E_{il\sigma}$ are the bare energies of local dot levels (with one exception (Fig. 8b) the energies of the dots are taken equal $E_{1l\sigma}=E_{2l\sigma}=E_{d}$). The next two terms parameterized by $\mathcal{U}$ describe intra and interorbital Coulomb interactions,  $V$ is the dot-lead hybridization amplitude and  the last term describes electrons in the electrodes.
Direct tunneling between the dots is assumed to occur between the orbitals of the same symmetry and is given   by:
\begin{eqnarray}
&&\mathcal{H}^{1-2}_{dir}=\sum_{l\sigma}(td^{\dag}_{1l\sigma}d_{2l\sigma}+h.c)
\end{eqnarray}
Marginally we will also  compare  in sec. IIIa  some results with the case of nonvanishing interdot-interorbital tunneling.
Mixing of states from different electrodes is then  given by:
\begin{eqnarray}
&&\mathcal{H}^{1-2}_{ind}=V'\sum_{l\alpha\sigma}(c^{\dag}_{k1l\alpha\sigma}d_{2l\sigma}+c^{\dag}_{k2l\alpha\sigma}d_{1l\sigma}+h.c)
\end{eqnarray}
The last two coupling terms, interdot interaction and the exchange, are discussed in the text only for the single-level dots and therefore the corresponding Hamiltonian are given here only for these cases:
\begin{eqnarray}
&&\mathcal{H}^{1-2}_{int}=\mathcal{U'}\sum_{\sigma\sigma'}n_{1\sigma}n_{2\sigma'}\\
&&\mathcal{H}^{1-2}_{exch}=\mathcal{J}\sum_{\sigma\sigma'}d^{\dag}_{1\sigma}d_{1\sigma'}d^{\dag}_{2\sigma'}d_{2\sigma}
\end{eqnarray}
To discuss correlation effects we use finite-U slave boson mean field theory (SBMFT) approach developed by Kotliar and Ruckenstein (K-R)~\cite{Kotliar2}. This approximation concentrates on many-body resonances taking into account spin and orbital fluctuations and strictly applies close to the unitary Kondo limit, but due to its simplicity this method is also often used in discussion of linear conductance in a relatively wide dot-level range  giving results in reasonable agreement with experiments and with numerical renormalization group calculations ~\cite{Dong2}. At $T = 0$ K-R approach  reproduces the results derived by the well known analytical technique of  Gutzwiller-correlated wave function ~\cite{Gutzwiller}. In the finite-U slave boson approach a set of auxiliary bosons $e_{i}$, $p_{i\sigma}$, and $d_{i}$ are introduced for each  of the single-level dots. These operators act as projection operators onto empty, singly occupied (with spin up or down) and doubly occupied states of quantum dots respectively. For two-level dots bosons are specified additionally by the orbital indices and further new boson operators representing triple occupied states $t_{il\sigma}$ and fully (quadruple) occupied states ($f_{i}$) are introduced. The single occupation projectors $p_{il\sigma}$,are labeled by indices specifying  the corresponding single electron states, $t_{il\sigma}$ operators by indices of the state occupied by a hole, and six  $d_{i}$ operators of the i-th dot  denote projectors onto  $(\uparrow\downarrow,0)$ and $(0,\uparrow\downarrow)$ ($d_{il=1,2}$) and $(\uparrow,\uparrow)$, $(\downarrow,\downarrow)$, $(\uparrow,\downarrow)$, $(\downarrow,\uparrow)$ ($d_{i\sigma\sigma'}$). In order to eliminate unphysical states the completeness relation for these operators and the correspondence between fermions and bosons have to be imposed, for brevity  both these conditions included in (7), are written  here  only for the more complicated case of  two-level dots (the analogous formalism for single dot with the use of only $e$, $p$, $d$ operators can be found e.g.  in ~\cite{Dong2}). The constraints can be enforced by introducing Lagrange multipliers $\lambda_{i}$, $\lambda_{il\sigma}$ and the effective SB Hamiltonian for $\mathcal{J} = 0$  then reads:
\begin{eqnarray}
&&\mathcal{H}^{K-R}=\sum_{l=1,2\sigma}E_{il\sigma}n^{f}_{il\sigma}+\mathcal{U}\sum_{il}d^{\dag}_{il}d_{il}\nonumber\\
&&+\mathcal{U'}\sum_{i\sigma\sigma'}d^{\dag}_{i\sigma\sigma'}d_{i\sigma\sigma'}+(\mathcal{U}+2\mathcal{U'})t^{+}_{il\sigma}t_{il\sigma}\nonumber\\
&&+(2\mathcal{U}+4\mathcal{U'})f^{+}_{i}f_{i}+\sum_{il\sigma}\lambda_{il\sigma}(n^{f}_{il\sigma}-Q_{il\sigma})\nonumber\\
&&+\sum_{i}\lambda_{i}(\mathcal{I}_{i}-1)+V\sum_{kl\sigma\alpha=L,R}(c^{\dag}_{kil\alpha\sigma}z_{il\sigma}f_{il\sigma}+h.c)\nonumber\\
&&+\sum_{kil\sigma\alpha}\varepsilon_{kil\alpha\sigma}n_{kil\alpha\sigma}+\sum_{l\sigma}(tz^{\dag}_{1l\sigma}d^{\dag}_{1l\sigma}z_{2l\sigma}f_{2l\sigma}+h.c)\nonumber\\
&&+V'\sum_{l\alpha\sigma}(c^{\dag}_{k1l\alpha\sigma}z_{2l\sigma}f_{2l\sigma}+c^{\dag}_{k2l\alpha\sigma}z_{1l\sigma}f_{1l\sigma}+h.c)
\end{eqnarray}
with $Q_{il\sigma} = p^{+}_{il\sigma}p_{il\sigma}+d^{+}_{il}d_{il}+d^{+}_{i\sigma\sigma}d_{i\sigma\sigma}
+d^{+}_{i\sigma\overline{\sigma}}d_{i\sigma\overline{\sigma}}+t^{+}_{il\sigma}t_{il\sigma}
+t^{+}_{i\overline{l}\sigma}t_{i\overline{l}\sigma}+t^{+}_{i\overline{l}\overline{\sigma}}t_{i\overline{l}\overline{\sigma}}
+f^{+}_{i}f_{i}$, $\mathcal{I}_{i}=\sum_{l\sigma\sigma'}(e^{+}_{i}e_{i}+p^{+}_{il\sigma}p_{il\sigma}
+d^{+}_{il}d_{il}+d^{+}_{i\sigma\sigma'}d_{i\sigma\sigma'}
+t^{+}_{il\sigma}t_{il\sigma}+f^{+}_{i}f_{i})$  and $z_{il\sigma}=(e^{+}_{i}p_{il\sigma}+p^{+}_{il\overline{\sigma}}d_{il}
+p^{+}_{i\overline{l}\overline{\sigma}}(\delta_{l,1}d_{i\sigma\overline{\sigma}}+\delta_{l,2}d_{i\overline{\sigma}\sigma})+p^{+}_{i\overline{l}\sigma}d_{i\sigma\sigma}+
d^{+}_{i\overline{l}}t_{il\sigma}+d^{+}_{i\overline{\sigma}\overline{\sigma}}t_{i\overline{l}\overline{\sigma}}
+(\delta_{l,2}d^{+}_{i\sigma\overline{\sigma}}+\delta_{l,1}d^{+}_{i\overline{\sigma}\sigma})t_{i\overline{l}\sigma}
+t^{+}_{il\overline{\sigma}}f_{i})/\sqrt{Q_{il\sigma}(1-Q_{il\sigma})}$. $z_{il\sigma}$ renormalize interdot hoppings and dot-lead hybridization (7). For polarized electrodes the bare coupling strengths between the QD and the leads given by $\Gamma_{il\sigma\alpha}=2\pi|V|^{2}\sum_{k}\delta(E-\varepsilon_{kil\alpha\sigma})$ are spin dependent due the spin dependence of the density of states. One can express coupling strengths for the spin-majority (spin-minority) electron bands introducing polarization parameter $P$ as $\Gamma_{il\sigma\alpha} = \Gamma_{il\alpha}(1+\sigma P)$  with $\Gamma=\Gamma_{i l}=\sum_{\alpha}\Gamma_{il\alpha}$. The stable SBMFA are then found from the minimum of the free energy with respect to the mean values of boson operators and Lagrange multipliers. The spin dependent linear conductances read ${\cal{G}}_{i\sigma}=\frac{e}{h^{2}}{\cal{T}}_{i\sigma}=\widetilde{\Gamma}_{L\sigma}G^{R}_{i\sigma i\sigma}\widetilde{\Gamma}_{R\sigma}G^{A}_{i\sigma i\sigma}+\widetilde{\Gamma}_{L\sigma}G^{R}_{i\sigma \overline{i}\sigma}\widetilde{\Gamma}_{R\sigma}G^{A}_{\overline{i}\sigma i\sigma}$, where $\mathcal{T}$ denotes transmission matrix, $\widetilde{\Gamma}_{\alpha\sigma}$ are SB renormalized coupling strengths to electrodes and  $G^{R(A)}_{i\sigma i\sigma}$ are retarded (advanced) Green's function of the dot. The polarization of conductance is given by $PC_{i}=({\cal{G}}_{i\uparrow}-{\cal{G}}_{i\downarrow})/({\cal{G}}_{i\uparrow}+{\cal{G}}_{i\downarrow})$. The numerical results discussed below are presented with the use of energy unit defined by its relation to the band-width ($2D=100$).

\begin{figure}
\includegraphics[width=5 cm,bb=0 0 415 520,clip]{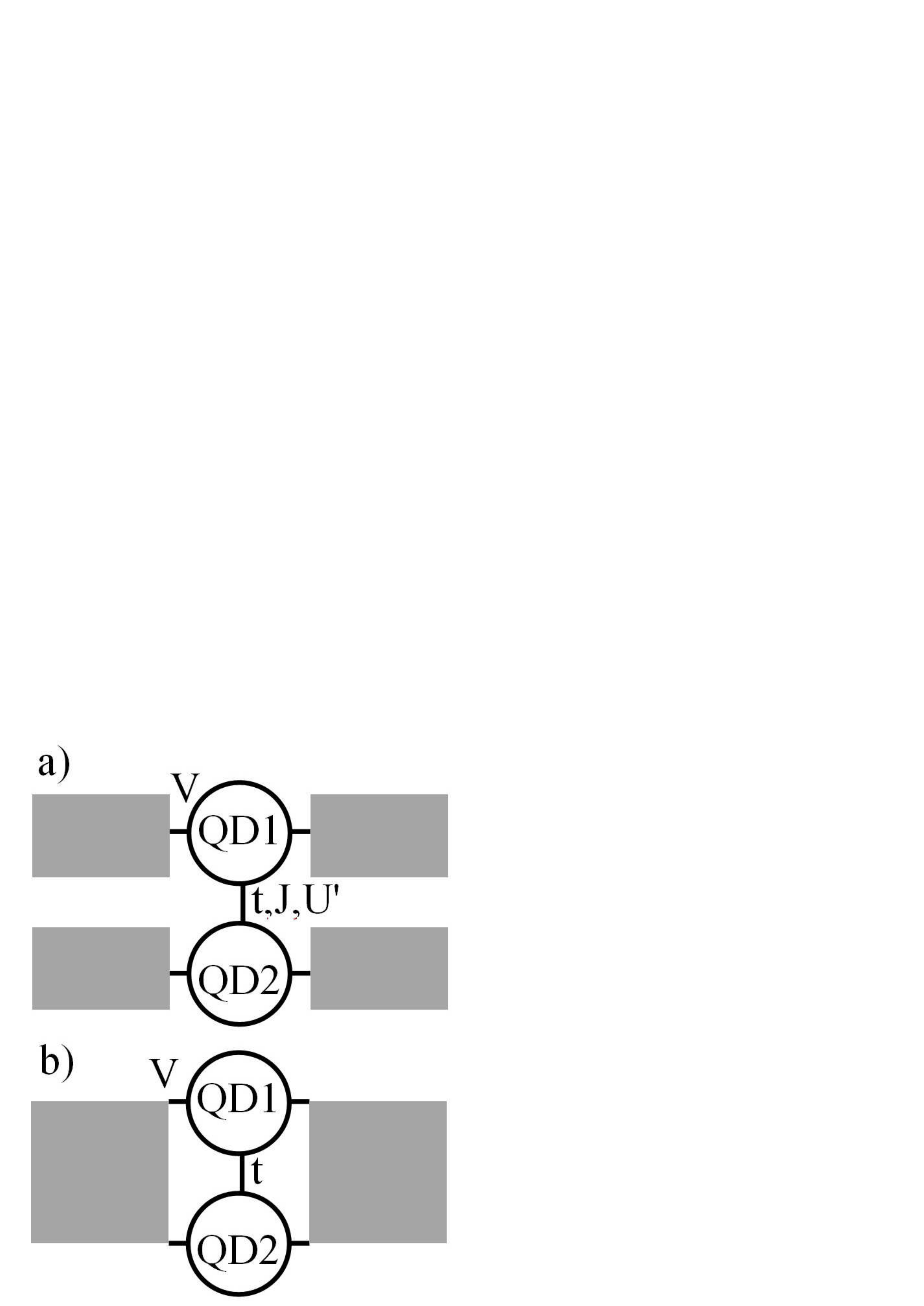} \hspace{4cm}
\caption{\label{fig:epsart} Schematic of the parallel double-quantum dot setup attached to separate leads (a) and to common electrodes (b).}
\end{figure}

\section{Different many-body regimes}
\subsection{Interdot coupling}

The coupling between the dots formed in semiconductor electron gas system can be changed freely by applying the split gate voltage between them. For CNTQDs and graphene dots similar gate control is much more difficult, but also in these systems the methods of  tuning the  interdot potential barrier through the gates have been elaborated  ~\cite{Graber, Wang, Wei}. In the absence of interdot tunneling, the screening processes occurring on the dots are independent. For single orbital SU(2) dots the observed plateau of conductance centered around particle-hole (p-h) symmetry point $E_{d} = -\mathcal{U}/2$ (Fig. 2a)  reflects spin Kondo effect occurring for single occupancy.
\begin{figure}
\includegraphics[width=4 cm,bb=0 0 290 290,clip]{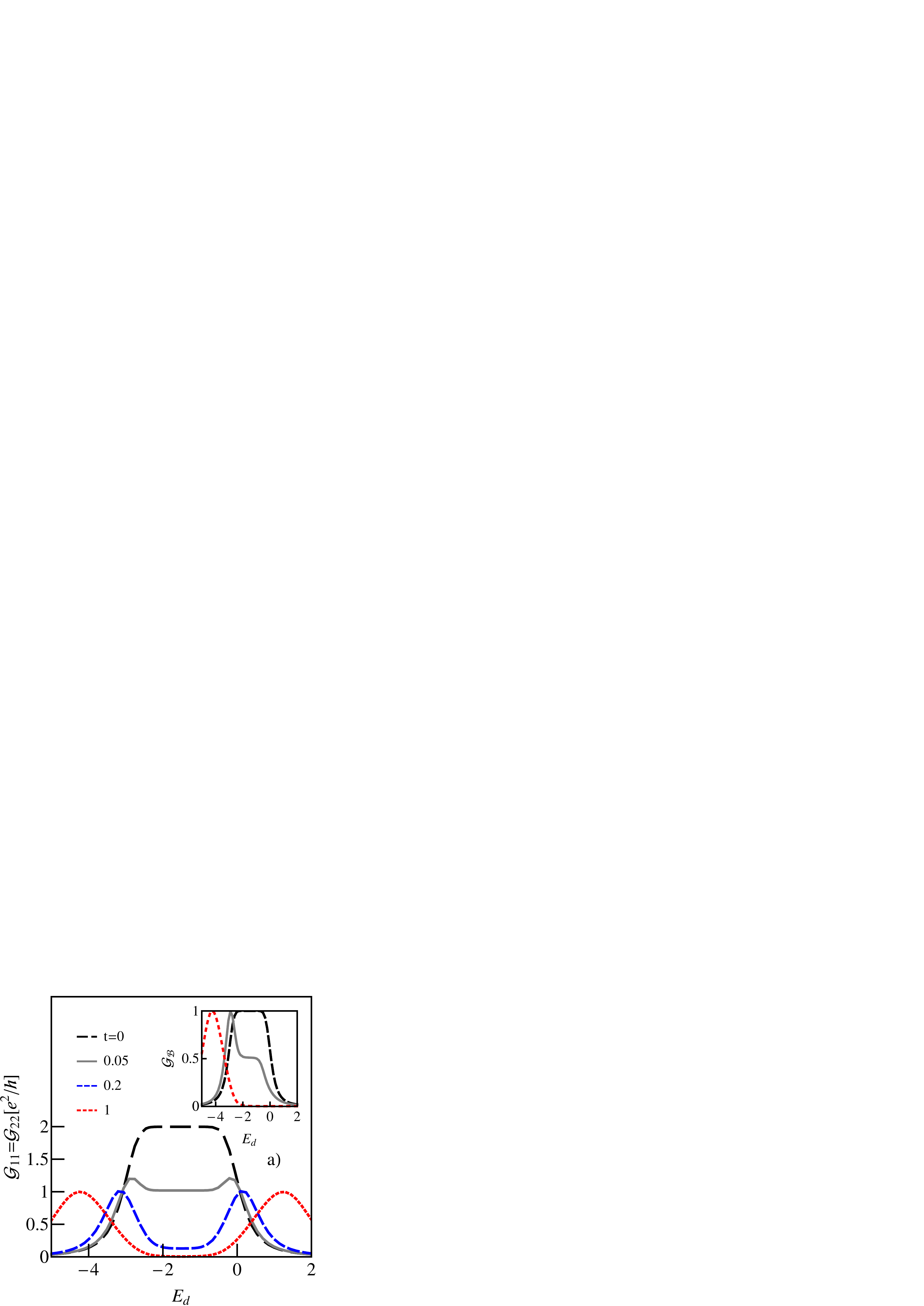}
\includegraphics[width=4 cm,bb=0 0 405 410,clip]{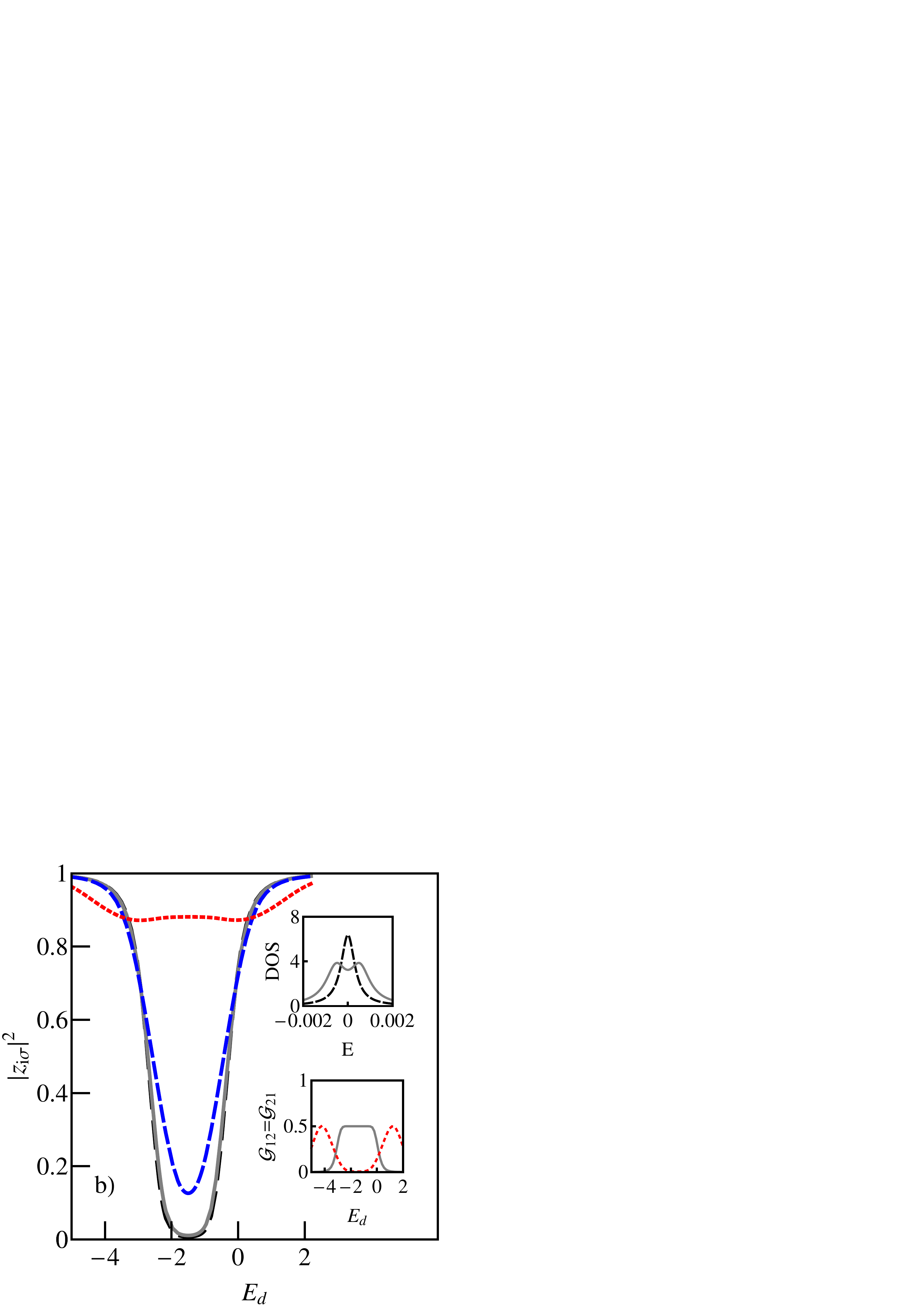}
\includegraphics[width=4.5 cm,bb=0 0 410 350,clip]{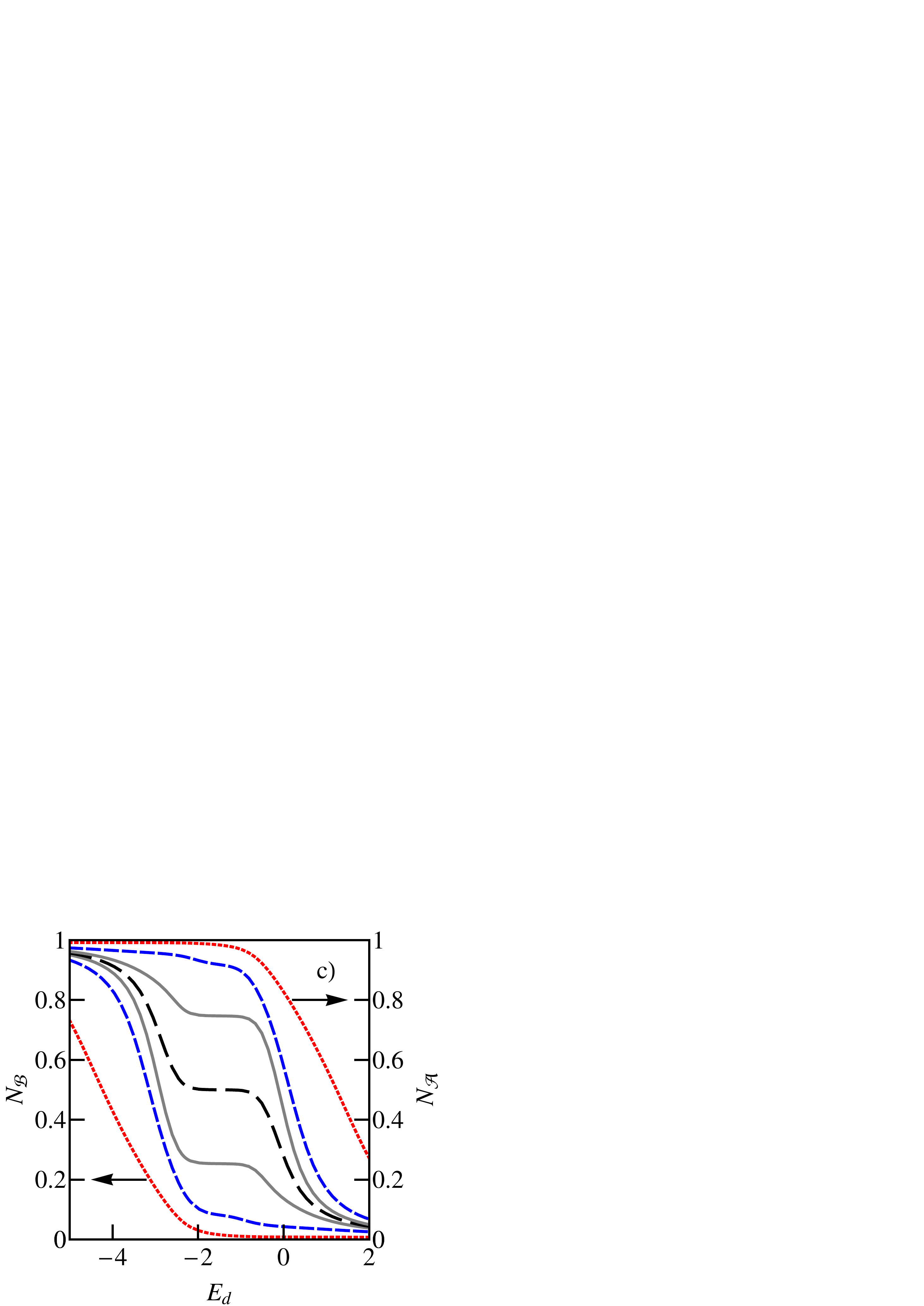}
\includegraphics[width=4 cm,bb=0 0 405 400,clip]{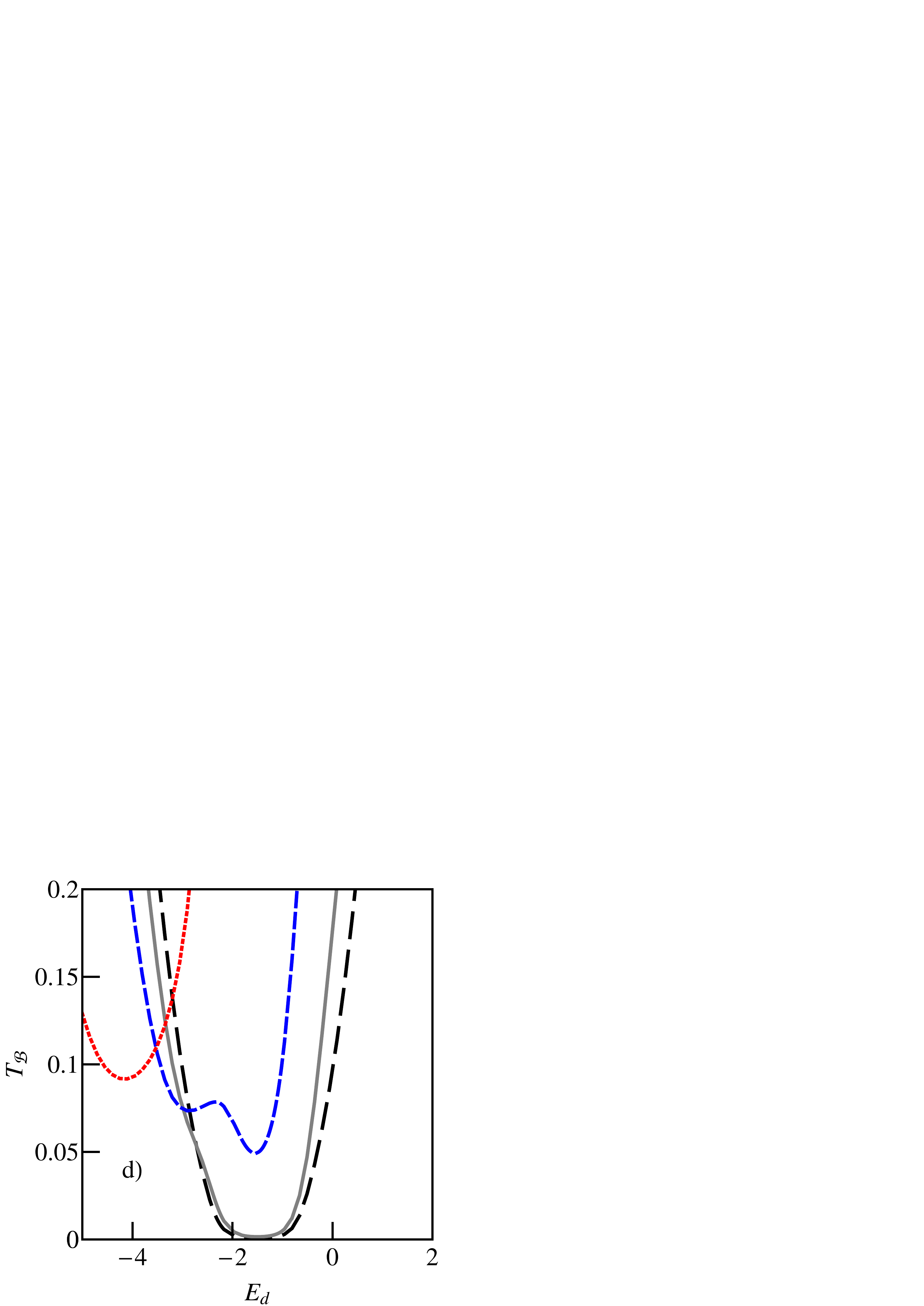}
\caption{\label{fig:epsart} (Color online) SU(2) PDQD, $\mathcal{U} = 3$, $\Gamma = 0.05$ (if not specified differently  the same  parameters apply also to other pictures.(a) Conductances for different interdottunneling (the same assignment of the linesis valid also for pictures b, c, d). Inset shows selected conductances of bonding channel compared with conductance of isolated dot. (b) SB tunneling renormalization parameter $z_{i\sigma}$ presented  for several  hopping parameters and  selected densities of states for $E_{d} =-\mathcal{U}/2$   ( upper inset) and  interdot contributions to conductance (lower inset). (c) Occupations of bonding and antibonding coherent many-body states. (d) Characteristic temperatures of coherent many-body bonding state.}
\end{figure}
For deep dot level the Kondo peak is pinned at the Fermi level ($E_{F}$)  and the scattering phase at $E_{F}$ is $\delta_{SU(2)}=\pi/2$ and zero temperature linear conductance  reaches the unitary limit  $\mathcal{G}^{SU(2)}=2e^{2}/h$. In the spin-orbital SU(4) case (two-orbital dot), Kondo effect occurs not only for single (1e) and triple occupancies (3e,  single hole) of each dot, but also for double occupancy (2e) (Fig. 3a).  Spin and isospin fluctuations result from electron tunneling in and out of the dot, what in 1e (3e) cases corresponds to  transitions between four degenerate  states characterized by different spin or isospin of electron  (hole).
\begin{figure}
\includegraphics[width=4 cm,bb=0 0 290 305,clip]{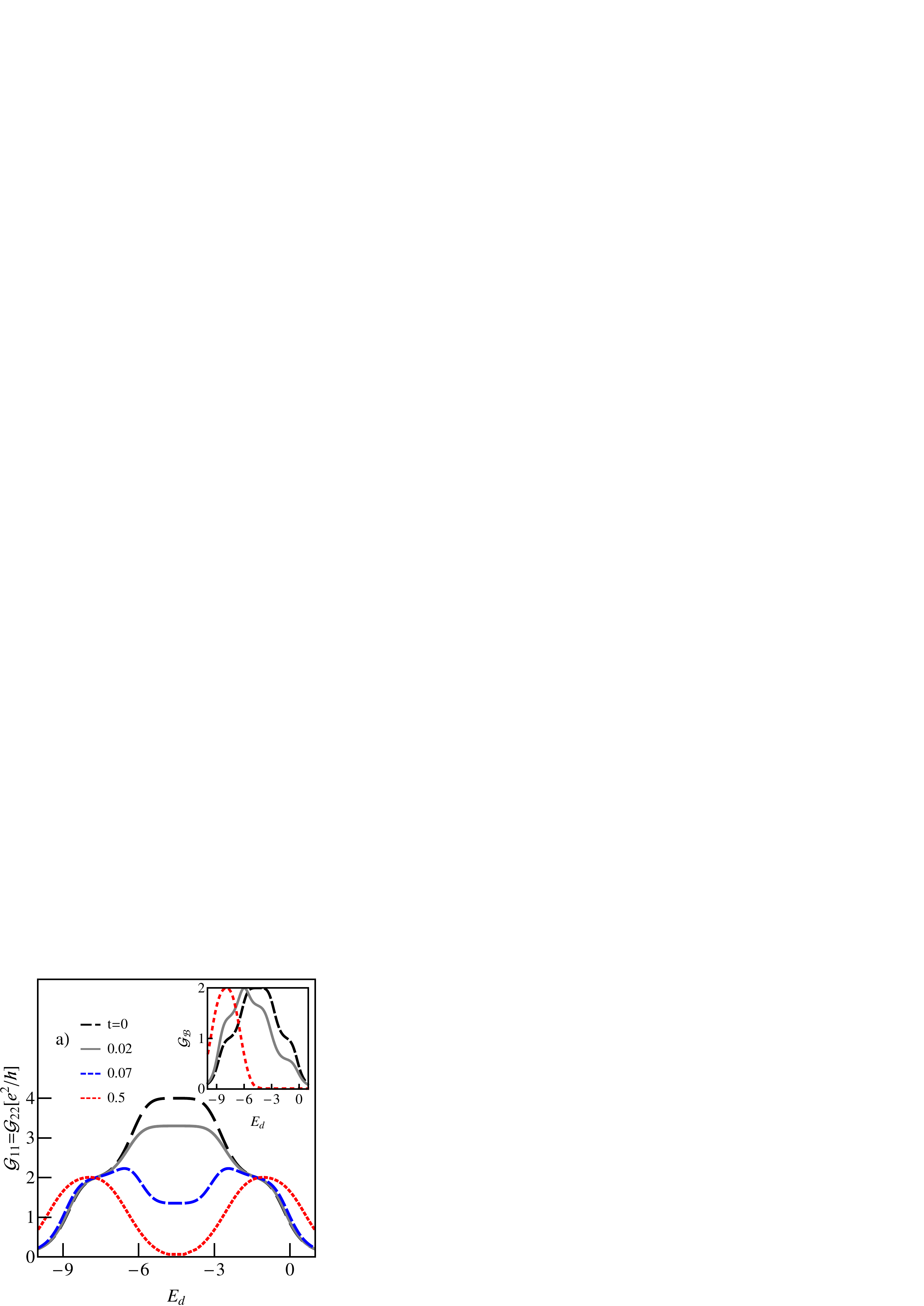}
\includegraphics[width=4 cm,bb=0 0 290 300,clip]{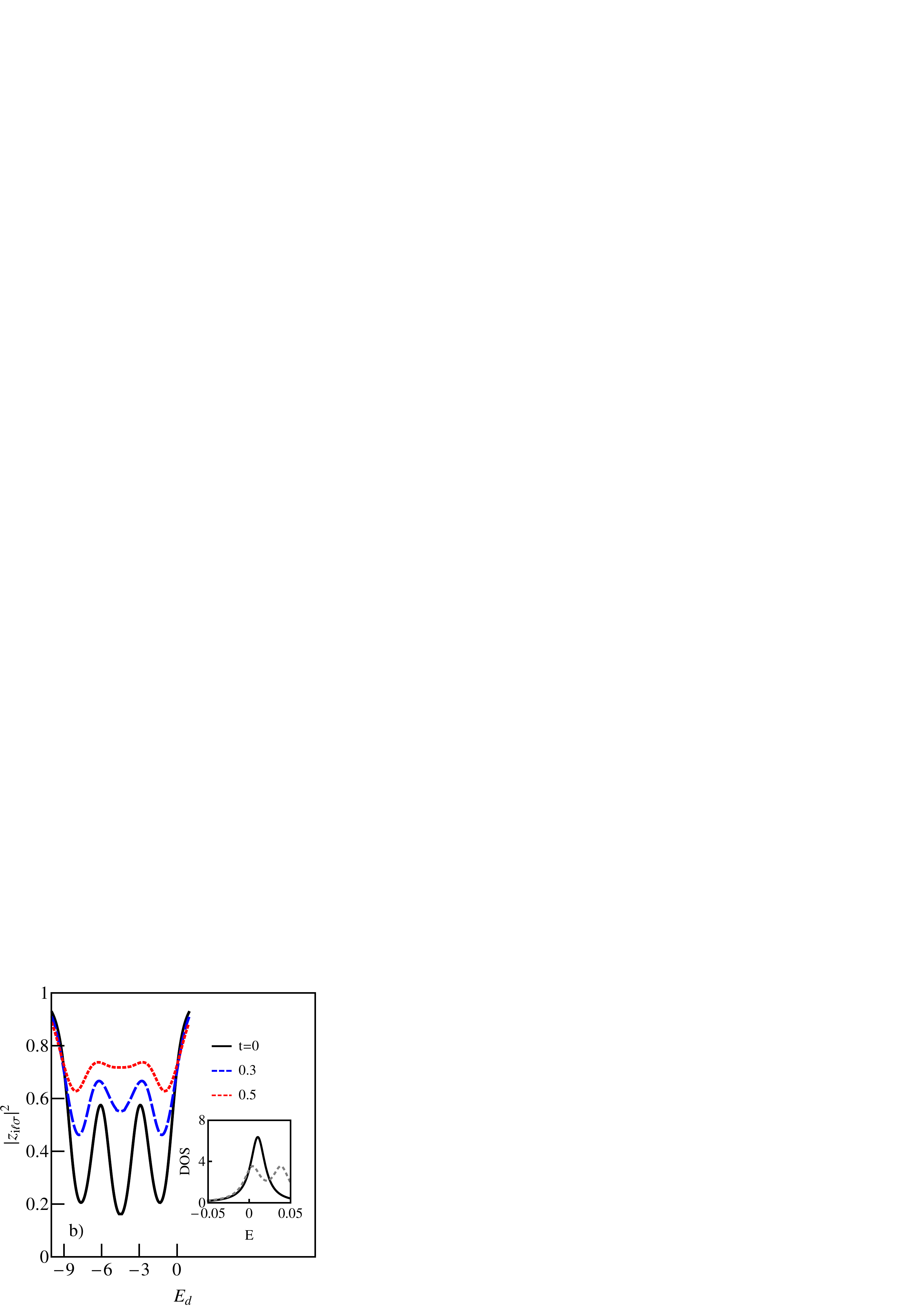}
\includegraphics[width=4.5 cm,bb=0 0 400 380,clip]{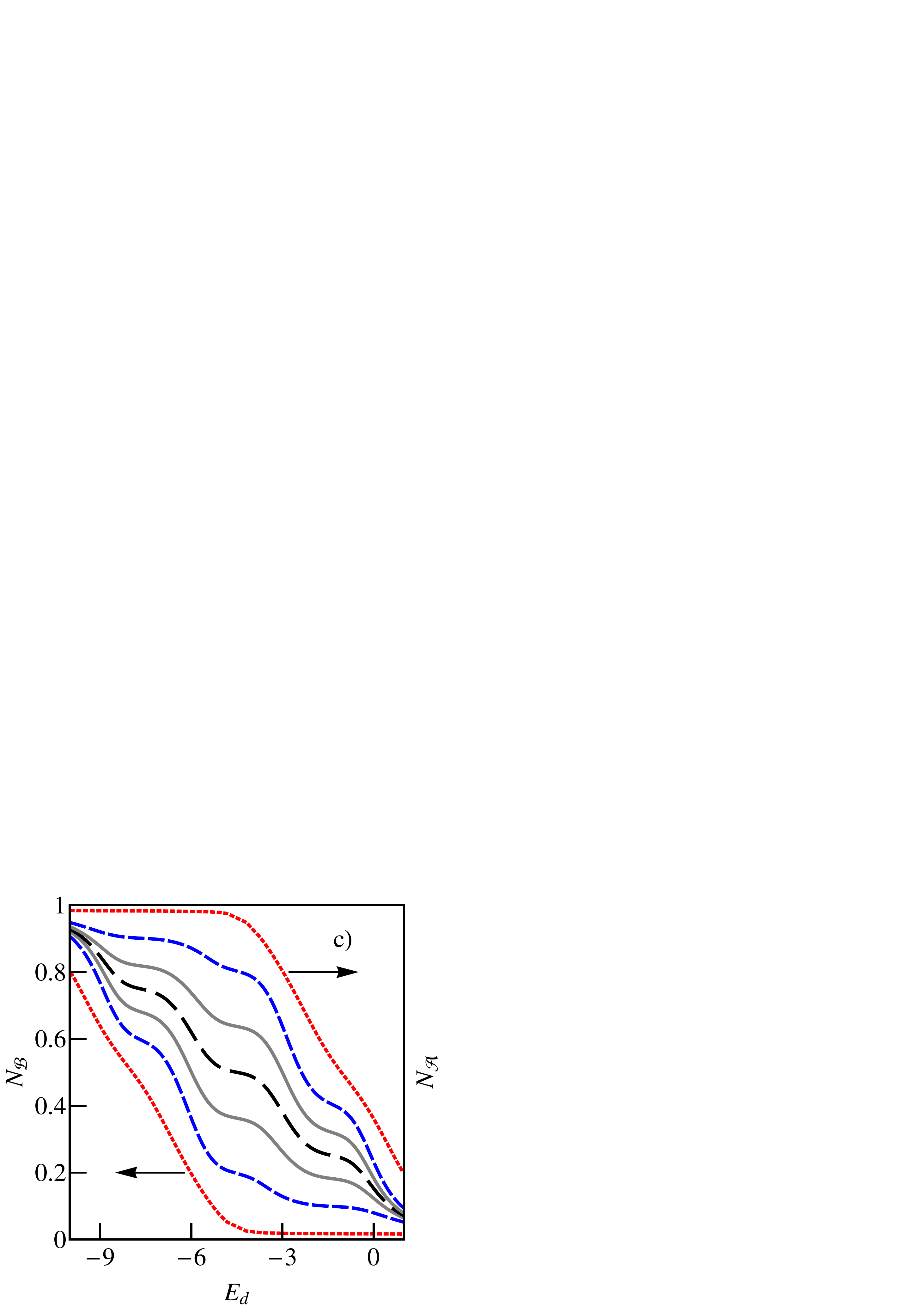}
\includegraphics[width=4 cm,bb=0 0 405 400,clip]{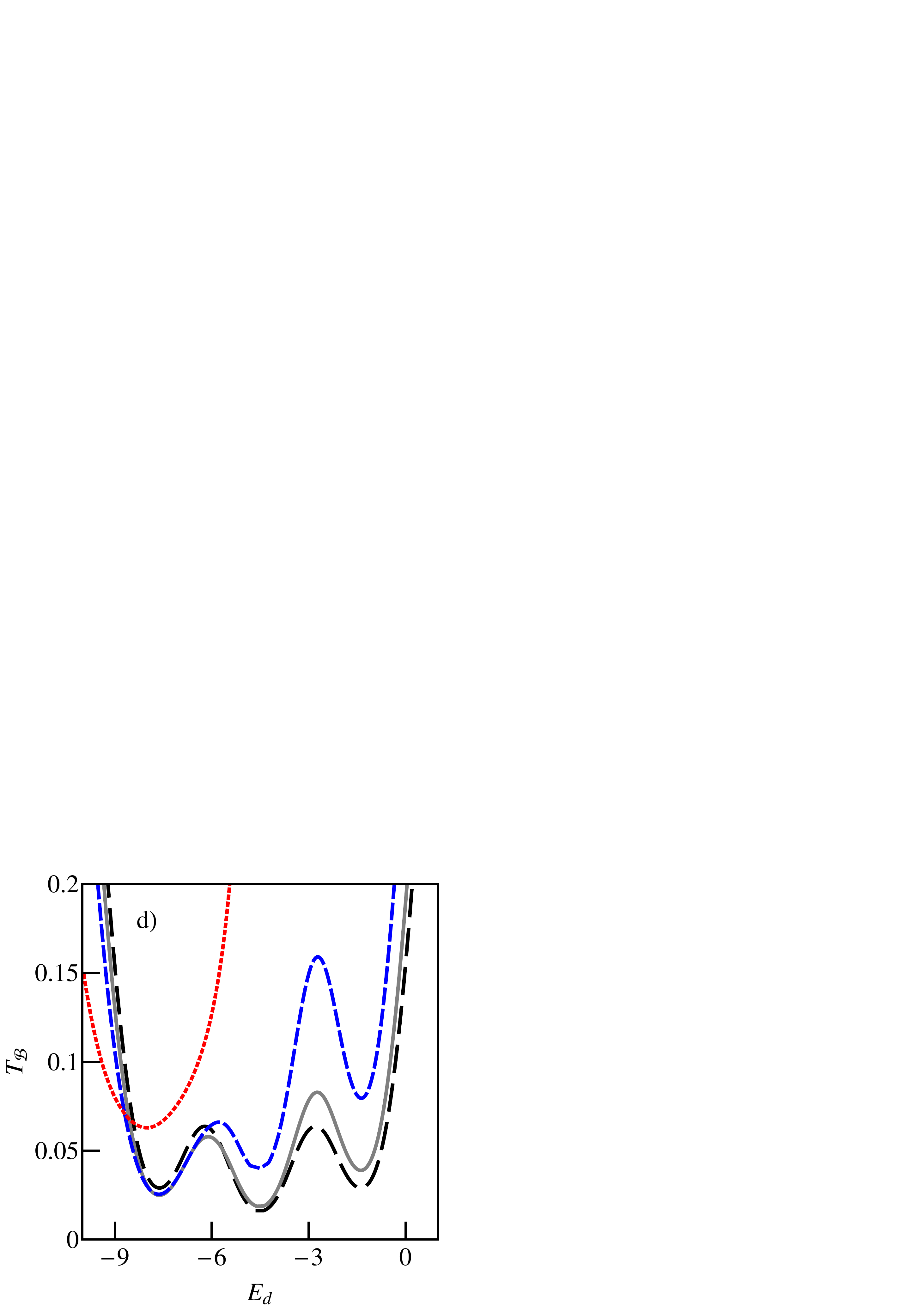}
\caption{\label{fig:epsart} (Color online) SU(4) PDQD(a) Conductances for different interdot tunneling (the same assignment of the linesis valid also for pictures c,d). Inset shows selected conductances of bonding channel compared with conductance of isolated dot. (b) SB tunneling renormalization parameter $z_{il\sigma}$  for several  hopping parameters and  in the inset  densities of states for $t = 0$ and $t = 0.09$ ($E_{d} =-\mathcal{U}/2$).  (c) Occupations of bonding and antibonding coherent many-body states. (d) Characteristic temperatures of coherent many-body bonding state}
\end{figure}
The spin-orbital many body peak is shifted above (1e) or below (3e) the Fermi level, what together with broadening of the peaks in comparison to SU(2) case, means enhanced Kondo temperature (see insets of Figs. 2b, 3b and Figs. 2d, 3d). The phase shifts for the deep levels are $\delta_{SU(4)}^{1e(3e)}=\pi/4(3\pi/4)$) and the total conductance at the dot reaches similarly to SU(2) case the limit $2e^{2}/h)$ (Fig. 3a). In 2e valley the SU(4) cotunneling processes  differ from the effects in 1e and 3e valleys, because now  six degenerate low-energy  two-electron states participate  in formation of Kondo resonance ~\cite{Zhukov}. Coherent tunneling among all these states corresponds to spin, orbital  pseudospin and spin-orbital fluctuations  and these processes  lead to a formation of  a  broad resonance centered at $E_{F}$, the corresponding   phase shift is $\delta_{SU(4)}^{2e}=\pi/2$ ~\cite{Zhukov} and the conductance is doubled in comparison to the  standard odd occupation Kondo resonances (Fig. 3a). Figures 2, 3 illustrate  the evolution of  many body processes with the increase of coupling between the dots.
For coupled dots ($t\neq0$)  the cotunneling processes include apart from direct tunneling  to the adjacent  electrodes  also the interdot hopping and  indirectly also higher order tunneling form the leads not directly attached to the given dot.  The interdot hopping increases a coherent superposition of the Kondo states of each of the dots. For the case of single orbital at the dot (SU(2)) for  half filling of  PDQD ($N_{tot} = 2$) $2\times$SU(2)   Kondo resonance splits (upper inset of Fig. 2b)  reflecting a formation of coherent bonding and antibonding many-body states.  In consequence conductance drops at half filling and for strong enough coupling suppression of the Kondo state is observed.
\begin{figure}
\includegraphics[width=4 cm,bb=0 0 290 300,clip]{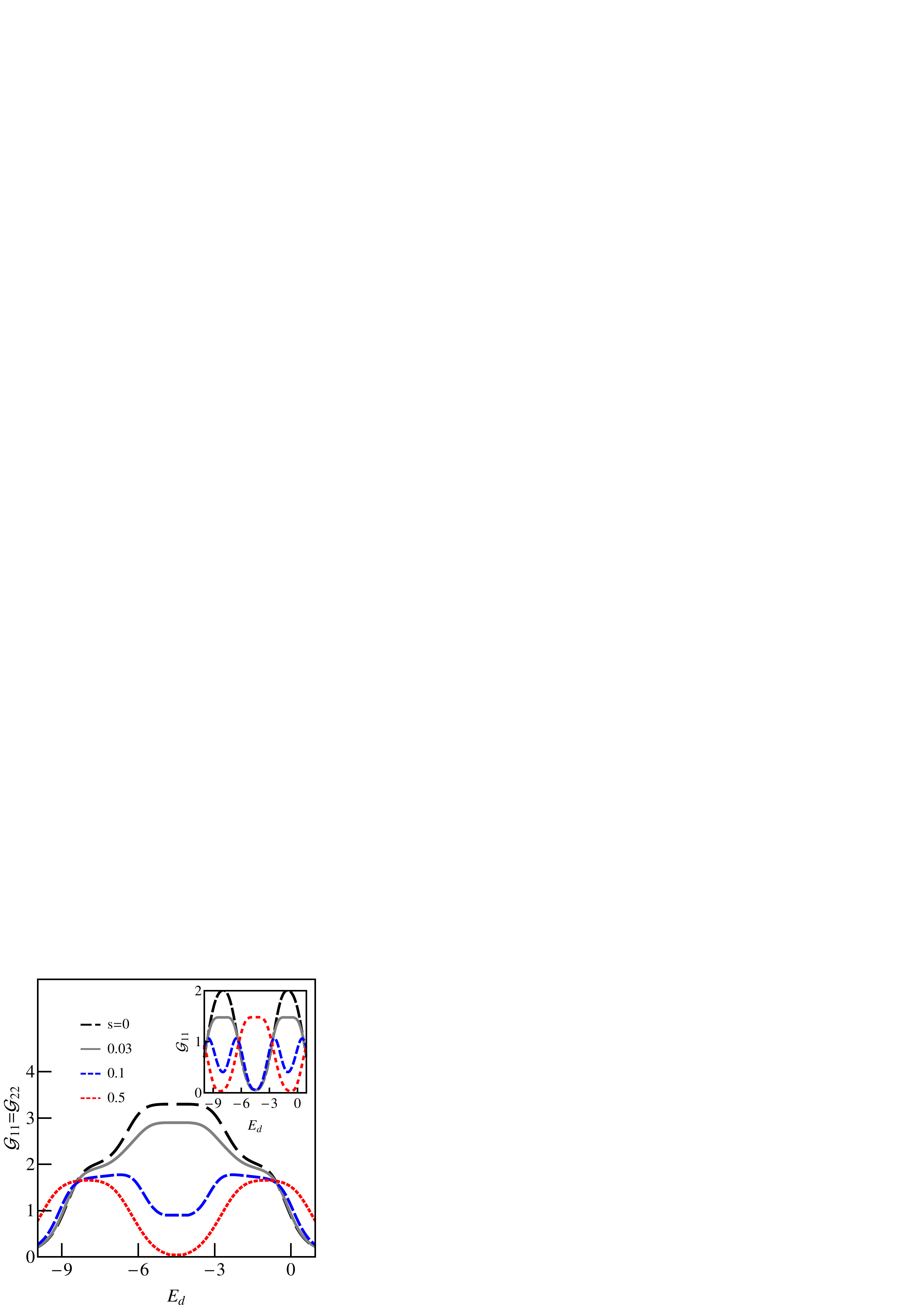} \hspace{4cm}
\caption{\label{fig:epsart} (Color online) Conductance of SU(4) PDQD system with interorbital tunneling s: weak off-diagonal tunneling (main picture),  strong off-diagonal tunneling (inset).}
\end{figure}
The strong renormalization of the coupling between the dots occurring for small values of  $t$ weakens for stronger coupling  and  this is represented in SB MFA formalism  by the gradual increase of $z$ (Fig. 2b). In one and three electron regimes of PDQD in the strong interdot coupling regime, the new many-body molecular like resonances emerge opening new transmission channels what leads to the  enhanced conductance in these ranges.
\begin{figure}
\includegraphics[width=3.8 cm,bb=0 0 405 430,clip]{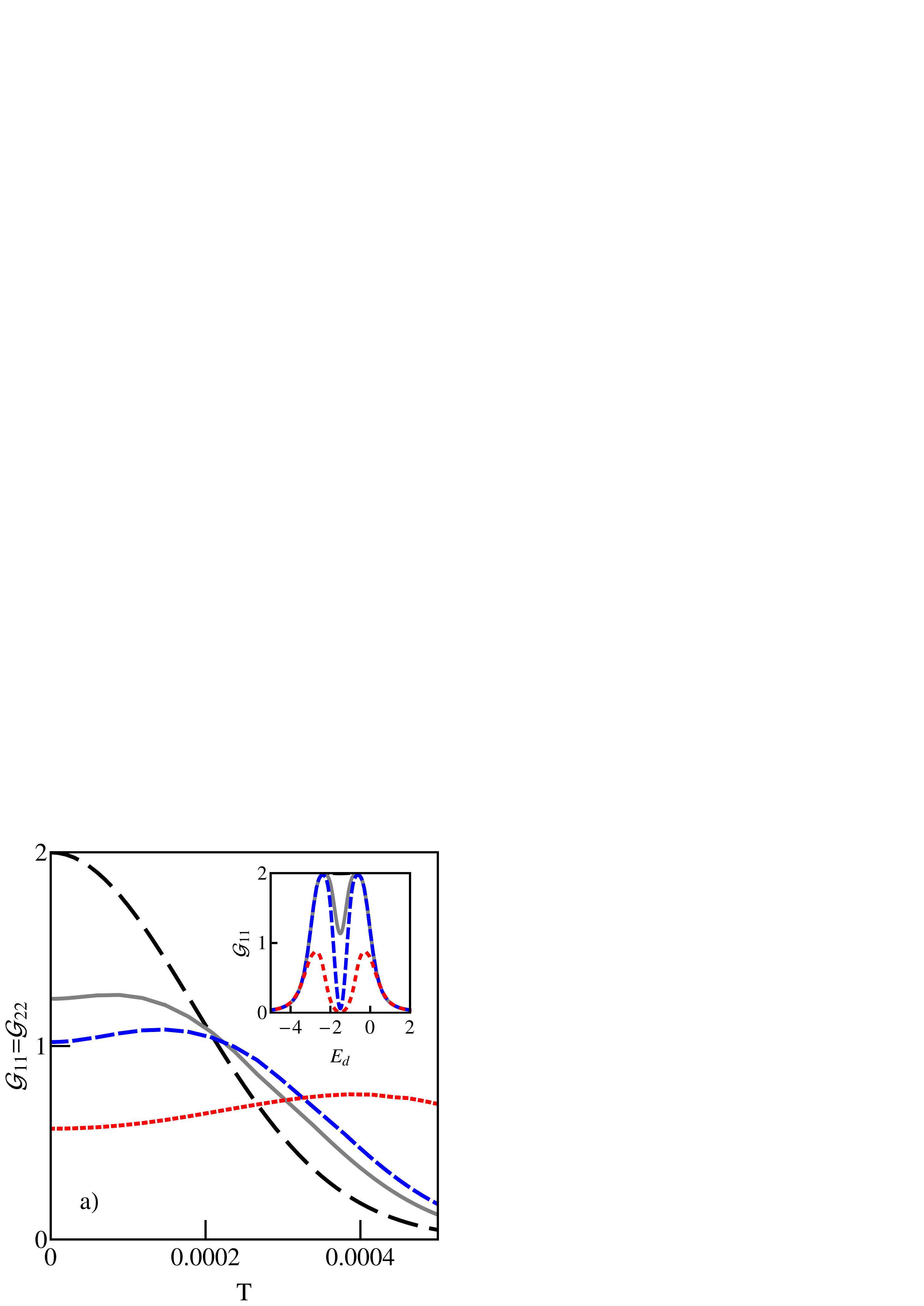}
\includegraphics[width=4 cm,bb=0 0 405 410,clip]{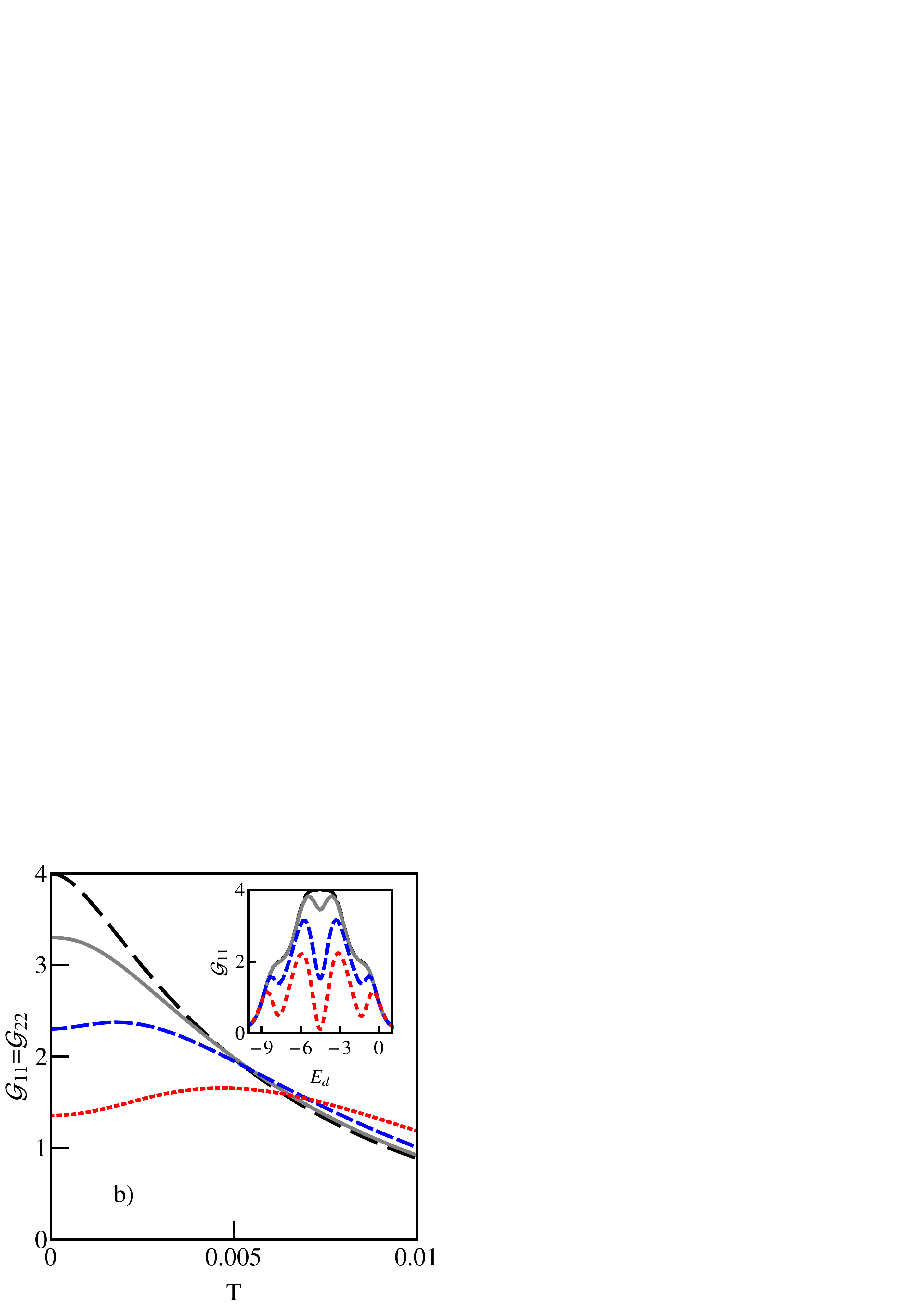}
\includegraphics[width=4.2 cm,bb=0 0 405 400,clip]{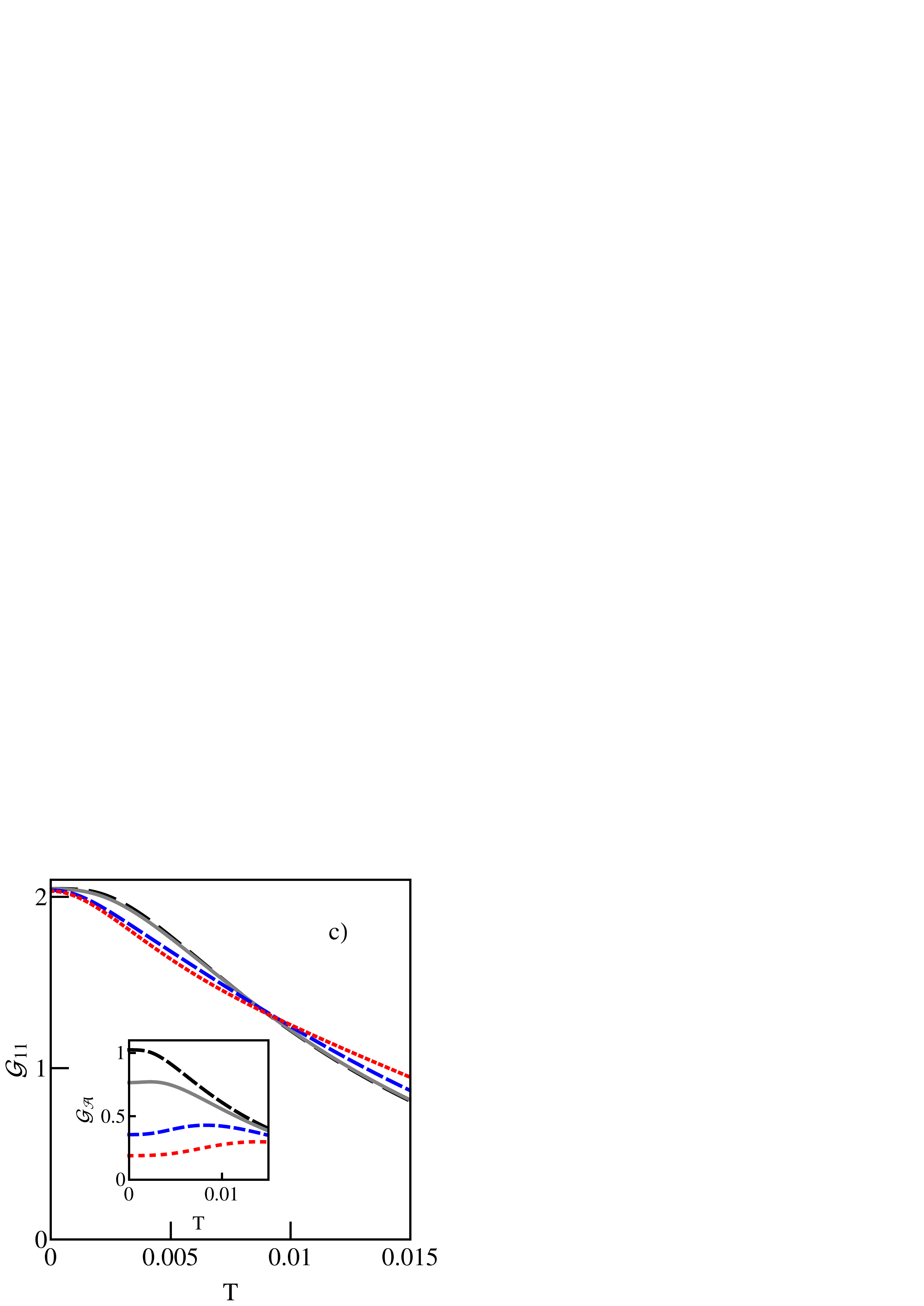} \hspace{4cm}
\caption{\label{fig:epsart} (Color online) Temperature dependencies of conductances of tunnel coupled dots (a) SU(2) ($E_{d} = -\mathcal{U}/2$) for $t = 0$ (broken black line), $t = 0.04$ (solid grey), $t = 0.05$ (broken blue), $t = 0.08$ (dotted red). Inset presents conductances of decoupled dots ($t = 0$) vs. dot energy for several temperatures: $T = 0$ (solid black line), $T = 0.0002$ (solid grey), $T = 0.005$ (broken blue) and $T = 0.05$ (broken red). (b) Conductance of  SU(4) PDQD  for $Ed = -\mathcal{U}/2 - \mathcal{U}$ (half filling) plotted for $t= 0$ (broken black line), $t= 0.02$ (solid grey), $t = 0.04$ (broken blue), $t = 0.07$ (dotted red).Inset presents conductances of decoupled dots  vs. dot energy for several temperatures: $T = 0$ (solid black line), $T = 0.002$ (solid grey), $T = 0.008$ (broken blue) and $T = 0.02$ (broken red). (c) Conductance of  SU(4) PDQD  for $E_{d} = -\mathcal{U}/2 - 2\mathcal{U}$  plotted for the same choice of tunneling parameters as in (b).  Inset shows the corresponding temperature dependencies of many-body antibonding state contributions to conductance.}
\end{figure}
For strong interdot coupling direct conductance of the dot ($\mathcal{G}_{ii}$) becomes equal to interdot conductance ($\mathcal{G}_{ij}$) (lower inset of Fig. 2b). The  character of resonances can be easily understood looking at the corresponding occupations of coherent bonding and antibonding states presented in Fig. 2c. Depending on the strength  of  hybridization with the leads, these resonance can be interpreted as Kondo bonding ($N_{tot}\approx1$) or antibonding  ($N_{tot}\approx3$) for weak hybridization  or corresponding  mixed valence (MV) resonances for  stronger hybridization  (case presented in Fig. 2). The characteristic temperatures of bonding many-body resonances are shown in Fig. 2d, the antibonding temperature curve can be obtained by a mirror reflection with respect to $E_{d} = -\mathcal{U}/2$. The described evolution with the increase of $t$ can be summarized as transition from $2\times$SU(2)  many-body state (separate resonances at the dots) occurring at half filling to SU(2)double dot coherent many-body  states (bonding) for $N\approx1$ or (antibonding) for $N\approx3$. A similar evolution for SU(4) dots (Fig. 3) is much richer due to larger number of many body resonances present already for the case of vanishing coupling. Increase of interdot hopping, which is assumed to bind only the states of the  same symmetry, results in splitting of the corresponding resonance  into two independent  degenerate bonding and degenerate  antibonding coherent states.  The $2\times$SU(4) Kondo resonance is split and gradually suppressed with the increase of $t$ and the  conductance drops around half filling ($E_{d}=-\mathcal{U}-\mathcal{U}/2$) (Fig. 3a). The conductance plateaus  for $N_{tot} = 2$ (single occupation on the dot) and $N_{tot} = 6$ (triple occupation) first asymmetrically deform for weak hopping and then,  for larger interdot coupling  the  two broad conductance peaks emerge.  The  enhancement above the plateau values  at intermediate coupling reflects  the interplay of bonding and antibonding states from two successive resonances, both these coherent molecular states contribute to conductance  in these regions. The conductance peaks in the strong coupling limit and single minima in the curves of characteristic temperatures  represent pure bonding or antibonding many body resonances respectively.  The $2\times$SU(4)  symmetry characterizing the  decoupled dots  with resonance at half filling  breaks and $2\times$SU(2) symmetry results (orbital degeneracy is preserved)   for strongly coupled dots with single occupations of each of two  bonding or antibonding states.  For the sake of completeness  we also present in  Fig. 4 examples of  conductance of SU(4) dots  when additionally interorbital hopping is present between the dots.  The interorbital part of the tunneling hamiltonian is expressed by $\sum_{l\sigma}(sd^{+}_{1l\sigma}d_{2\overline{l}\sigma}+h.c)$, where $s$ is the strength of the coupling between orbitals of different symmetries. Whereas for weak  intraorbital tunneling  evolution of conductance with the increase of interorbital tunneling is similar to the evolution with $t$ (Fig. 4),  for large intraorbital coupling ($t = 0.5$)  increase of conductance is observed at half filling with plateau of conductance  for $s = t$ (inset of Fig. 4). The latter observation is a consequence of restoration of degeneracy of states. Figure 5 presents temperature dependencies of conductance for tunnel coupled SU(2) and SU(4) dots. As a reference we show in the insets the corresponding gate voltage characteristics for isolated dots for several temperatures. At half filling  transmissions (not presented) are symmetrically located around $E_{F}$ and therefore due to thermal smearing deepening of conductances are observed around  corresponding  e-h symmetry points ($E_{d}=-\mathcal{U}/2$ for SU(2) or $E_{d}=-\mathcal{U}-\mathcal{U}/2$ for SU(4)). For odd occupancies of single SU(4) dot ($N\approx1$, $N\approx3$)  transmissions are shifted from the Fermi level and consequently weaker temperature dependencies are seen (Fig. 5b).  For  coupled dots the temperature dependences of conductance at half filling of tunnel perturbed $2\times$SU(2) and $2\times$SU(4) are not monotonous (Figs. 5a, 5b). Transmissions split with the increase of $t$ and for temperatures exceeding  this splitting  clear maxima are observed. For odd occupations of the single dots of tunnel perturbed $2\times$SU(4) system ($N_{tot}\approx2$ or  $N_{tot}\approx6$) (Fig. 5c)  no similar maxima are observed in temperature dependencies of total conductances. This is a consequence of a shift of transmission peaks from $E = 0$ already present for $t=0$. Maxima are still visible however in the  corresponding  partial conductances of Kondo bonding ($N_{tot}\approx2$) or antibonding ($N_{tot}\approx6$) states (inset of Fig. 5c).

\subsection{Interdot interaction}

Parallel dots can be fabricated to have both electrostatic and interdot couplings.  Here we focus on $t = 0$ case, but for comparison we also plot conductance for finite tunneling. We do not restrict to the analysis of the impact of  capacitive coupling alone, but we also present some  results for effective attractive interaction.
\begin{figure}
\includegraphics[width=4 cm,bb=0 0 290 310,clip]{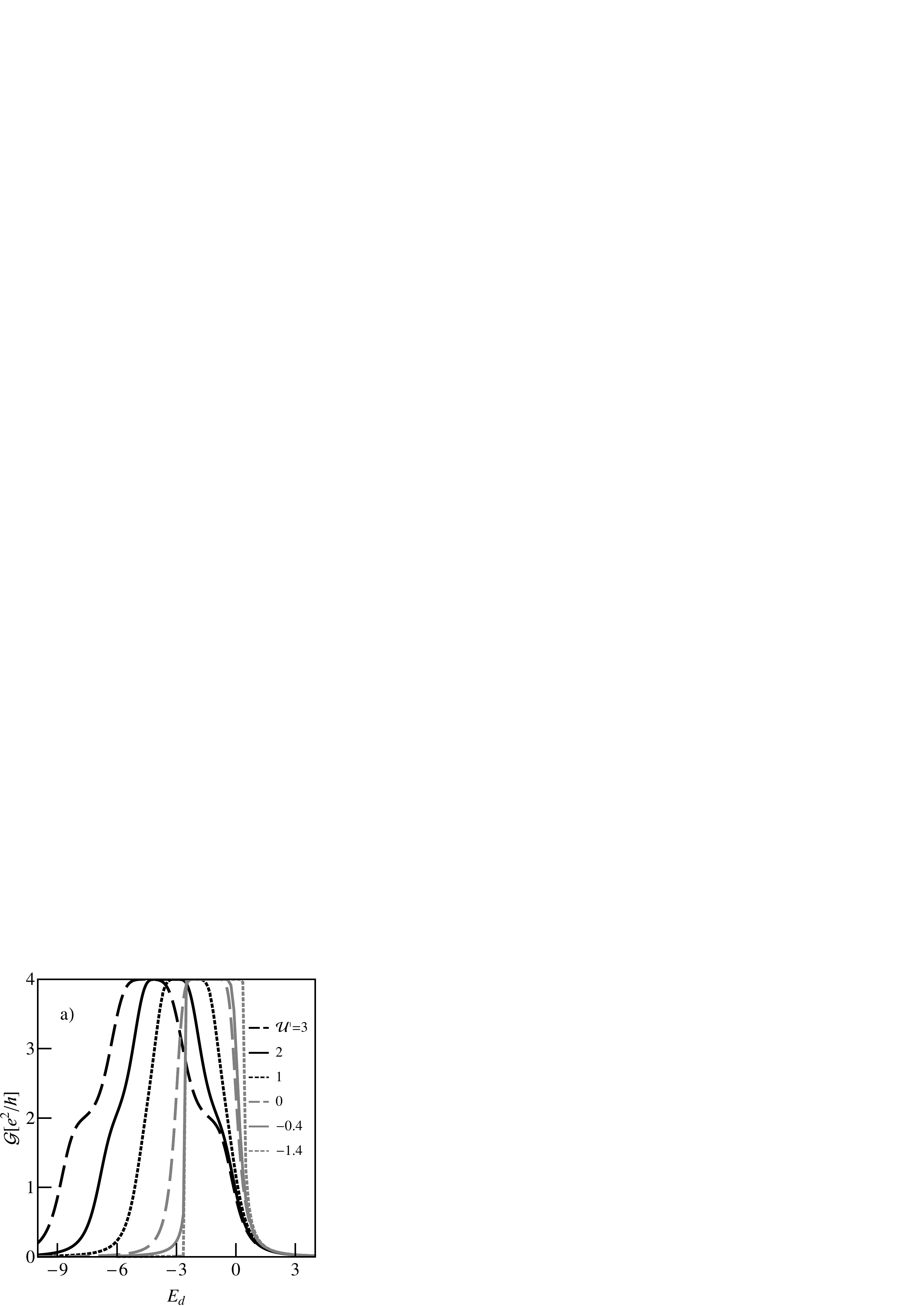}
\includegraphics[width=4.2 cm,bb=0 0 405 410,clip]{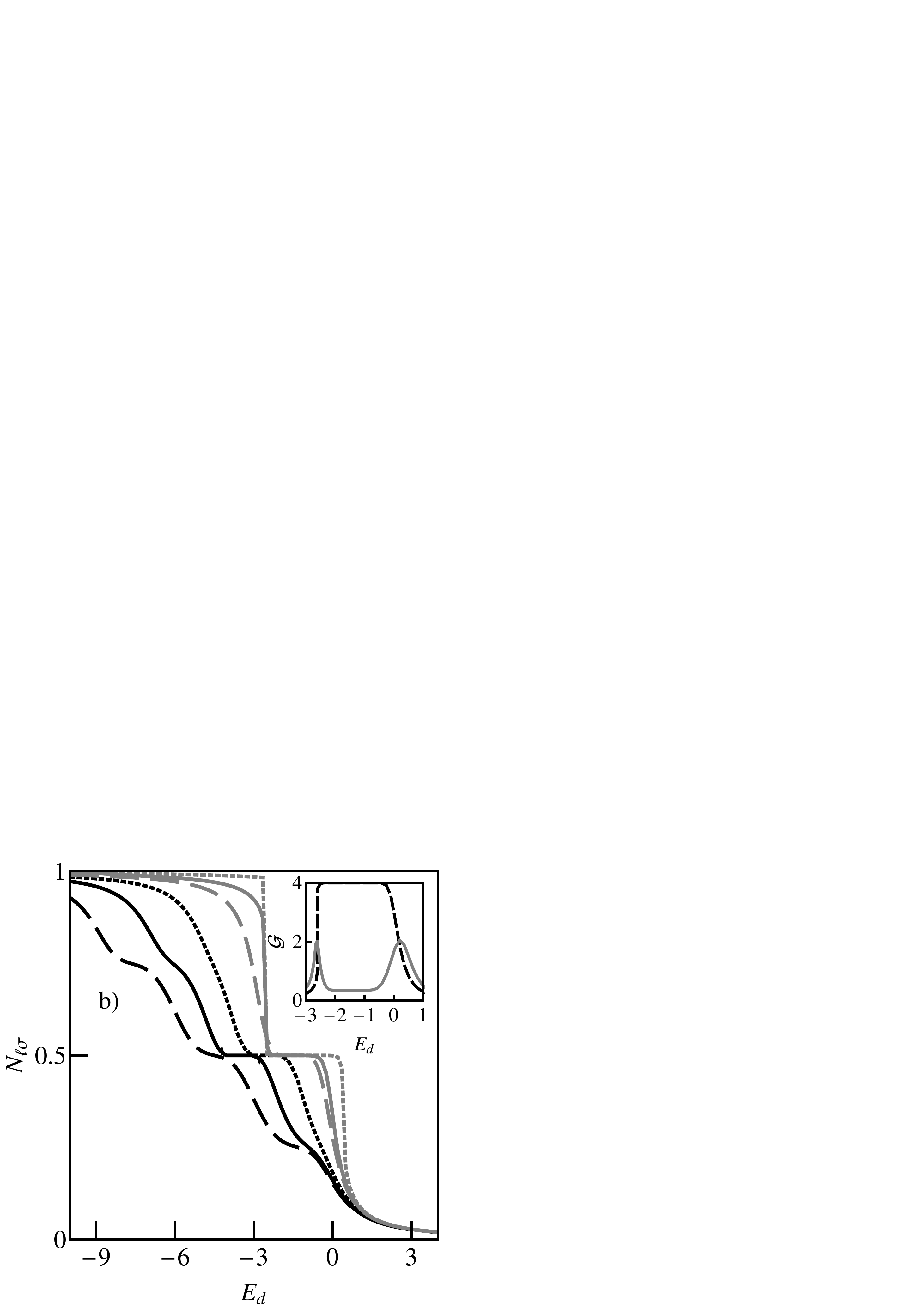}
\caption{\label{fig:epsart} (a) Conductance of SU(2) dots coupled by interdot interaction $\mathcal{U}'$, $t = 0$. (b) Corresponding dot occupations for the same choice of  interdot interaction parameters as in Fig (a). Inset - conductance for $\mathcal{U}' = -0.4$ and $t =0, 0.05$.}
\end{figure}
The  attractive interaction can be understood as the effect of coupling with phonons or other boson excitations which  allows to overcome  Coulomb repulsion ~\cite{Alexandrov}. The considerations for negative $\mathcal{U'}$ are addressed mainly to molecular systems.  In order not to prolong the discussion, we restrict in this section to the case of SU(2) symmetry,  the effects in SU(4) case are similar.   Fig. 6a presents the evolution of  conductance with the change of the  dot-dot interaction  and Fig. 6b the corresponding gate dependencies of occupations.  $\mathcal{U'}>0$ corresponds to the effective deepening of the dot energy and $\mathcal{U'}<0$ shifts $E_{d}$ closer to $E_{F}$.  In consequence, in the former case  the unitary Kondo range narrows and shifts to lower gate voltages and mixed valence range extends. Special attention requires the case of $\mathcal{U'} = \mathcal{U}$, where system reaches higher symmetry - SU(4), charge and spin degrees of freedom become entangled and both  spin and  charge pseudospin fluctuations are active in the formation of Kondo resonance. This case has been  already discussed in this paper (compare Fig. 3a for $t = 0$). For attractive interaction ($\mathcal{U'}<0$) the Kondo ranges extend and move towards shallower levels and transitions from MV to Kondo states sharpen. The asymmetric shape of gate voltage dependence of conductance is also seen for coupled dots, where the central Kondo plateau is suppressed, but  two conductance peaks at MV borders have in opposite to cases presented in Fig. 2a, different widths (inset of Fig. 6b).

\section{Separate electrodes versus common electrodes}

Predominant part of the discussion carried out in this article applies to the case of  separated electrodes  ($\Gamma_{i,j \alpha } = \Gamma_{i,\alpha}\delta_{i,j} = \Gamma_{\alpha}$, $i, j = 1, 2$, $\alpha = L, R$) and we assume  $\Gamma_{L} = \Gamma_{R} = \Gamma$.   In this section we discuss for comparison how transport of strongly correlated  PDQD is modified by the change of interference conditions caused by mixing of the electrode channels. We consider the case when matrix of hybridization  is nondiagonal in dot indices.
\begin{figure}
\includegraphics[width=4 cm,bb=0 0 405 410,clip]{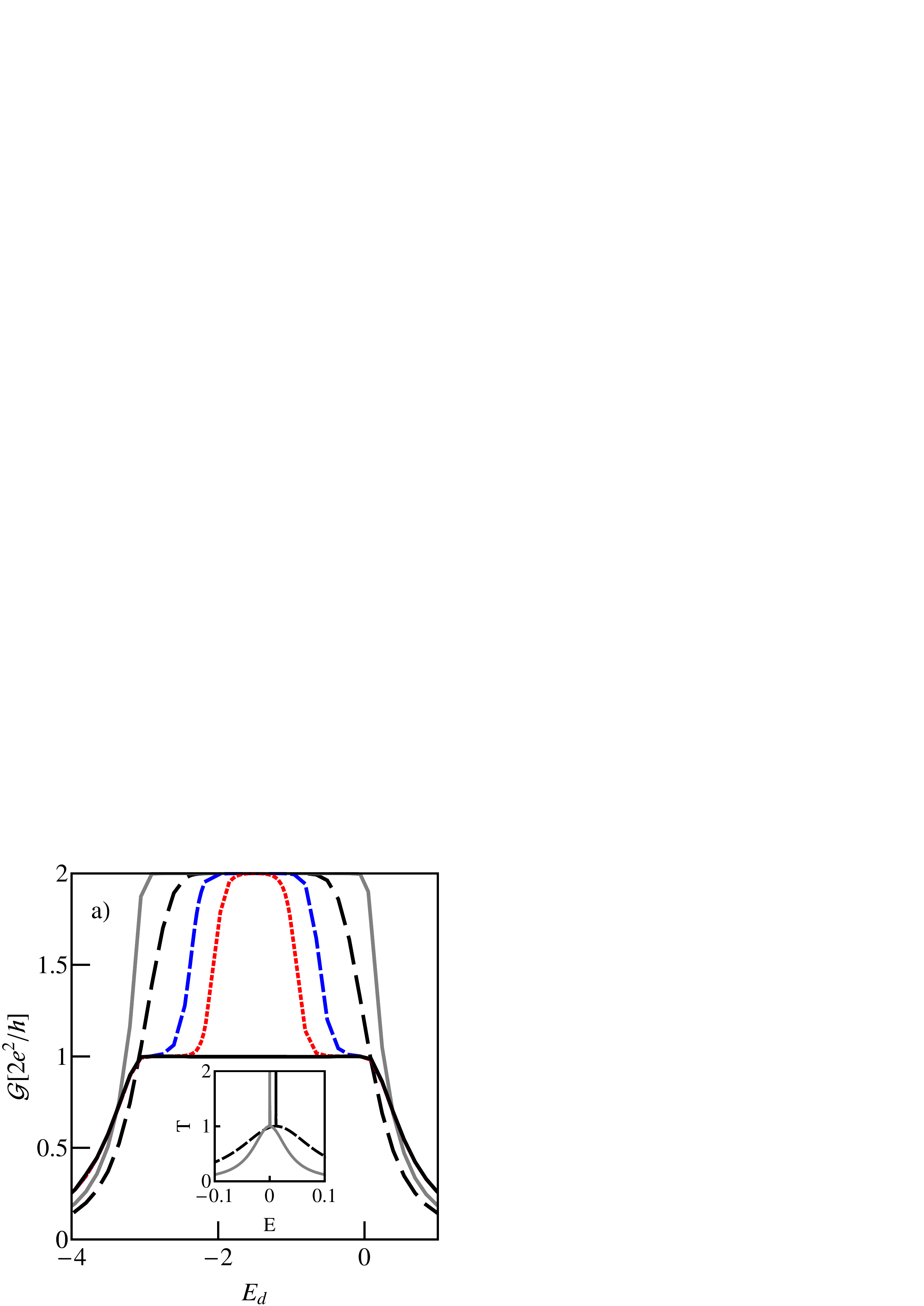}
\includegraphics[width=4 cm,bb=0 0 405 410,clip]{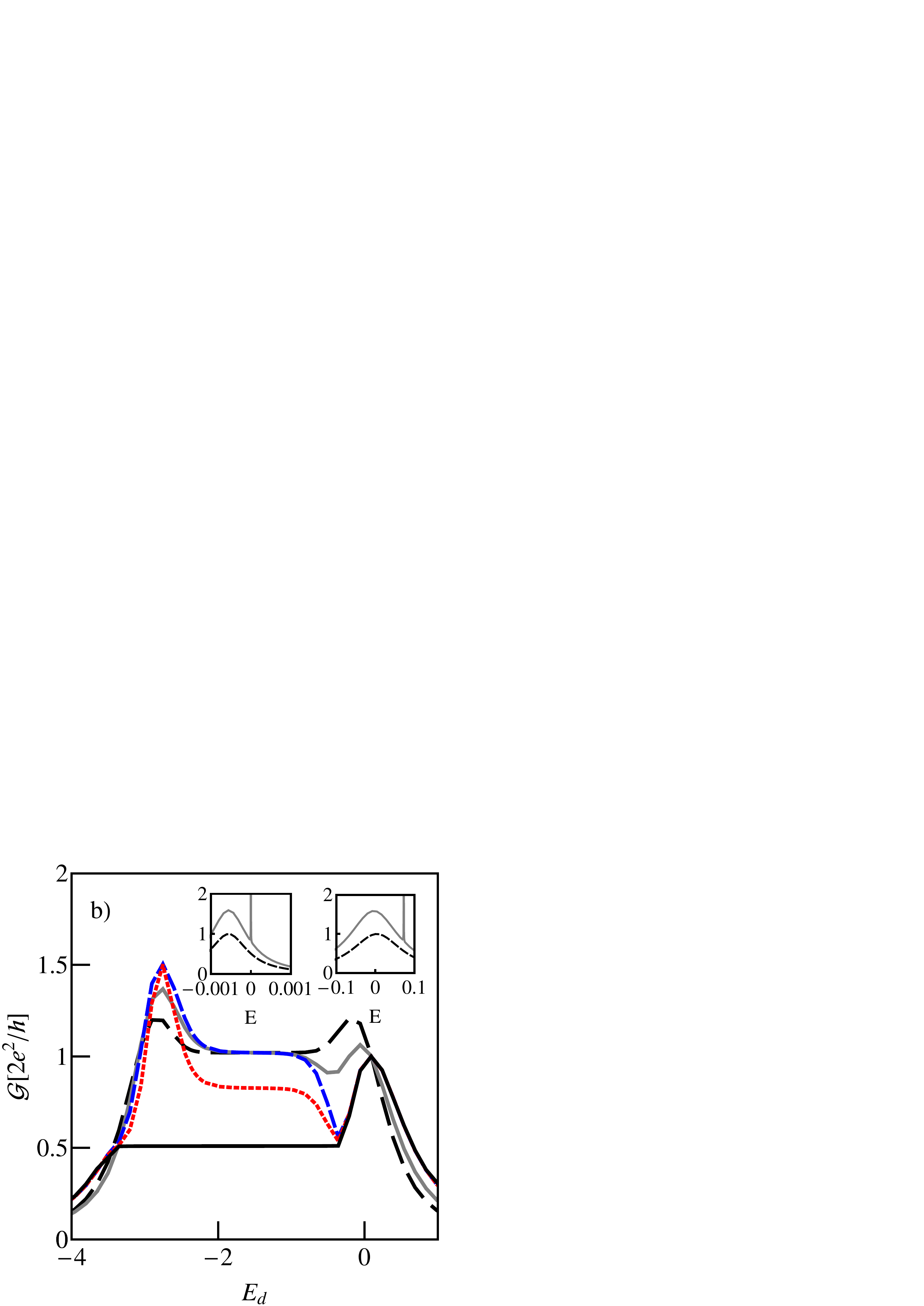}
\caption{\label{fig:epsart} (Color on line)  Conductances of  SU(2) PDQD system with hybridizations of each of the dots to both pairs of electrodes (a)  ($t = 0$) - plots for different strengths of  the off-diagonal hybridization: $q = 0$ (grey solid line), $q = 0.99$ (broken black), $q = 0.999$ (broken blue), $q = 0.9999$ (dotted red ) and  $q=1$ (solid black).  Inset shows transmission for $t=0$, $q = 0.999$ and $E_{d} = -\mathcal{U}/2, 0$   (b) Conductance  of tunnel coupled  dots  ($t = 0.05$) for $q = 0$ (broken black), $q = 0.5$ (grey solid line), $q = 0.99$ (broken blue), $q = 0.994$ (dotted red) and $q = 1$ (solid black).  Insets show DOS and transmissions for completely symmetric case of common electrodes ($q =1$), $t = 0.05$ for $E_{d} = -\mathcal{U}/2$ (left inset) and $E_{d}  = 0$ (right).}
\end{figure}
The off-diagonal elements are usually taken in the form $\Gamma_{12}=\sqrt{\Gamma_{1}\Gamma_{2}}=\Gamma$ ~\cite{Nahm}.  The case of common electrodes corresponds to  equal diagonal and off-diagonal couplings,  we discuss also the case of reduced off-diagonal coupling $\Gamma_{12}=q\Gamma$, $q\leq1$ ~\cite{Trocha3}.  Fig. 7a presents conductance of two SU(2) dots for vanishing direct tunneling ($t=0$) plotted  for different mixing parameters $q$. Completely symmetrical parallel configuration i.e. the case of common reservoirs is described by  $q = 1$ and the case of separate electrodes by $q = 0$. In the latter situation  the Kondo processes take place for each of the dots independently and the total conductance of PDQD doubles the conductance of the single dot.
\begin{figure}
\includegraphics[width=4 cm,bb=0 0 410 420,clip]{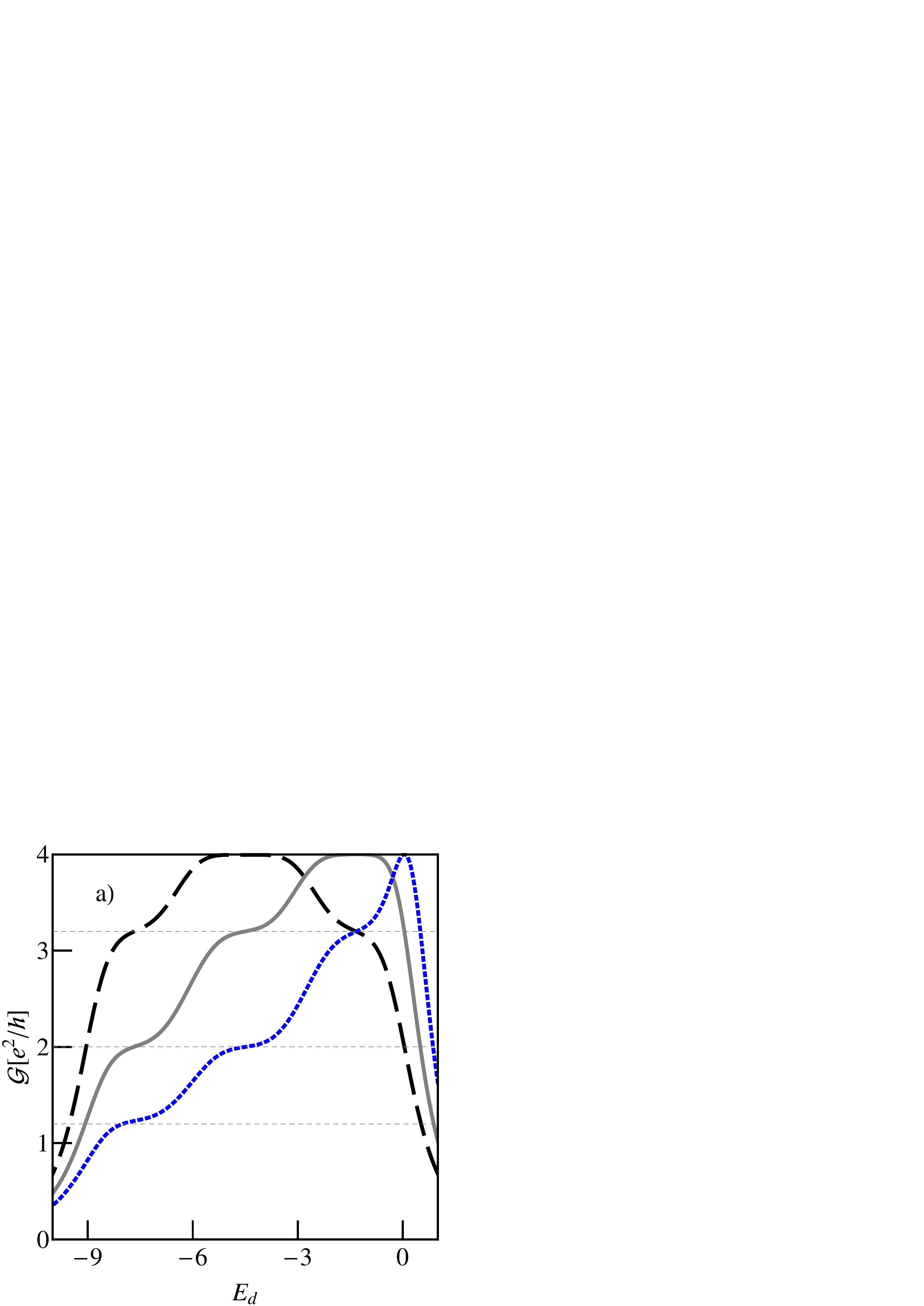}
\includegraphics[width=4.2 cm,bb=0 0 405 400,clip]{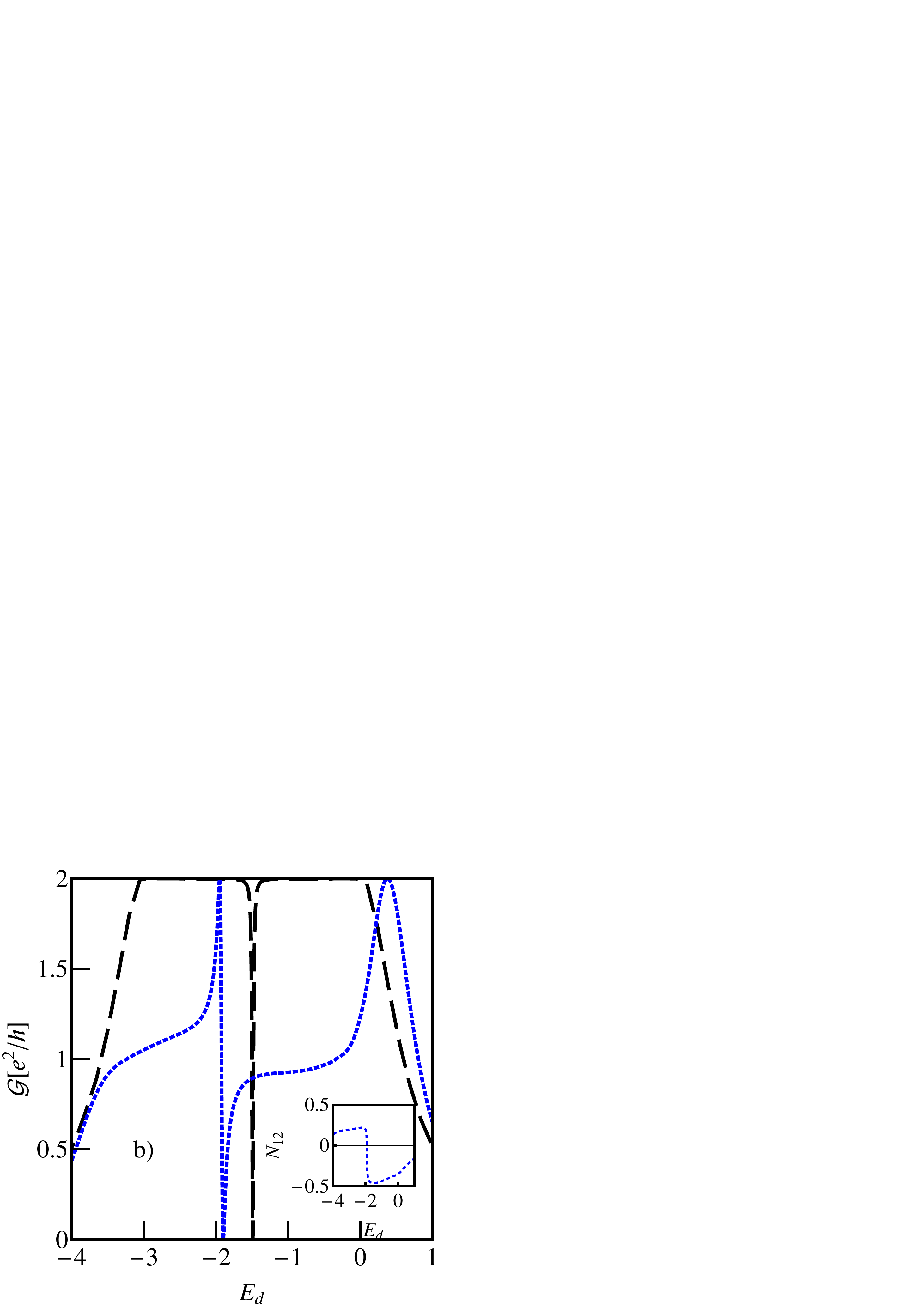}
\caption{\label{fig:epsart} (Color on line) (a) Conductance of SU(4) PDQD with common electrodes ($q =1$) for three values of interdot tunneling: $t = 0$ (broken black line), $t = 0.002$ (solid grey) and $t = 0.05$ (dotted  blue).  (b) Conductance of  SU(2) PDQD  with different dot site energies $\Delta E = E_{1} - E_{2}  = 0.2$  plotted vs. average energy of the dots for $t = 0$ (black dotted line ) and $t = 0.05$ (dotted blue).  Inset shows interdot correlator  $N_{12} = \langle d^{+}_{1\sigma}d_{2\sigma}\rangle$.}
\end{figure}
For $q\neq0$,  in addition to the direct cotunneling processes also the indirect interdot tunneling via the states of electrodes comes into play. Interplay of interference and  many-body processes reflects in this case  in the  increase of regions of suppressed conductance.  In the unitary range conductance of the dots attached to common electrodes ($q=1$) becomes doubly suppressed compared to the case of separates electrodes ($q=0$). Inset of Fig. 7a shows transmissions for  heavily mixed states of electrodes ($q =0.999$). The lines are the sums of two peaks, the broad coherent bonding  Kondo resonance characterized by Lorentzian shape and narrow antibonding Kondo resonance. For $E_{d} = -\mathcal{U}/2$ both are located at $E = 0$ and unitary limit of conductance is preserved for $q\neq1$, for other gate voltages the peaks are shifted from $E = 0$, which results in depression of conductance.  The spectral structure results from the constructive and destructive interference processes for electrons transmitted through bonding and antibonding channels. In the limit  of $q = 1$ the antibonding  resonance does not contribute to transmission, but is still visible in density of states as Dirac $\delta$ peak (inset of Fig. 7b). This resonance is totally decoupled from the leads, which results in the  double suppression of conductance. Fig. 7b presents conductance for tunnel coupled dots. The gate dependencies of conductance curves  become asymmetric  with respect to $E_{d} = -\mathcal{U}/2$.
\begin{figure}
\includegraphics[width=4.4 cm,bb=0 0 405 350,clip]{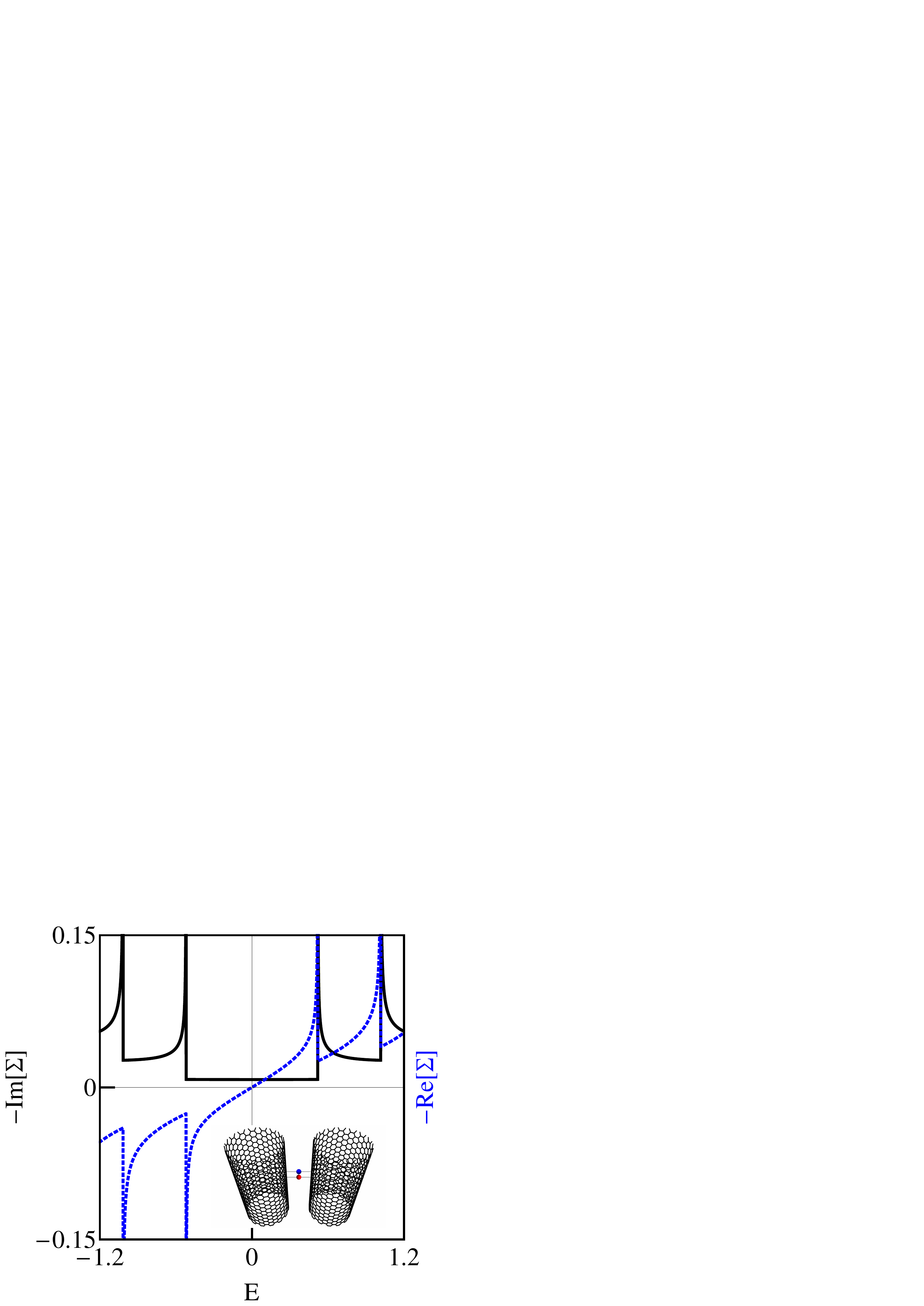} \hspace{4cm}
\caption{\label{fig:epsart} (Color on line) Hybridization of  the dot coupled to armchair carbon nanotube $C = (15,15)$.  Inset is schematic view of  tunnel  coupled dots or impurities placed in the top positions above carbon atoms from the same graphene sublattice.}
\end{figure}
Upon increasing direct tunneling between the dots the many body  peaks move away from each other. The  bonding many-body resonance locates at  $\widetilde{E}_{d} + \widetilde{t}$  and antibonding peak  at  $\widetilde{E}_{d} - \widetilde{t}$, where $\widetilde{E}_{d}$ and $\widetilde{t}$ denote renormalized (SB) site energy and hopping  (inset of Fig. 7b). Around $E_{d}\approx-\mathcal{U}/2$ the conductance is suppressed and the observed plateau structure corresponds to the contribution of bonding Kondo resonance. For $q = 1$ only bonding  resonance contributes to transmission and consequently conductance is double suppressed in this case. For $q < 1$  antibonding resonance is not totally decoupled from the reservoirs. The corresponding conductance   exhibits the  two peak structure. The higher peak reflects crossing the Fermi level by  bonding resonance and lower by antibonding. Fig. 8a presents linear conductance for $2\times$SU(4) system with interdot tunneling and common electrodes attached ($q = 1$). One can roughly visualize the gate dependence of conductance as  a superposition  of  three asymmetric patterns (plateau and a  peak at higher energies) for each regions of occupations. Again responsible for these shapes of conductance are interference induced radical difference of many body bonding and antibonding states and effective decoupling of the antibonding states.

\section{Electrodes with singular electron spectrum}

Commonly used approximation in the description of electronic states of electrodes, adopted also by us so far, is the wide-band approximation, in which details of the electrode band structure are ignored and the electronic energy distribution is assumed to be uniform. However in  real low dimensional systems,  the density of states  may exhibit one or more kinks, commonly referred as Van Hove singularities (VHs) ~\cite{vanHove}.  When gate voltage shifts the Fermi level into one of these singularities Kondo physics changes dramatically and interference conditions are strongly modified.
\begin{figure}
\includegraphics[width=4.8 cm,bb=0 0 460 380,clip]{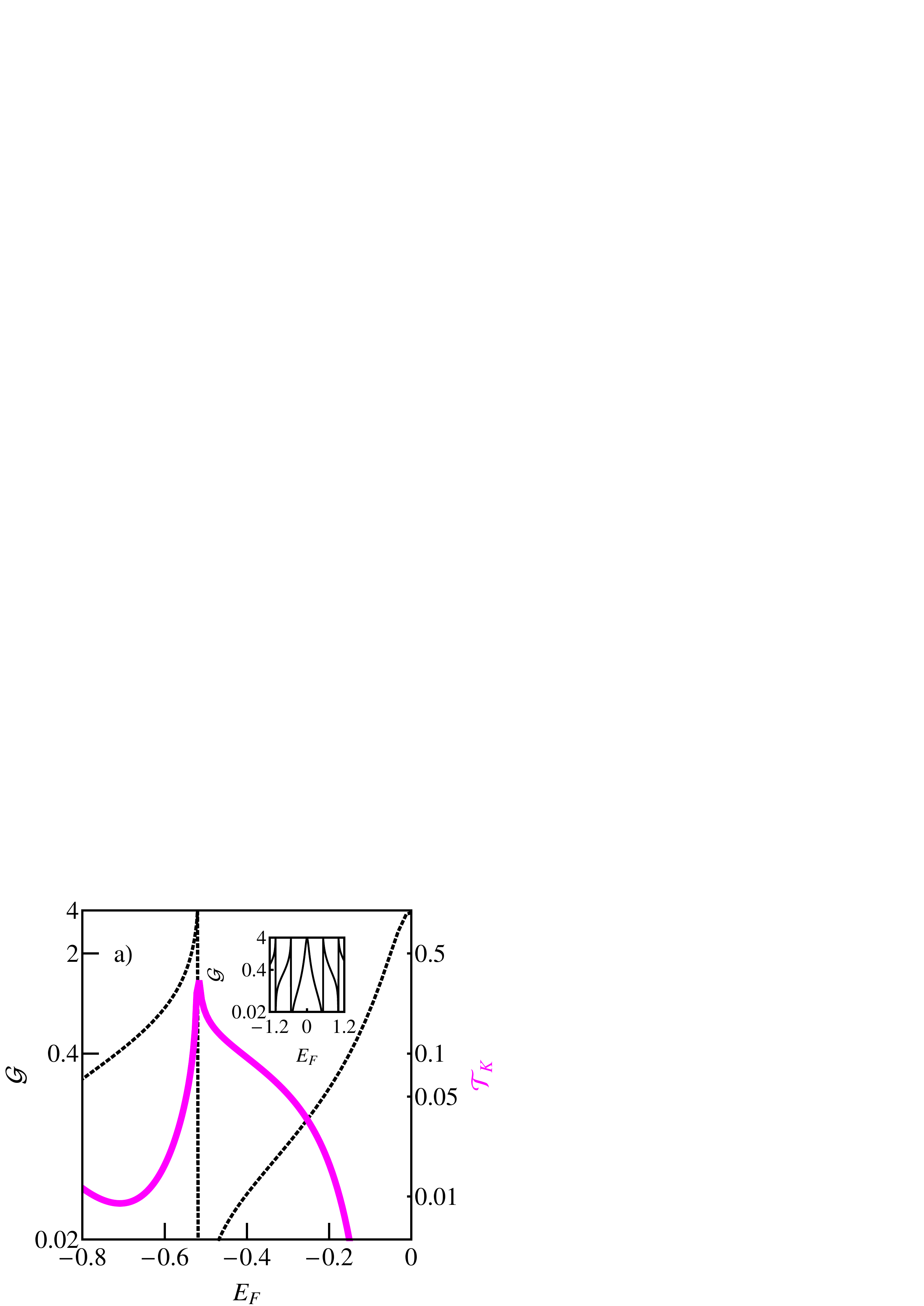}
\includegraphics[width=3.7 cm,bb=0 0 405 420,clip]{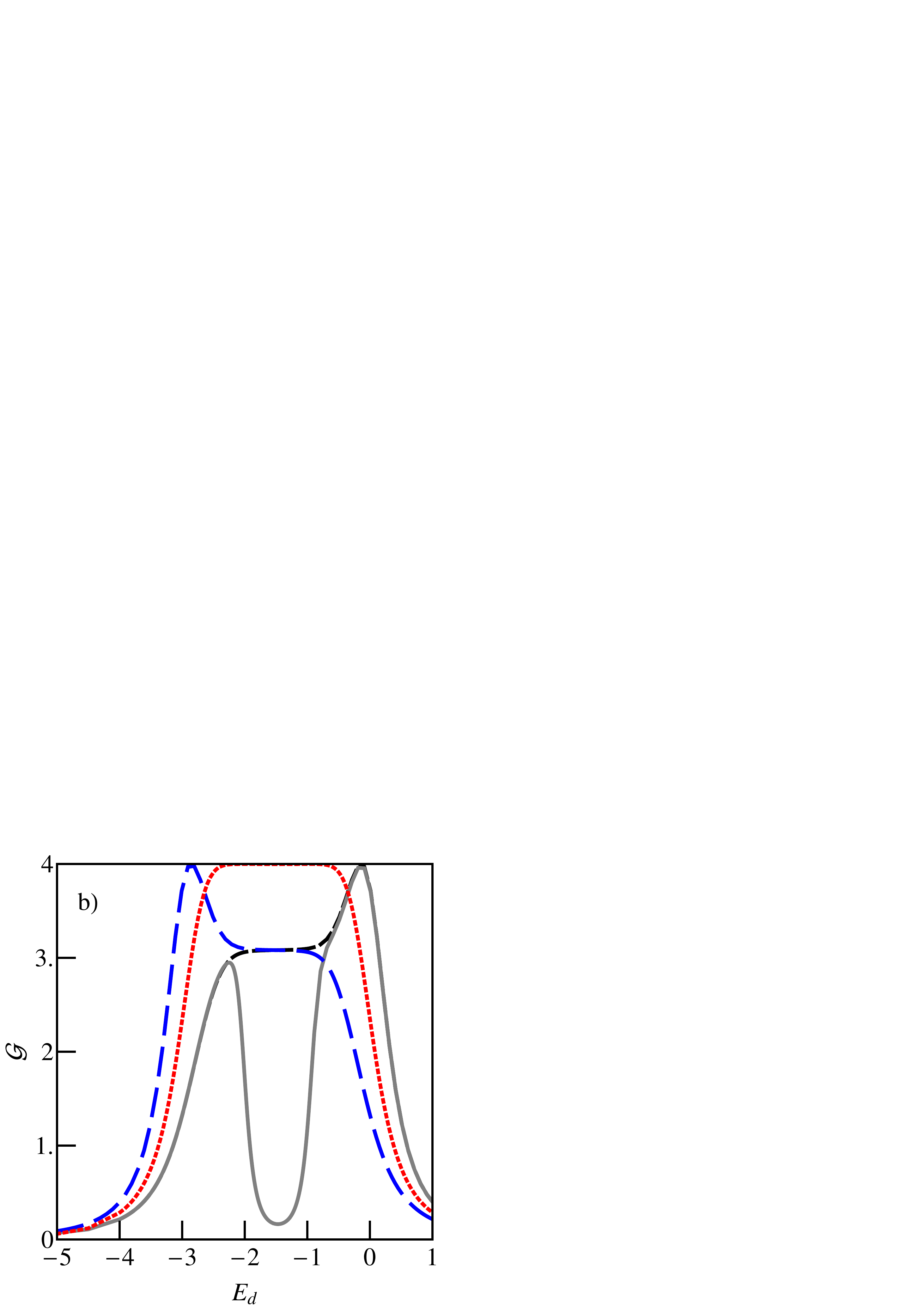}
\includegraphics[width=3.8 cm,bb=0 0 405 420,clip]{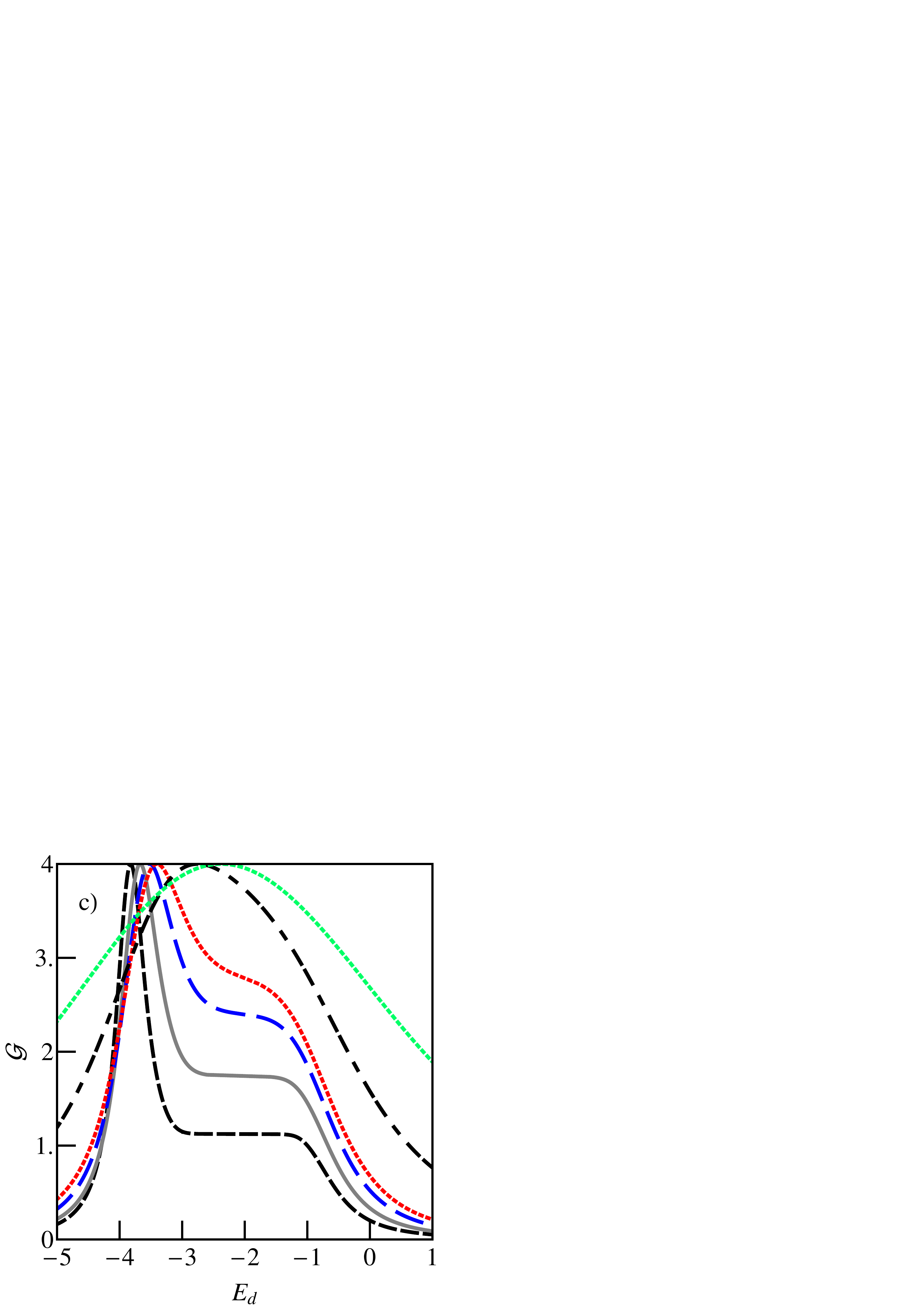} \hspace{4cm}
\caption{\label{fig:epsart} (Color online) (Color on line) (a) Conductance (dotted grey line) and characteristic temperature of many-body resonance (solid pink line) of SU(2) dot coupled to a pair of CNTs $C(15,15)$  presented as a function of position of the Fermi level. Hybridization amplitude $V = 0.5$ and $E_{d}=-1$. Inset show conductance in the extended range encompassing several VH singularities. (b) Gate dependence of  conductance in the range of constant density of states of CNT: $E_{F} = 0$ (red solid line), $E_{F} =-0.003$ (broken blue), $E_{F} = 0.003$ (broken grey) compared with the  conductance  of  tunnel coupled SU(2) dots  (PDQD)  with $t = 0.002$ for $E_{F} = 0.003$ (grey solid line). (c) Gate dependence of conductance close to the  first Van Hove singularity  ($VH_{-1} =-0.519$). Curves plotted for $E_{F} =-0.6$ (short dashed line), $E_{F} = -0.55$ (solid line), $E_{F} =-0.54$ (long dashed line), $E_{F} = -0.53$ (red dotted line), $E_{F} =-0.525$ (double dashed line) and  $E_{F} = - 0.52$ (green dotted line).}
\end{figure}
As the illustrative example of electrodes with singular DOS we discuss carbon nanotubes (CNTs). They exhibit excellent ballistic transport capabilities with mean free paths of order of microns ~\cite{Soldano, Goss}. The numerical results are presented for armchair metallic CNT with chiral vector $\mathbf{C} = (15,15)$ ($\mathbf{C}$ is written in the basis of unit vectors of graphene ~\cite{Guclu}). Graphene  lattice consists of two interpenetrating A and B triangular sublattices. We focus on the case when both impurities are in top positions above the sites from the same sublattice, say A (inset of Fig. 9). The corresponding  diagonal parts of  hybridization function  $\Sigma(E) =  \sum_{kN\alpha\sigma}\frac{|V|^{2}}{E-\varepsilon_{kN\alpha\sigma}}$ are presented in  Fig.  9. The off diagonal parts, which are not  presented  here, are much smaller. Figure 10a shows conductance and characteristic many-body temperature $T_{K}$ as a function of position of the Fermi level. Due to the  symmetries of dot-CNT  hybridization functions ($Im[\Sigma(E)] = Im[\Sigma(-E)]$ and  $Re[\Sigma(E)] = -Re[\Sigma(-E)]$) conductance and gate dependencies of Kondo temperature are also symmetric $\mathcal{G}(E_{d},E_{F}) = \mathcal{G}(-E_{d},-E_{F})$ (inset of Fig. 10a), $T_{K}(E_{d},E_{F}) = T_{K}(-E_{d},-E_{F})$. The  peaks of imaginary parts of hybridization  mean the strong enhancement of effective coupling and consequently the increase of characteristic temperatures of many-body resonances.
\begin{figure}
\includegraphics[width=4.1 cm,bb=0 0 460 430,clip]{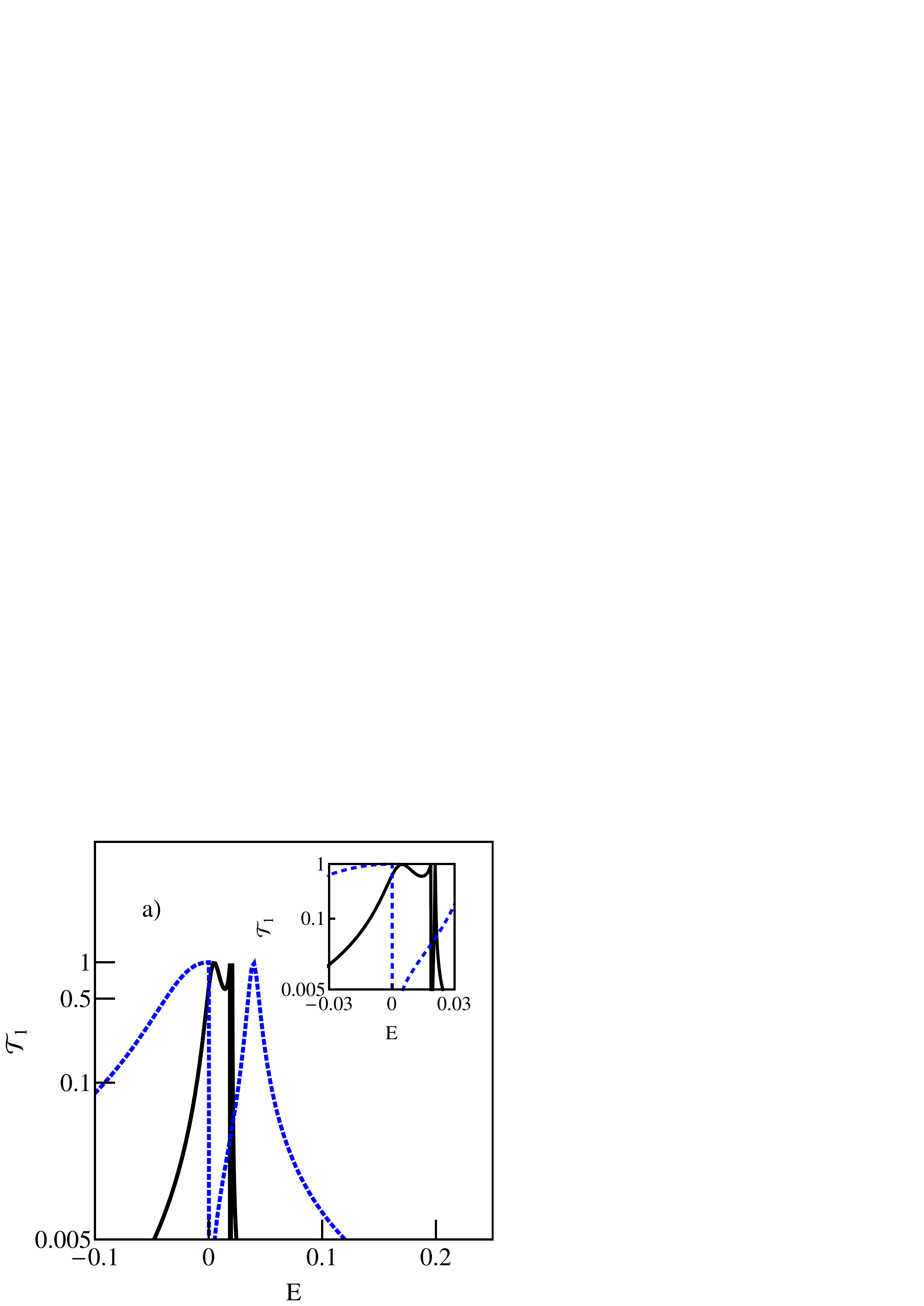}
\includegraphics[width=4.1 cm,bb=0 0 460 420,clip]{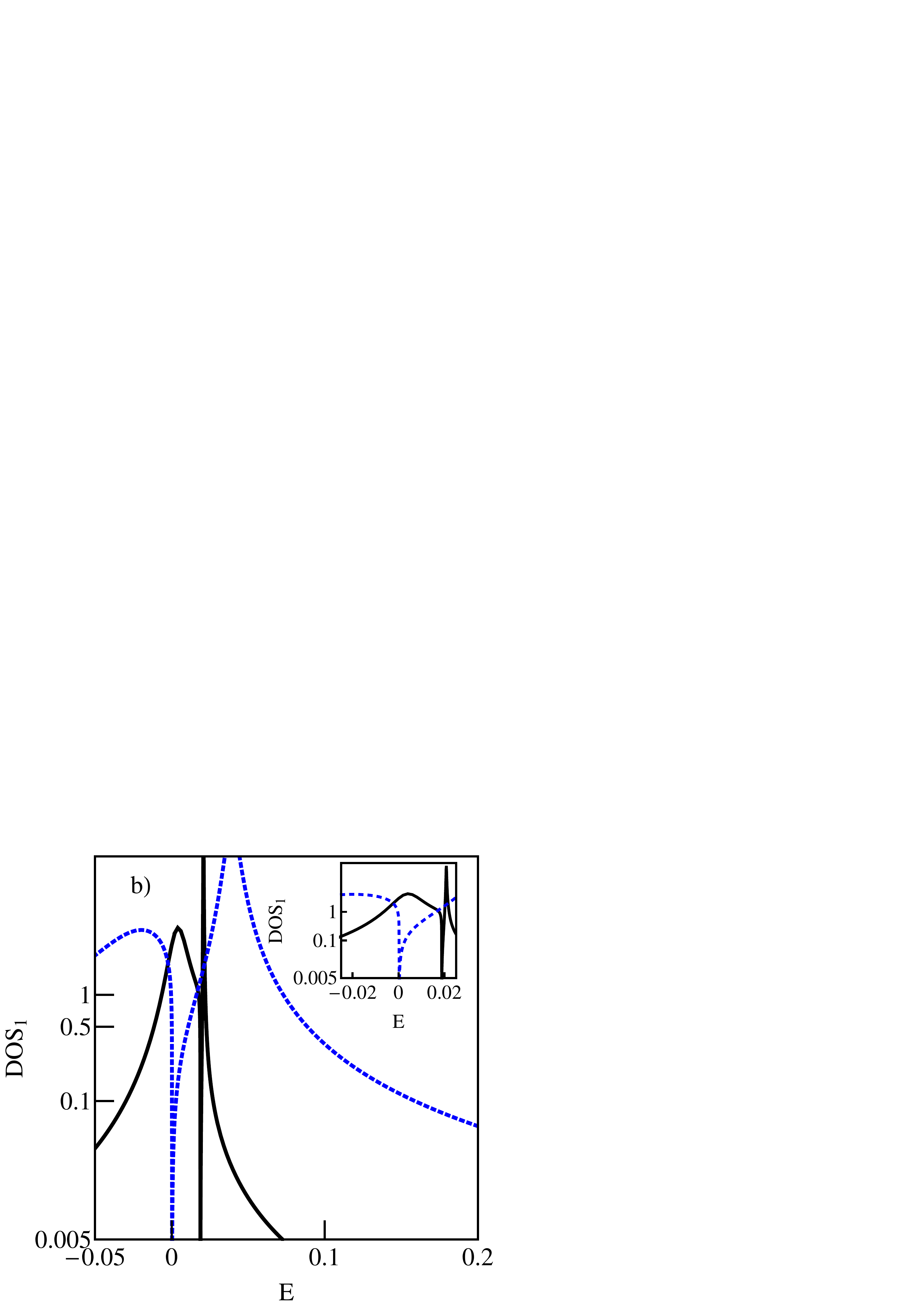}
\includegraphics[width=4.1 cm,bb=0 0 460 430,clip]{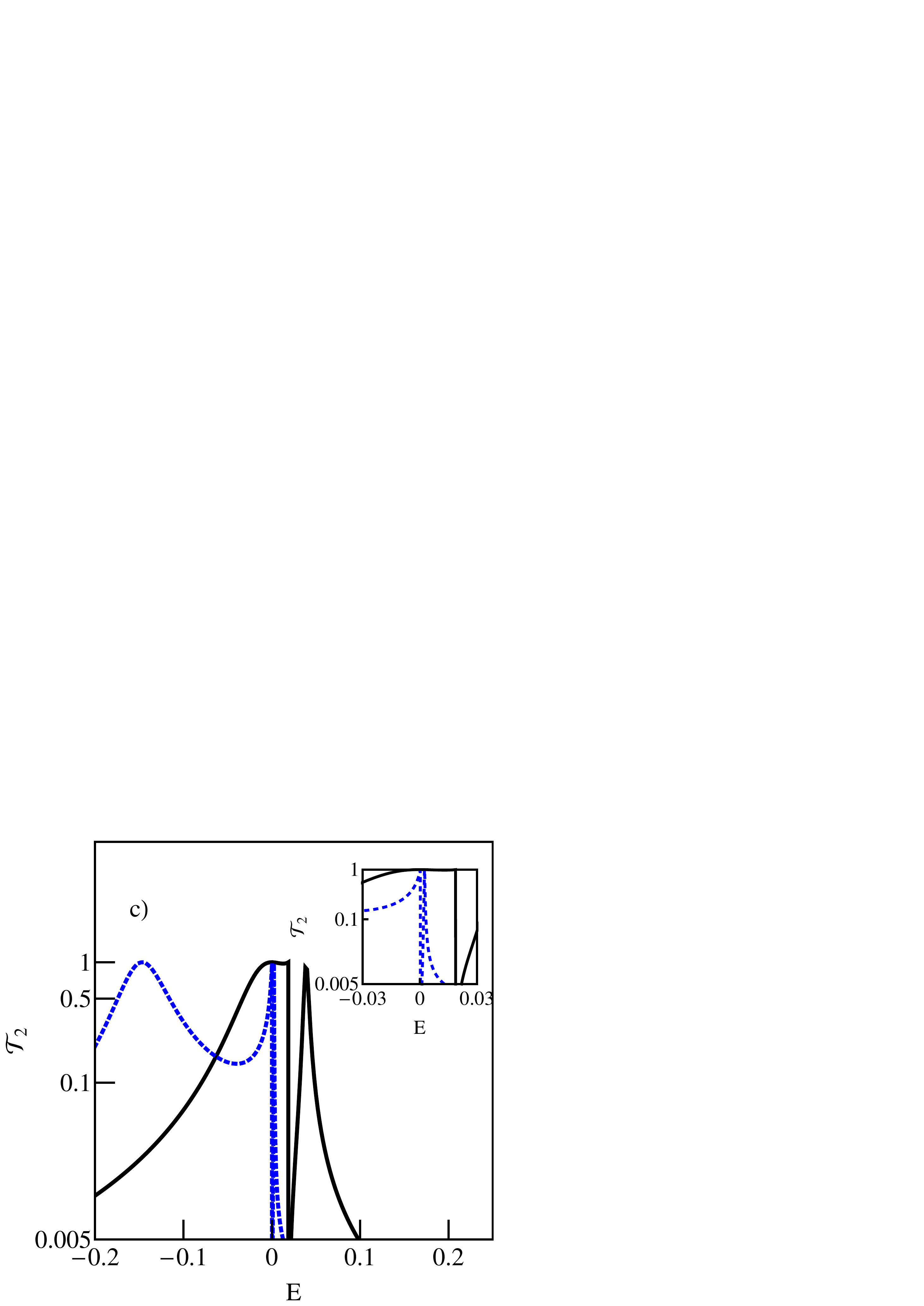}
\includegraphics[width=4.1 cm,bb=0 0 460 420,clip]{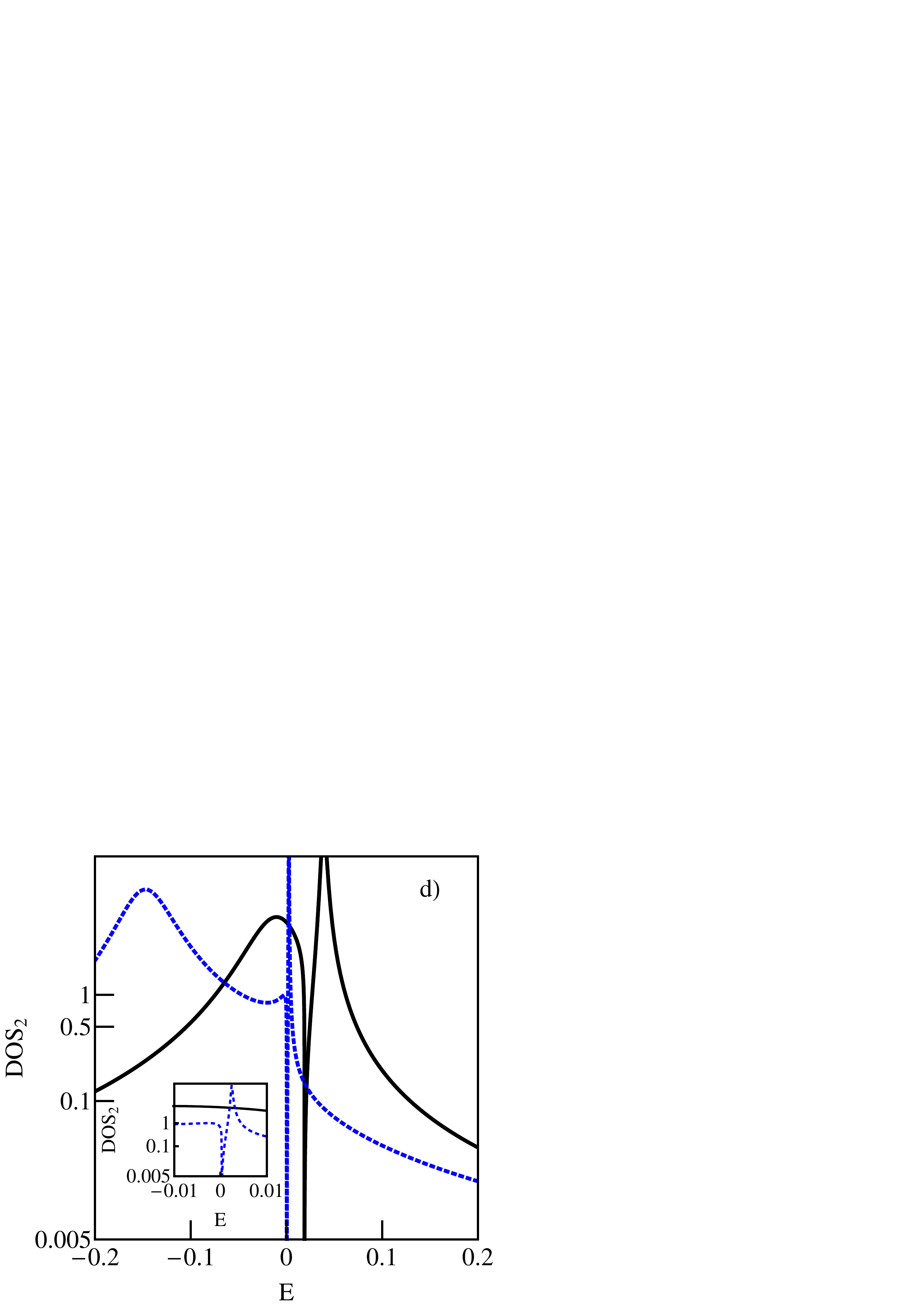}
\caption{\label{fig:epsart} (Color on line) Transmissions and corresponding DOS (right) a,b)for $E_{d}=-3.52$ for $E_{F}$  close to singularity $VH_{-1}$, $E_{F} = -0.54$  (black solid line) and $E_{F} = -0.52$ (blue dashed line). c,d) transmission and DOS for the same  $E_{F}$ values but for $E_{d} = -2.3$.}
\end{figure}
The resonances in the vicinity of VHs take the mixed valence character. When Fermi level enters VHs the dips of conductance are observed. The asymmetric shape of $\Sigma$ around singularities dictates significantly different behavior of conductances and $T_{K}$ on opposite sides of VHs. The gate dependences of linear conductance for different positions of the Fermi level are presented in  Figures 10b, c.   Case around $E_{F} = 0$  (Fig. 10b) reproduces the earlier discussed symmetric dependence for electrodes with constant density of states  with unitary limit around electron - hole symmetry point $E_{d} = -\mathcal{U}/2$ (real part of hybridization vanishes). For $E_{F}\neq0$ this symmetry is broken due to asymmetry of $Re[\Sigma]$. For $E_{F}<0$ lying above the first VH singularity, but not to close to divergence region of DOS the suppressed plateau in the center,  a peak for deep dot levels and reduced conductance at the right edge  are observed (Figs. 10b, c). This behavior results from the shift of transmission peaks towards higher  energies for lower Fermi levels (Fig. 11). Analogous shift towards positive energies reflects in mirror reflection of gate dependence of conductance (inset of Fig. 10a). For small values of  $E_{F}$  i.e. in the region of weak hybridization Kondo effect is easily suppressed even by  small interdot tunneling and conductance plateau around single occupancy of the dot disappears (grey line in Fig. 10b). In the vicinity of singularities densities of the states of the dots are very sensitive to the position of the Fermi level (Fig. 10c). Interesting transmission, which is determined by  both imaginary and real parts of hybridization, does not  directly reflect the shape of DOS  (Fig. 11). Depending on the side $E_{F}$ enters the singularity, whether $Im[\Sigma]$ is  high or low, conductance  maintains  the shape with sharp peak or evolve into broad peak (Fig. 10c).  In close proximity to singularities, however, the presented results should be treated with caution, only as a visualization of tendencies due to a break of applicability of SBMFA in the range of of divergent self-energy ~\cite{Zhuravlev}.

\section{Parallel dots for spintronics}

In this section we present a few examples illustrating  how the unique properties of PDQDs in the strong correlation range modified by the presence of  magnetic electrodes  may be exploited in spintronic devices. First we show how the spin polarization of conductance is transferred between the dots by tunneling.
\begin{figure}
\includegraphics[width=4 cm,bb=0 0 290 300,clip]{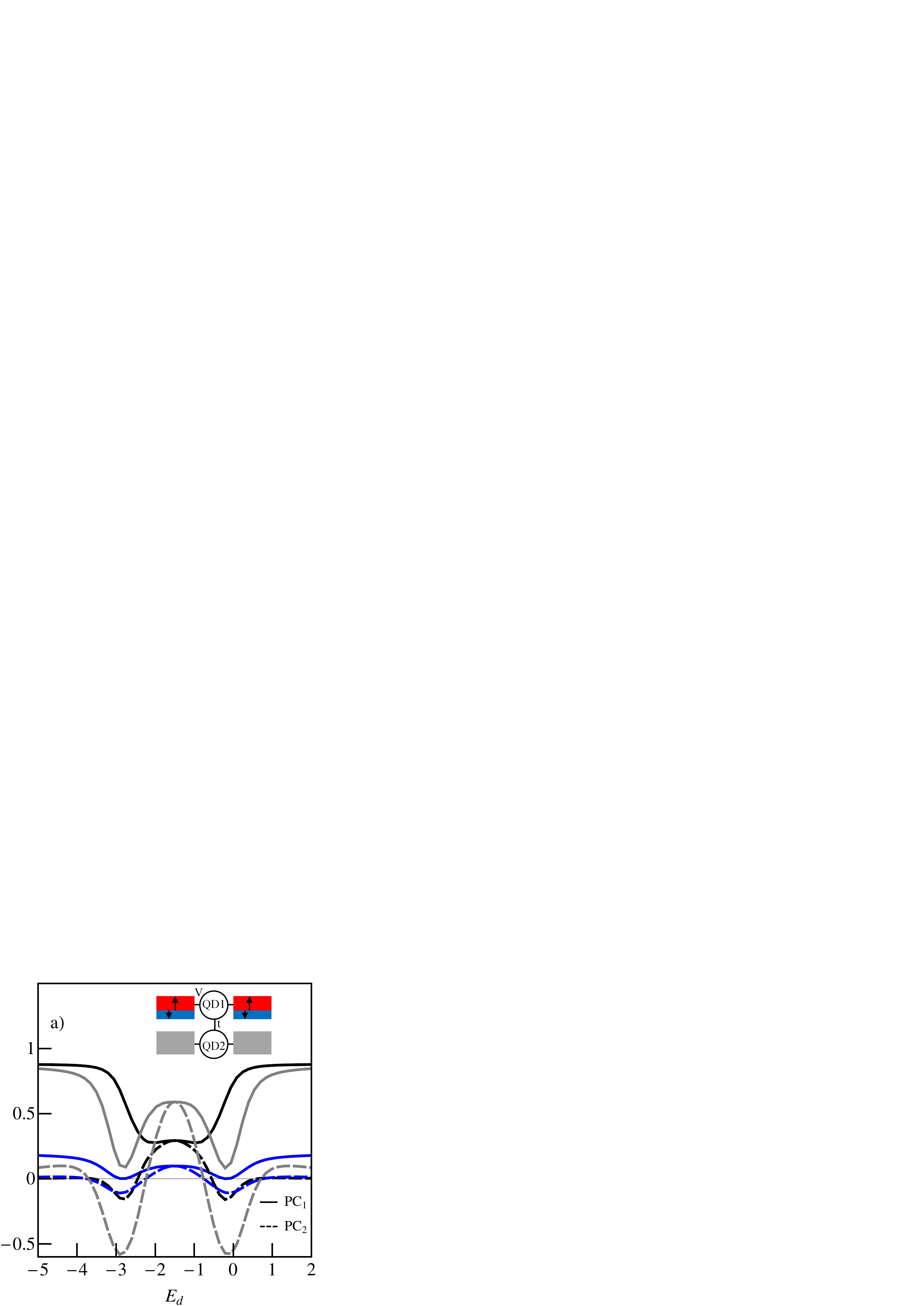}
\includegraphics[width=4.2 cm,bb=0 0 405 390,clip]{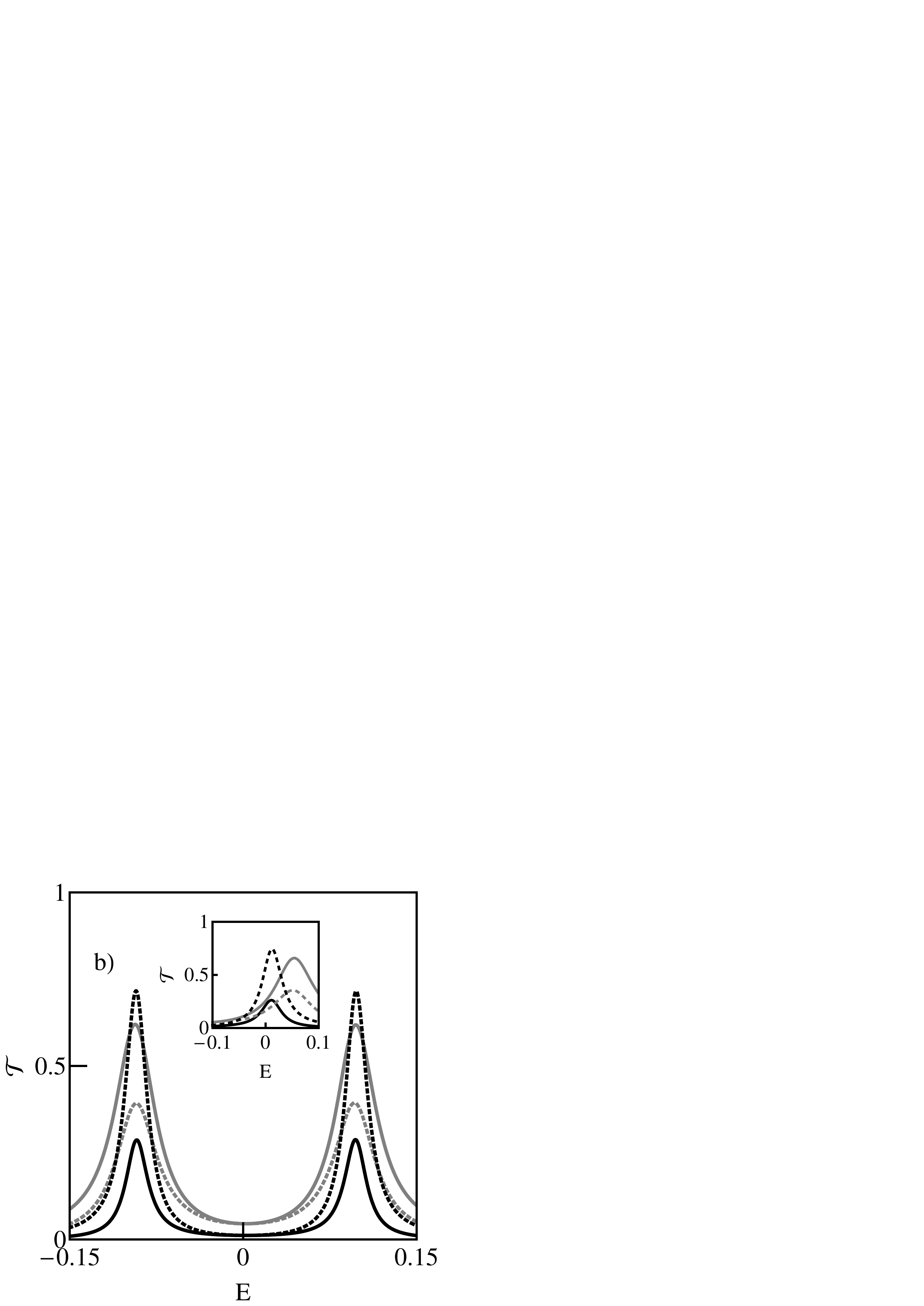}
\includegraphics[width=4.2 cm,bb=0 0 405 380,clip]{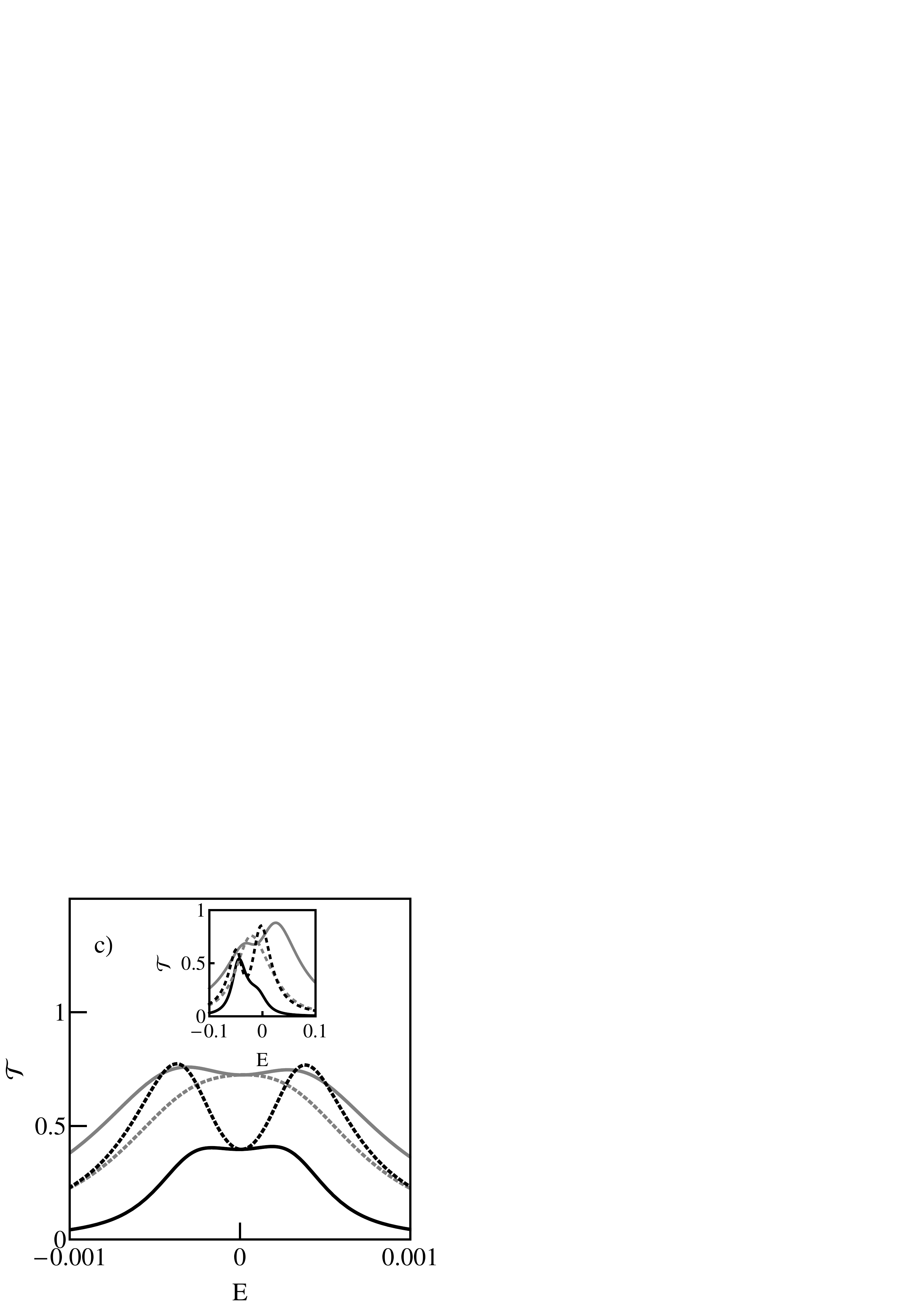} \hspace{4cm}
\caption{\label{fig:epsart} (Color on line)  SU(2) PDQD with upper dot attached to magnetic electrodes and lower to nonmagnetic (inset of Fig. a).  Polarization of conductance  of  the upper dot $PC_{1}$ (solid lines) and polarization of the lower dot $PC_{2}$ (broken lines) for strong tunneling ($t = 0.3$)  presented for   $P =  0.1$ (black) and $P = 0.6$ (grey). For comparison analogous polarization curves are also shown in the weak tunneling range ($t = 0.04$)  for $P = 0.6$ (blue lines). (b,c) Corresponding examples of  spin dependent transmissions (solid/dotted black and gray lines for QD1/QD2) for $P = 0.6$, $E_{d} = -\mathcal{U}/2$ and $E_{d} = \mathcal{U}$ (insets) for $t = 0.3$ (b) and $t = 0.04$ (c)}
\end{figure}
The considered system is presented in  the inset of Fig. 12a, where the upper dot (QD1) is coupled to polarized electrodes and the lower (QD2)  to nonmagnetic. The sign of  the induced spin polarization of conductance of  QD2 can be manipulated by gate voltage. In  the single occupancy region of the dots ($E_{d}$ around $-\mathcal{U}/2$) the induced spin polarization  takes the same sign as polarization of magnetic dot (QD1), this occurs both in weak coupling limit, where Kondo resonance is present and for strong coupling, where Kondo correlations are suppressed.  For the dots almost filled or almost empty  $N_{2}\approx2,0$, the opposite polarization of conductance is observed.
This can be understood looking at the corresponding spin resolved  transmissions (Figs. 12 b, c for strong tunneling and the corresponding insets for weak tunneling). For $E_{d}  = -\mathcal{U}/2$, where for $t = 0$ the many-body  peaks are centered  at $E = 0$  (not presented) sufficiently strong interdot coupling ($t = 0.3$)  splits the peaks and  in accordance with  polarization of the electrodes  majority peaks are higher. At QD2, the peaks, which for $t=0$ are identical  for both spin orientations,  evolve with increasing $t$ into spin dependent and the heights are reversed in comparison to QD1. This is a consequence of indirect contribution to the DOS of QD2 originating from QD1, which is significant in the range of the poles of the Green's functions of the first dot $G^{R}_{1\sigma1\sigma}$. This perturbation  introduces spin polarization opposite  to QD1.
Note that for $E_{d}  = -\mathcal{U}/2$ (Figs. 12b,c) the heights and widths of the transmission peaks become spin dependent, but the interdot  splitting does not depend on spin. For $E_{d}\neq-\mathcal{U}/2$ the splitting becomes spin dependent. Interesting property is that in e-h symmetry point ($E_{d} = -\mathcal{U}/2$, $E_{F} = 0$), the induced  polarization at QD2 is identical to polarization at QD1 independent of the strength of  coupling and polarization.  By changing the gate voltage the spin preference of QD2 might change, the center  of mass of  many body peaks move away from $E = 0$. For $E_{d} = -\mathcal{U}$ (insets of Figs. 12b, c) the shift of transmission spectrum towards lower energies is observed and opposite polarization of conductance results  (Fig. 12a).  As the next problem let us discuss tunnel magnetoresistance.  To examine this effect  the case with polarized  electrodes attached to both of the dots is considered.  This configuration allows the control of current based on  the dependence of conductance on the relative orientation of magnetic moments of the leads  - PDQD spin valve. We discuss the case, where instead of commonly used left and right spin asymmetry  the relative orientation of the upper and lower electrodes is changed. Tunnel magnetoresistance is then defined as  the difference in resistance between antiparallel and parallel arrangement of polarizations of upper and lower electrodes ($TMR =(\mathcal{G}_{\uparrow\uparrow}-\mathcal{G}_{\uparrow\downarrow})/\mathcal{G}_{\uparrow\downarrow}$, where $\mathcal{G}_{\uparrow\uparrow(\uparrow\downarrow)}$ are the total conductances for the corresponding magnetization configurations).
As was discussed for the case of unpolarized electrodes, in sec. IIIa, by changing the gate voltage one can move the system from half filling with the Kondo state (weak interdot coupling) or suppressed Kondo state (strong interdot coupling) to MV state for almost empty or filled dots. For PDQD with polarized electrodes positive, but suppressed TMR is observed around e-h symmetry point for weak coupling and enhanced for strong coupling. In the latter case it reaches Julli\`{e}re nonresonant limit $2P^2/(1-P^2)$ ~\cite{Juliere}. As it is illustrated in  the conductance dependences for parallel and antiparallel orientations of electrodes (Fig. 13b) this is a consequence of stronger suppression of Kondo correlations with the increase of interdot coupling for AP configuration than for P orientation. Outside single electron occupations of the dots $TMR$ changes sign (inverse $TMR$) and achieves large values for strong interdot coupling.
\begin{figure}
\includegraphics[width=4.3 cm,bb=0 0 405 390,clip]{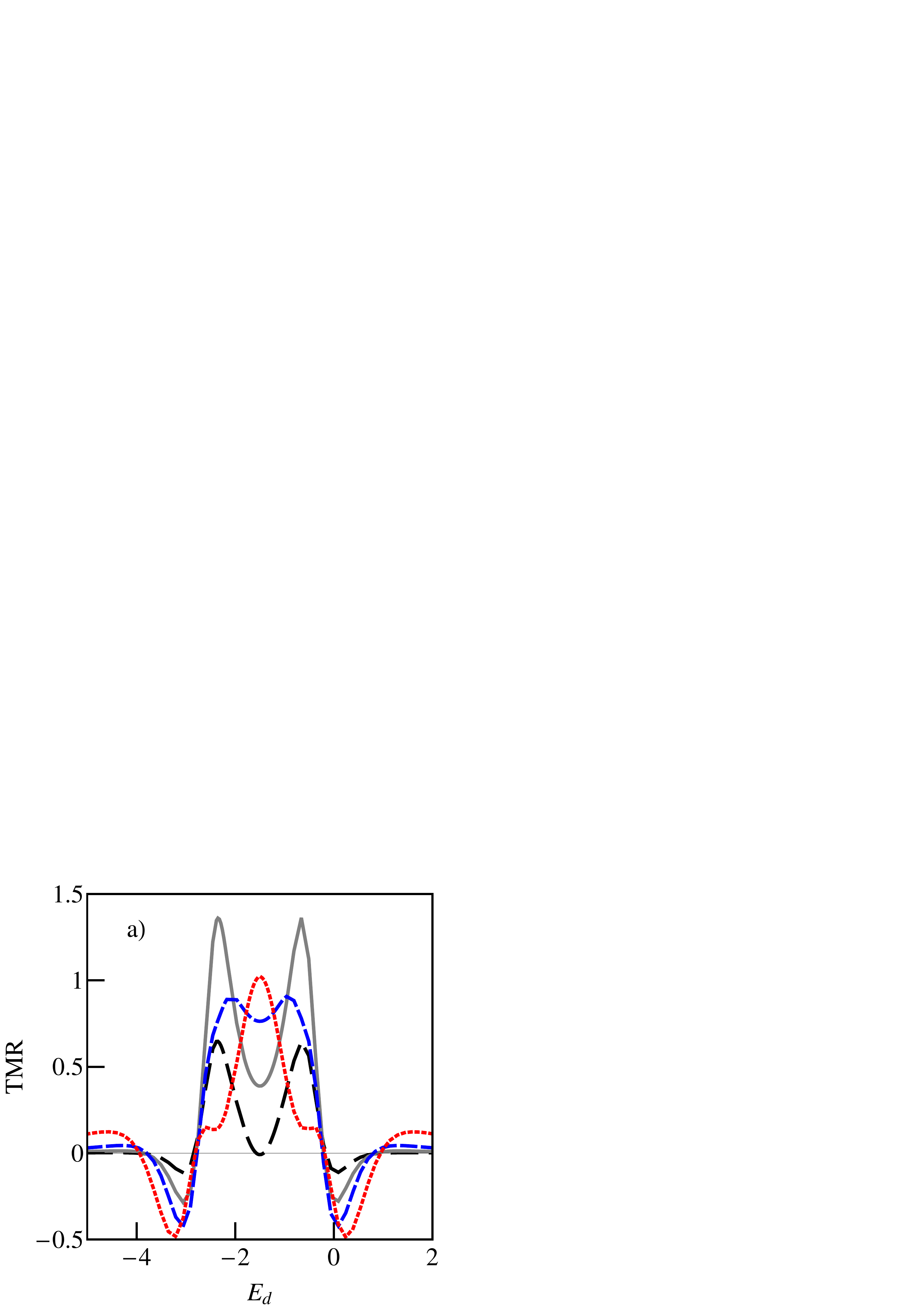}
\includegraphics[width=4.1 cm,bb=0 0 405 400,clip]{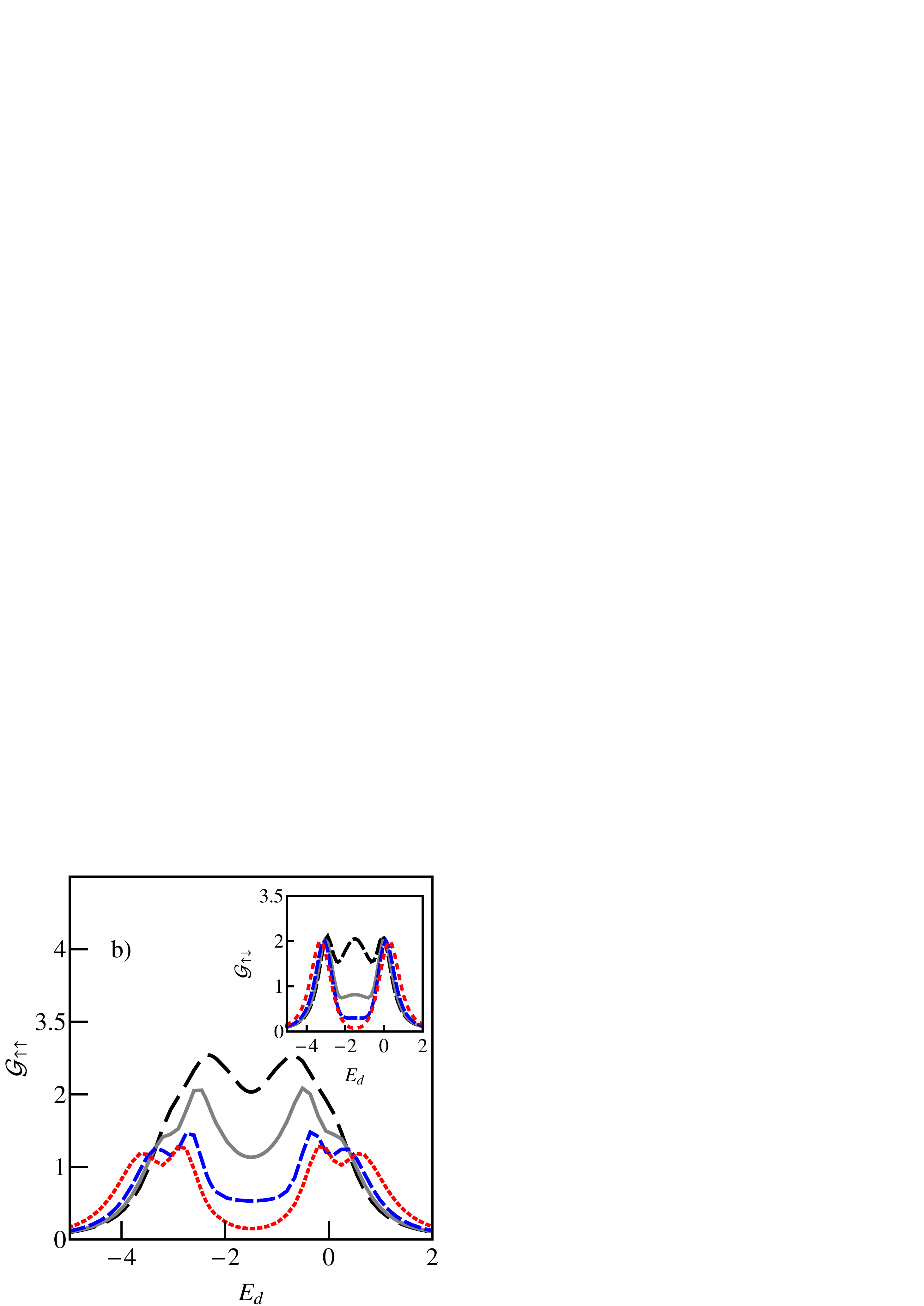}
\includegraphics[width=4.1 cm,bb=0 0 405 400,clip]{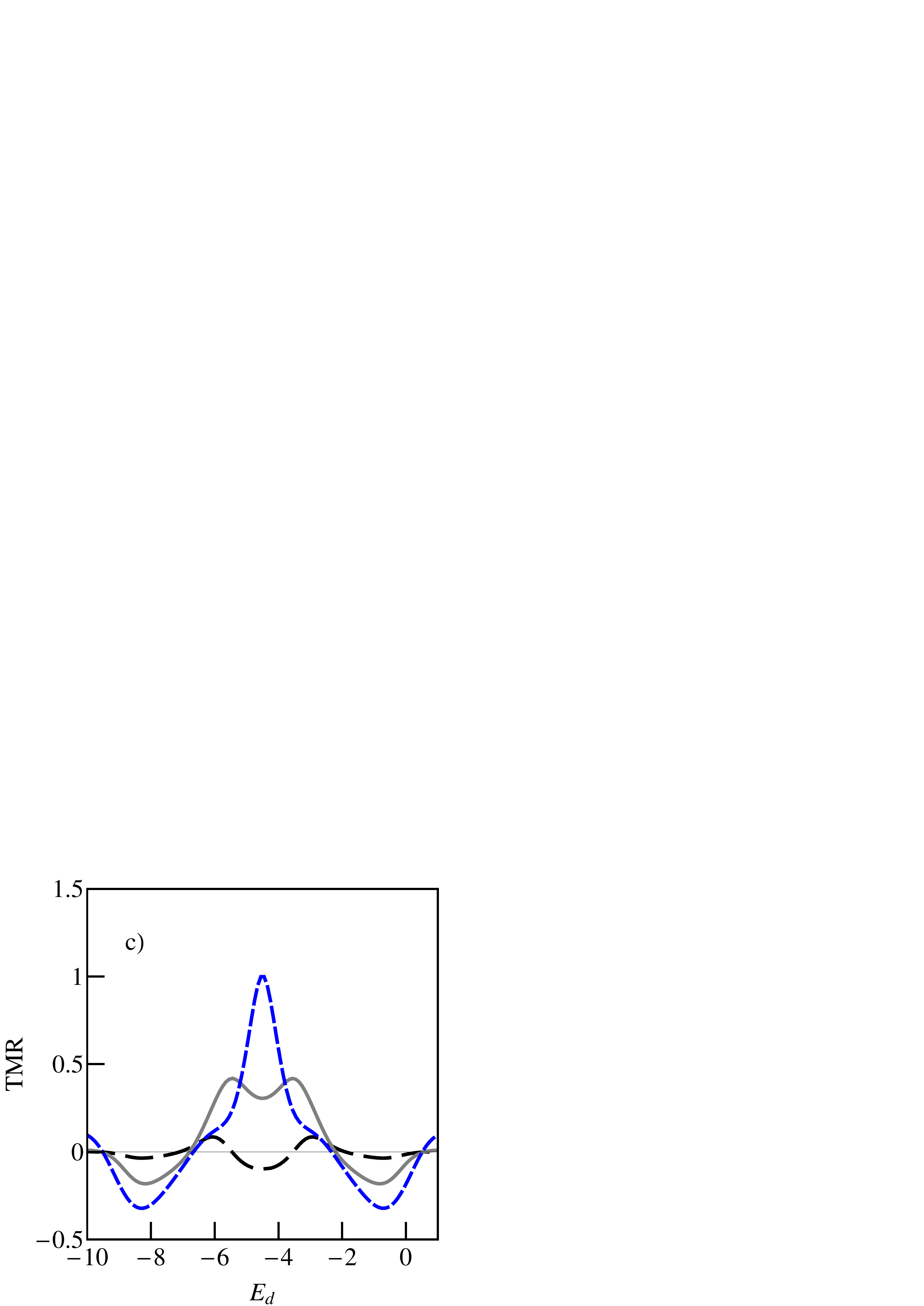}
\includegraphics[width=3.8 cm,bb=0 0 405 420,clip]{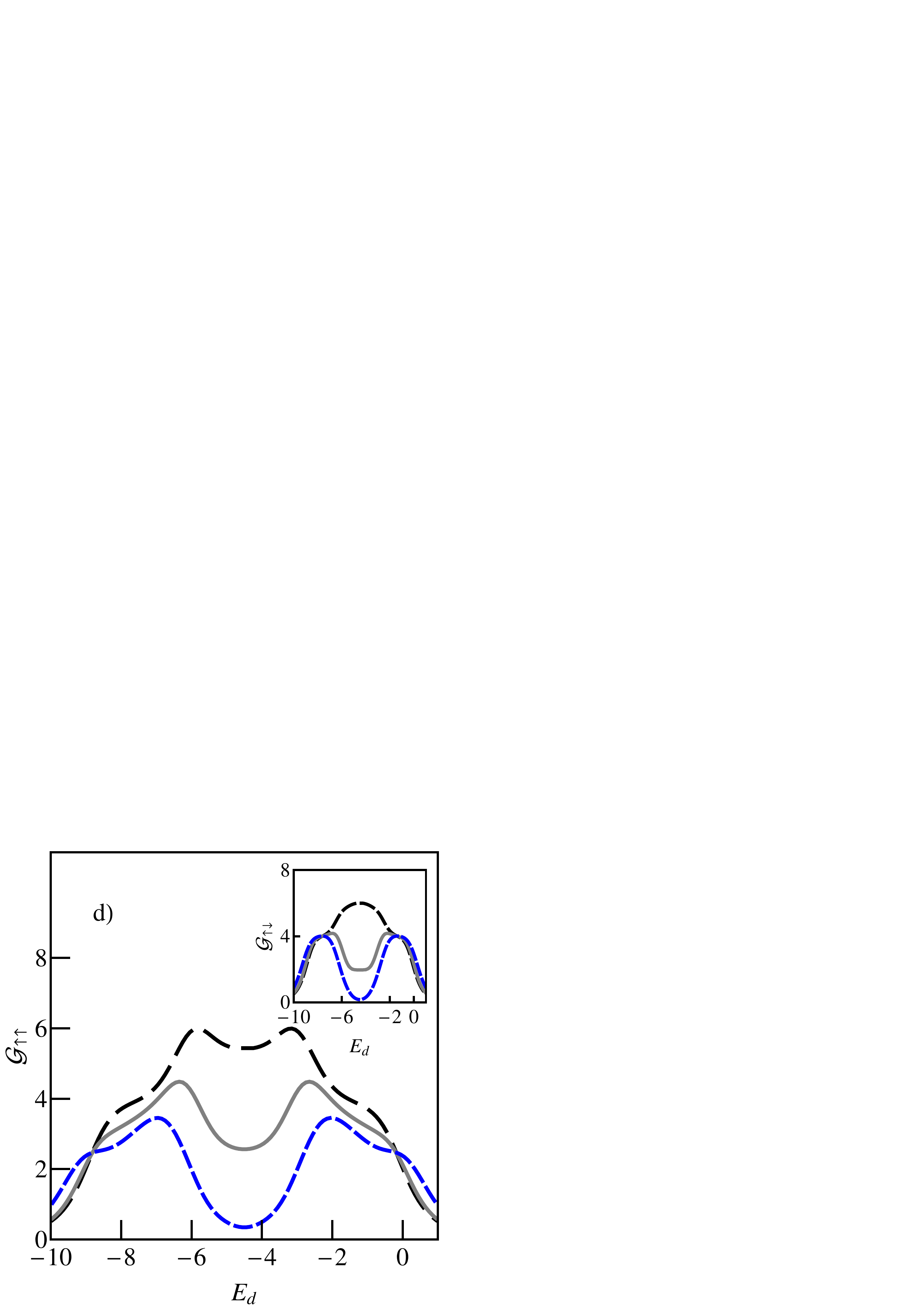}
\caption{\label{fig:epsart} (Color on line) Tunnel magnetoresistance of (a) SU(2) PDQD for $P = 0.6$ and $t = 0.04$  (black dashed line), $0.08$ (solid grey line), $0.14$ (short dashed blue line), $0.3$ (dotted red line) and (b) SU(4) PDQD for $P = 0.6$ and $t = 0.02$  (black dashed line), $0.07$ (solid grey line), $0.3$ (short dashed blue line. (b,d) Corresponding conductances for the same values of  $t$ as in (a,c) for parallel configuration of  polarization of electrodes and for antiparallel in the inset.}
\end{figure}
This in turn reflects double peak structure of conductance for strong coupling in this occupation  range for parallel configuration and single high peaks structure  for AP orientation.  Figures 13c, d present analogous gate voltage dependencies of TMR and conductances for coupled SU(4) dots.  For weak coupling inverse TMR is observed around e-h symmetry point, where AP narrow Kondo transmission peak dominates at $E_{F}$ over broader  Kondo peak for P configuration. For stronger coupling  the peaks split and Kondo correlations are suppressed, stronger suppression for P case than for AP, what results first in the change of sign of TMR and then in a gradual increase of TMR with the increase of tunneling.  In the ranges of odd occupations of the dots (molecular-like Kondo effects), in addition to tunnel induced splitting of Kondo peaks, also exchange splitting comes to the fore for P configuration, and in consequence the AP transmission peaks dominate and the inverse TMR results. Last example of spintronic application of parallel QD system discussed here  is  spin filter based on  spin dependent Dicke effect. The schematic view of the considered device is presented in the inset of Fig. 14a: parallel coupled quantum dots attached to a common  pair of polarized electrodes and directly coupled by tunneling.  Analogous system with nonmagnetic electrodes has been analyzed in  ~\cite{Nahm}, where different dot levels at the dots have been assumed $\Delta E = E_{1} - E_{2}$. To allow the reader a comparison with magnetic case discussed here, we also show in Fig. 8b corresponding plots of conductance for the dots with nonmagnetic electrodes, similar to these presented in ~\cite{Nahm}. Conductances of PDQD system with common reservoirs depicted as a function on average dots energy exhibit dips. Destructive interference leads to a complete suppression of conductance for some gate voltages. For $t =0$ the deep occurs at e-h symmetry point $\overline{E}= (E_{1}+E_{2})/2=-\mathcal{U}/2$ and for finite direct tunneling the conductance curve takes typical for the Fano resonance asymmetric form. The occurrence of the dip of conductance corresponds to passing of the correlator $\langle d^{+}_{1\sigma}d_{2\sigma}\rangle$ through zero (inset of Fig. 8b). If one  replaces the nonmagnetic electrodes by ferromagnetic the analogous   formation of interference induced antiresonances characterized by strictly zero transmission is possible  for each of the spin channels separately (Fig. 14). Spin-up and spin-down electrons  individually reach destructive interference for different gate voltages resulting in suppression of conductance in one of the spin channels i.e. $100\%$ spin polarization is achieved. Interesting this is achieved no matter how small polarization of electrodes is. Example presented in Fig. 14a   purposely illustrates the case for $P = 0.1$. One can reverse  the sign of polarization slightly varying  the gate voltage (bipolar spin filter - Fig. 14b). The gate voltage required to switch the polarization of conductance depends on polarization of electrodes, dot level energies and interdot coupling.

Let us close this section by a formal  remark on the use of SB approach  for a description of spin dependent effects. In Kotliar-Ruckenstein finite U formalism for polarized systems the introduced  auxiliary  bosons are spin dependent.
\begin{figure}
\includegraphics[width=4.4 cm,bb=0 0 400 350,clip]{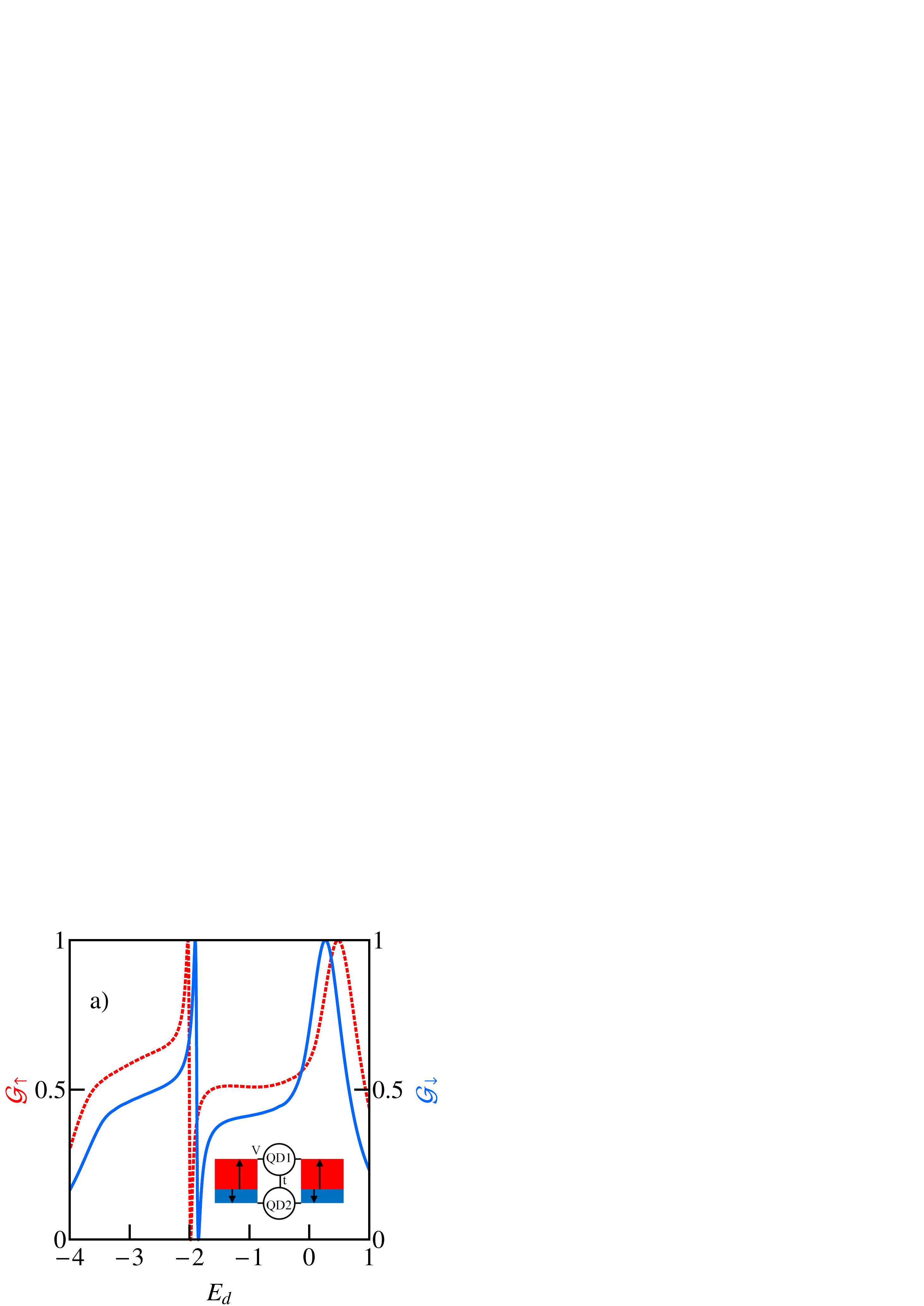}
\includegraphics[width=4 cm,bb=0 0 400 390,clip]{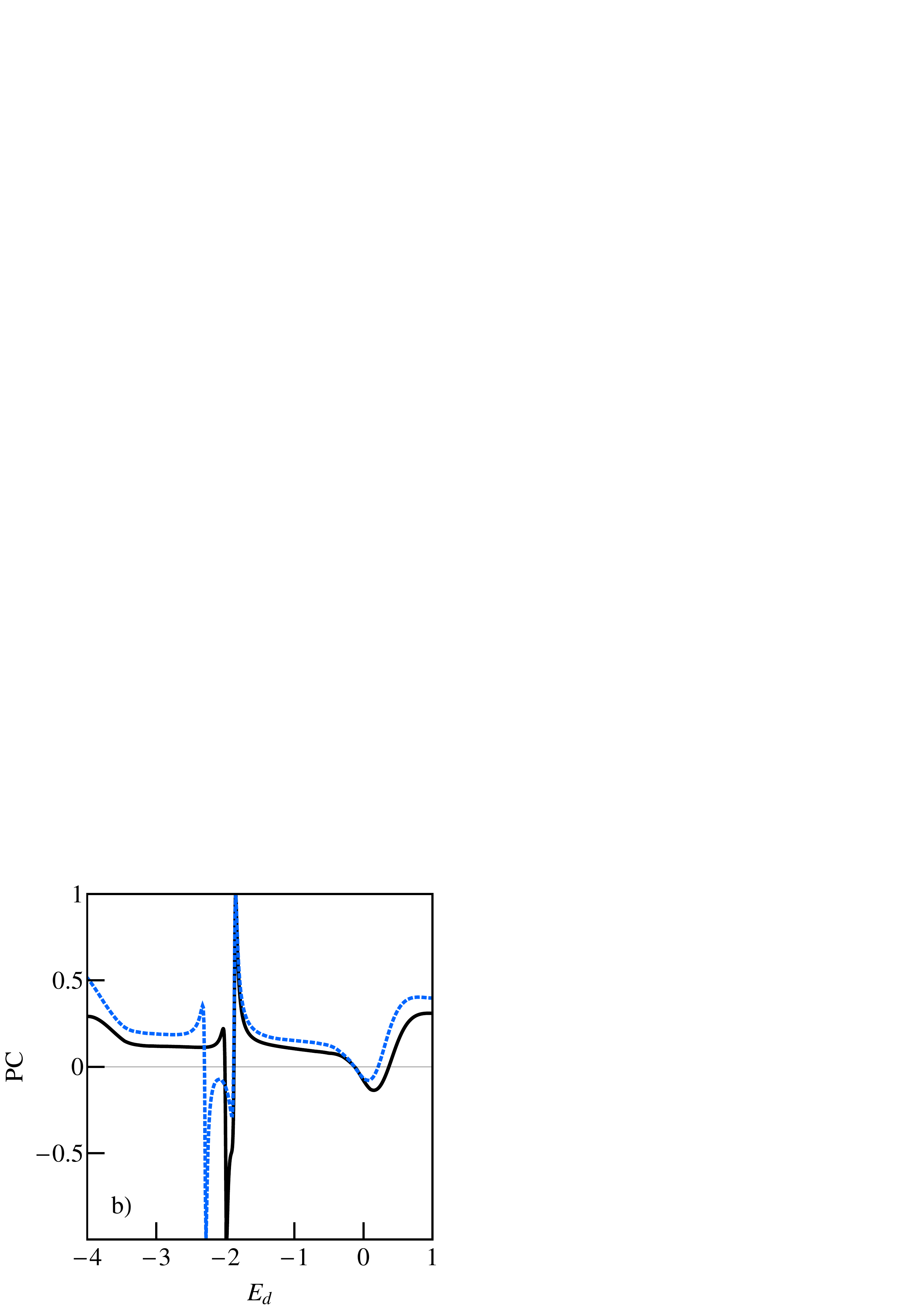}
\caption{\label{fig:epsart} (Color on line) PDQD with polarized common electrodes as a spin filter (a) Spin resolved conductances for $P = 0.1$ and  $t = 0.05$, $\mathcal{G}_{\uparrow}$ (dotted red line) $\mathcal{G}_{\downarrow}$ (solid blue). Inset shows schematic view of the device. (b) Polarization of conductance of PDQD as a function of gate voltage for $P = 0.1$ (black solid line) and $P = 0.4$ (blue dotted).}
\end{figure}
Due to spin dependent  hybridizations  not only the resulting  effective widths of many body peaks, but also the peak positions differ in different spin channels. The level renormalization in SBMFA  is spin dependent. The physical mechanisms causing spin splitting in K-R spin dependent formalism can not be identified with the effects of spin dependent charge fluctuations commonly invoked in more elaborate treatments of exchange effects e.g. in Haldane scaling approach ~\cite{Martinek2, Haldane}. Interesting the same sign of exchange field and the same gate voltage for vanishing of  this field are predicted in both approaches and thus both pictures agree qualitatively. For quantitative analysis of spin effects in SB approach it is necessary  to go beyond the mean field  treatment taking into account charge fluctuations. Such an extended  analysis, needed especially in  MV regions, will  be presented elsewhere ~\cite{Lipinski1}.

\section{Kondo effect versus antiferromagnetic coupling and a swap process between  the bonding and antibonding resonances}

In the foregoing discussion we have omitted the interdot exchange interactions, but as  pointed out e.g. by Aono and Eto ~\cite{Aono},  even small antiferromagnetic exchange  can significantly change behavior of conductance and compete with Kondo effect. In the tunnel coupled systems the dominant contribution to the exchange is superexchange, which in the limit of  strong Coulomb interaction takes the value of $4t^{2}/\mathcal{U}$. But there are other possible linking paths, depending on the system, which  can introduce another exchange mechanisms. For metallic link for instance  RKKY exchange is of importance, in which case the exchange could in principle be either ferromagnetic or antiferromagnetic. We restrict to antiferromagnetic coupling, but for the sake of generality we treat in the following the exchange coupling  $\mathcal{J}$ as effective independent parameter.
\begin{figure}
\includegraphics[width=4.3 cm,bb=0 0 370 330,clip]{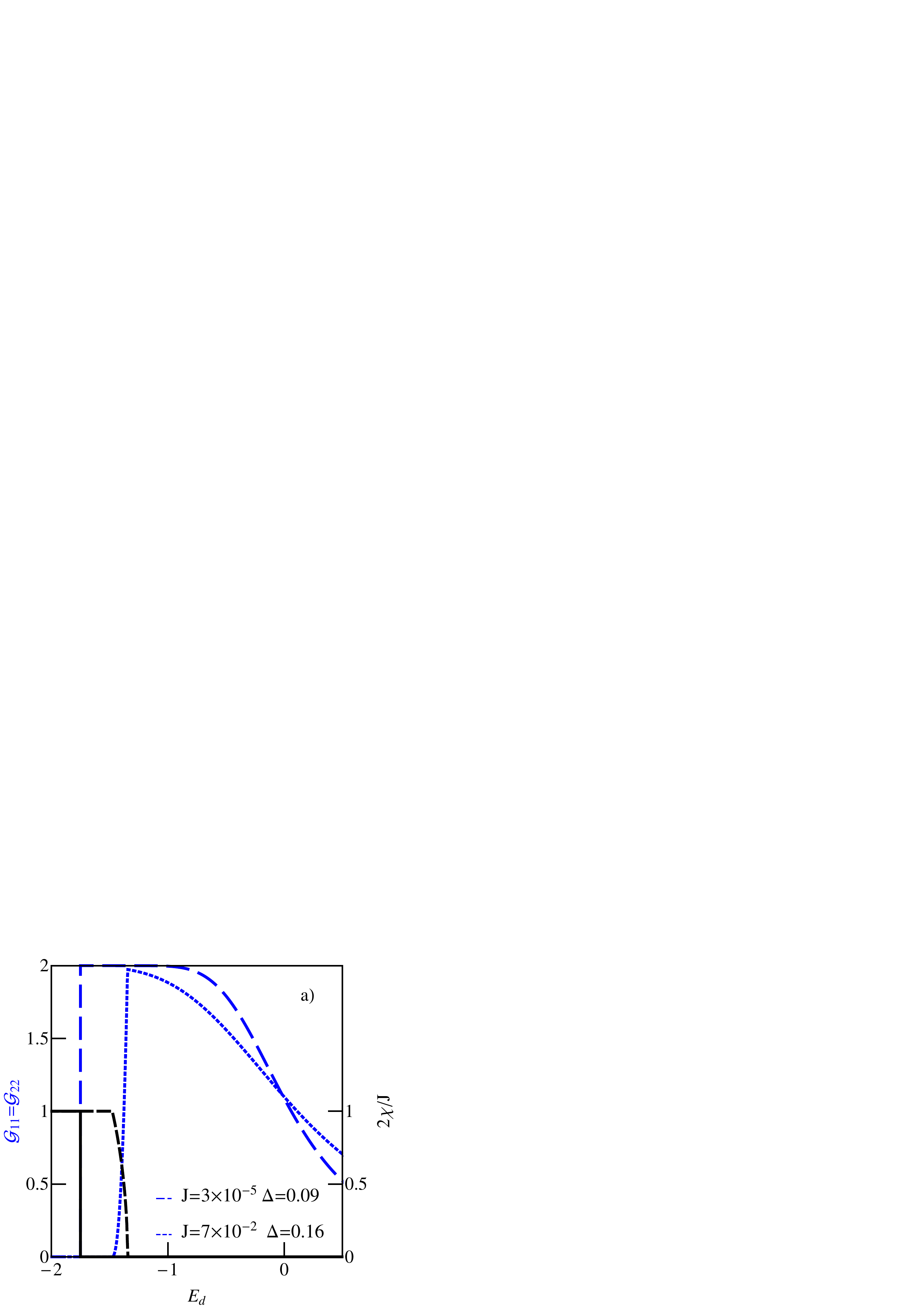}
\includegraphics[width=4 cm,bb=0 0 405 390,clip]{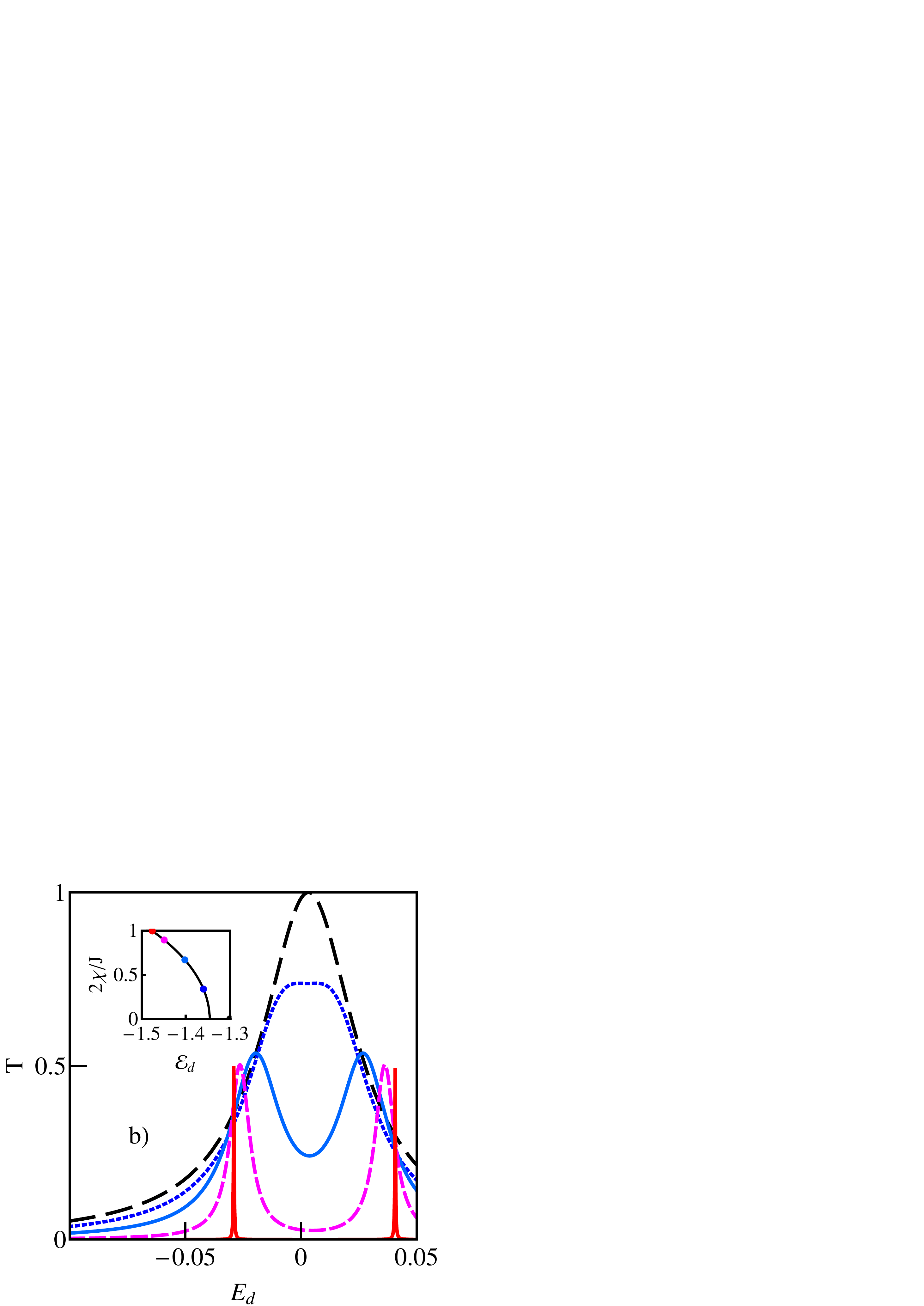}
\includegraphics[width=4.3 cm,bb=0 0 370 330,clip]{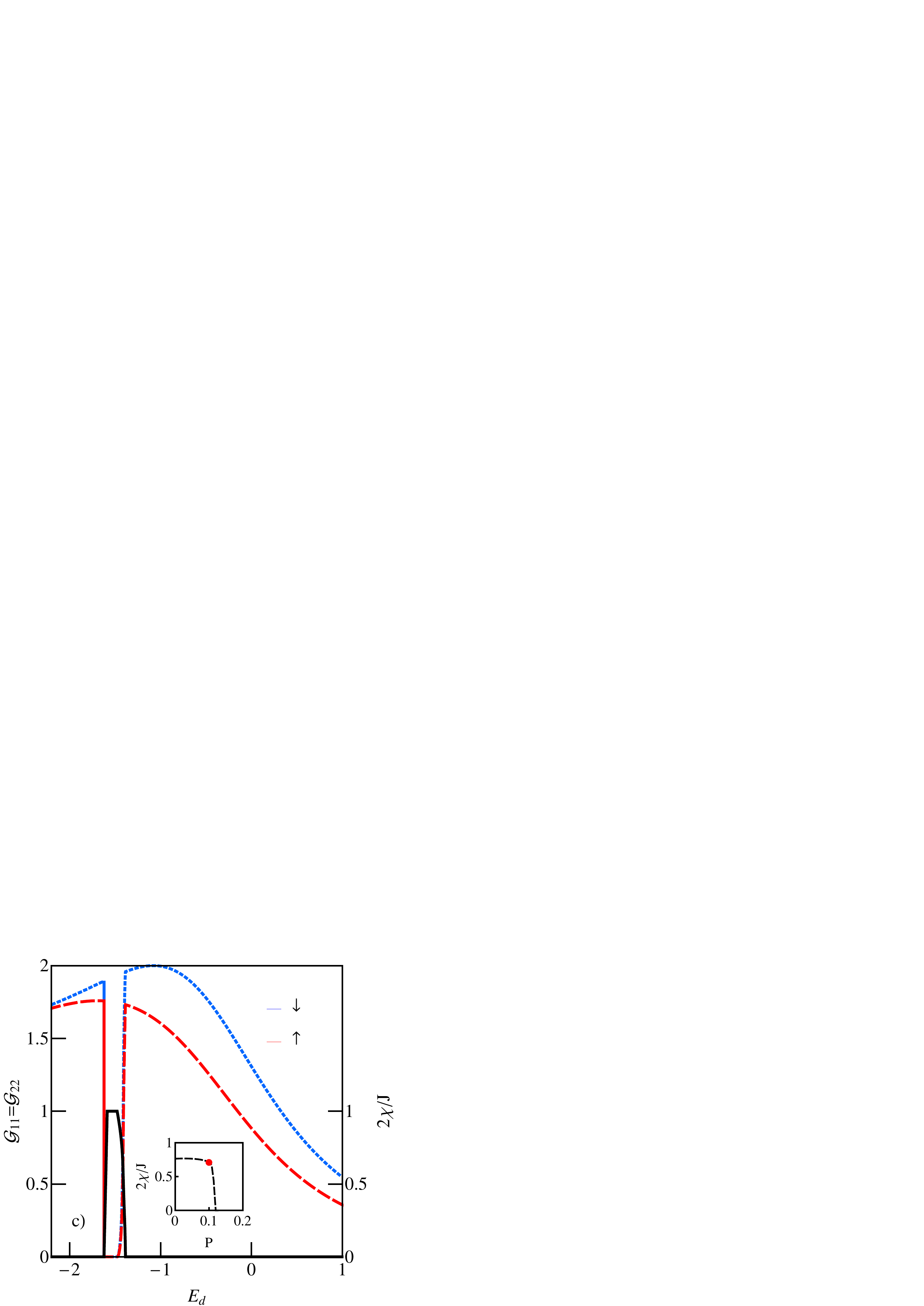}
\includegraphics[width=4 cm,bb=0 0 400 390,clip]{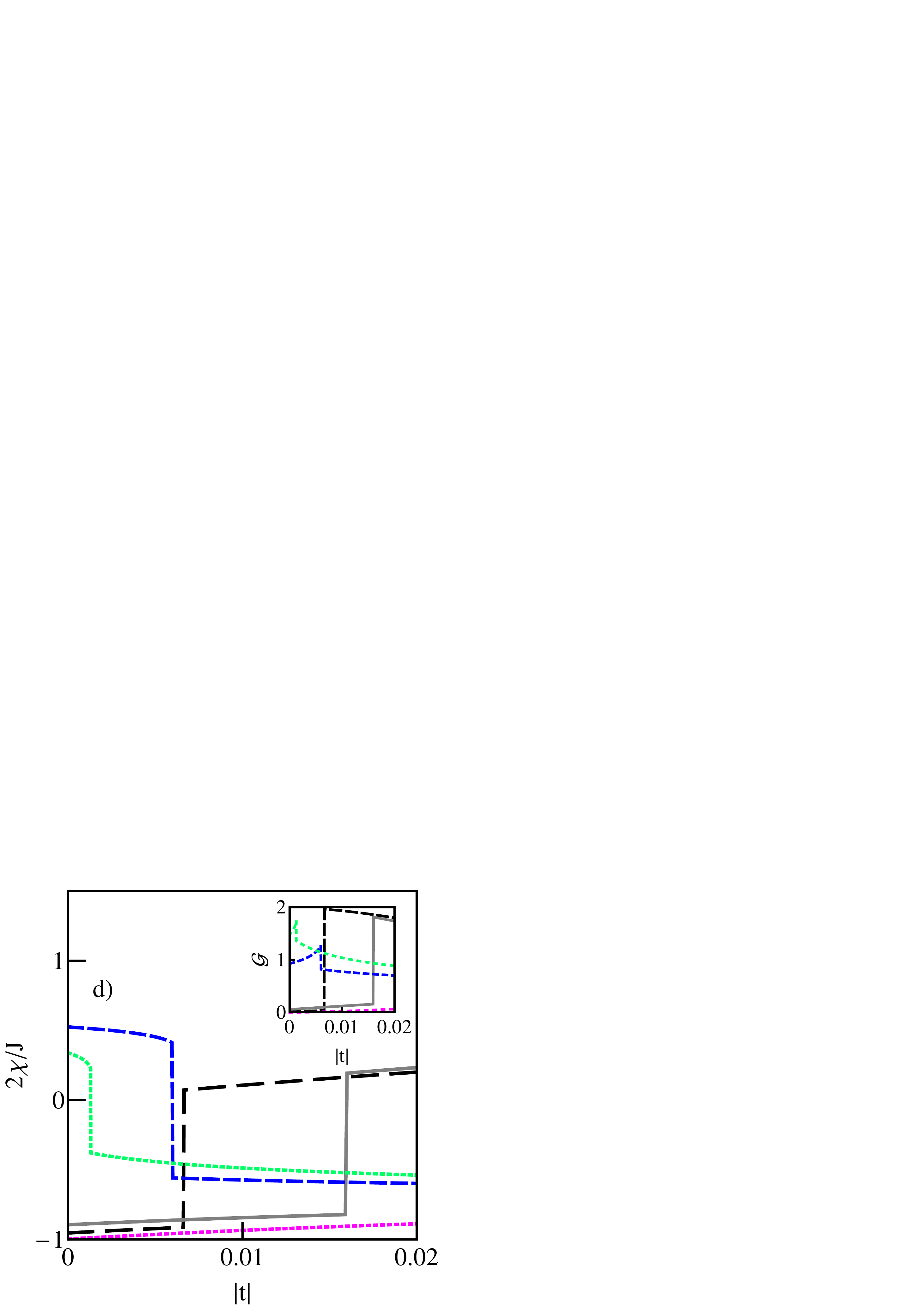}
\caption{\label{fig:epsart} (Color on line) PDQD with antiferromagnetic interdot exchange, $\mathcal{U}\rightarrow \infty$  (a) Conductance (blue long dashed and dotted lines) and expectation value of VB operator (black solid and dashed lines) for $\mathcal{J}=3\times10^{-5}$, $\Gamma = 0.09$ (blue long  dashed and black solid lines) and  $\mathcal{J}=3\times10^{-5}$, $\Gamma = 0.09$ (blue dotted and dashed lines). (b) Transmission for $\mathcal{J} = 0.07$ plotted for several values of  dot energy: $E_{d} = -1.36,-1.38,-1.45,-1.46,-1.48$. Inset shows the gate voltage  dependence of VB parameter. (c) Spin resolved conductances: $\mathcal{G}_{\uparrow}$ (red broken line) $\mathcal{G}_{\downarrow}$ (dotted blue) and VB parameter (solid black) for the case of polarized electrodes  ($P = 0.1$). Inset presents VB parameter as a function of polarization of electrodes. (d) Illustration of SWAP operations, VB operator as a function of interdot tunneling and in the inset  the corresponding plots of conductance.  Curves are  plotted for dot energies depicted in the inset of Fig. 14b maintaining the same types of drawing lines in both pictures. }
\end{figure}
It is known, that two opposing quantum many-body effects: Kondo screening  and magnetic ordering lead for an array of dots or impurities with p-h symmetry ($t =0$) to a quantum critical transition ~\cite{Varma}  or to a crossover in the case when this symmetry is broken  ~\cite{Lopez}. The antiferromagnetic interaction we discuss is taken here in the form (6) and we treat it in the mean field approximations introducing a valence bond (VB) operator ~\cite{Aono, Lopez} with the corresponding expectation value $\chi=-(\mathcal{J}/2)\sum_{\sigma}\langle f^{+}_{\nu2\sigma}f_{\nu1\sigma}\rangle$.
\begin{figure}
\includegraphics[width=4 cm,bb=0 0 290 300,clip]{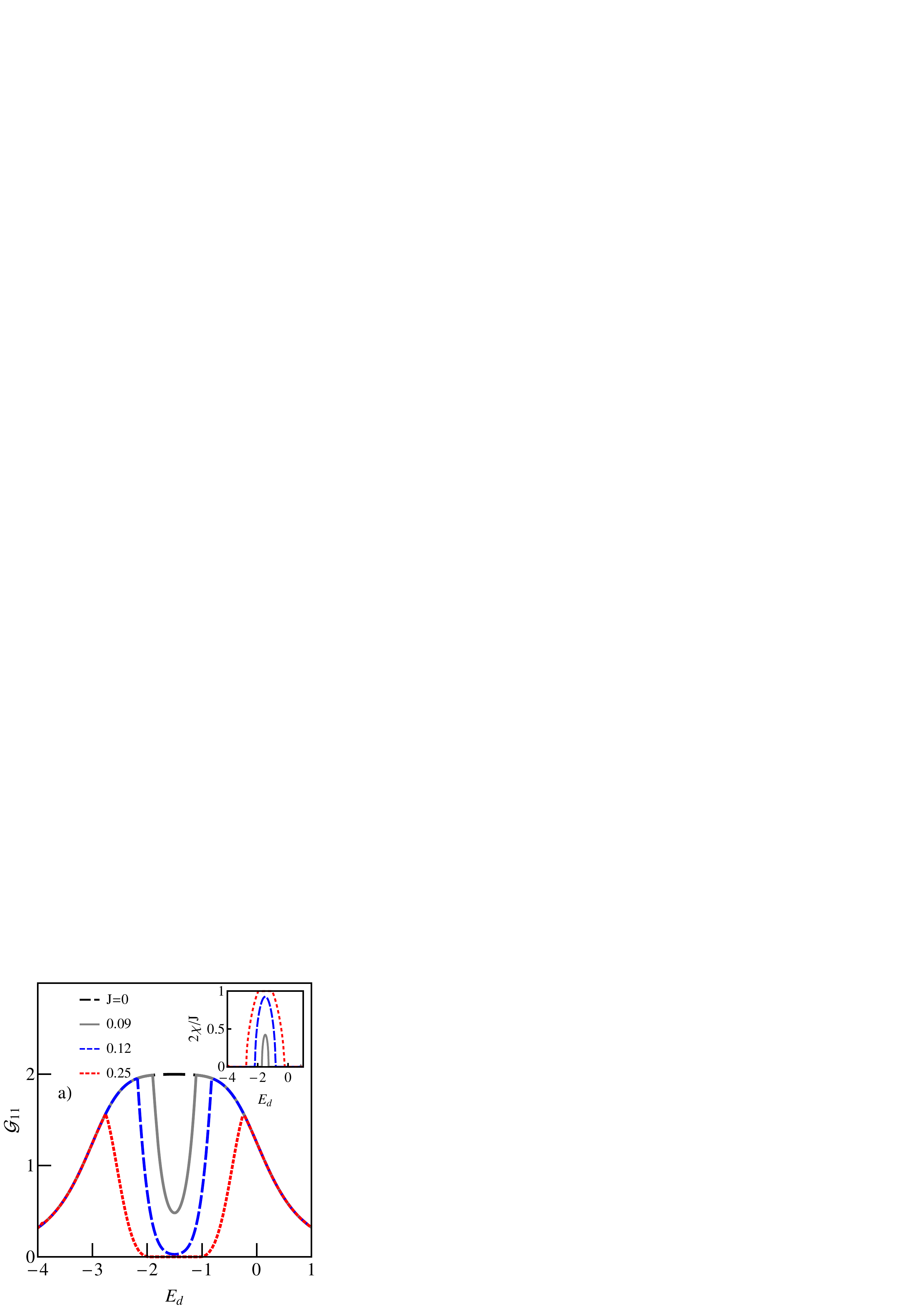}
\includegraphics[width=4 cm,bb=0 0 405 420,clip]{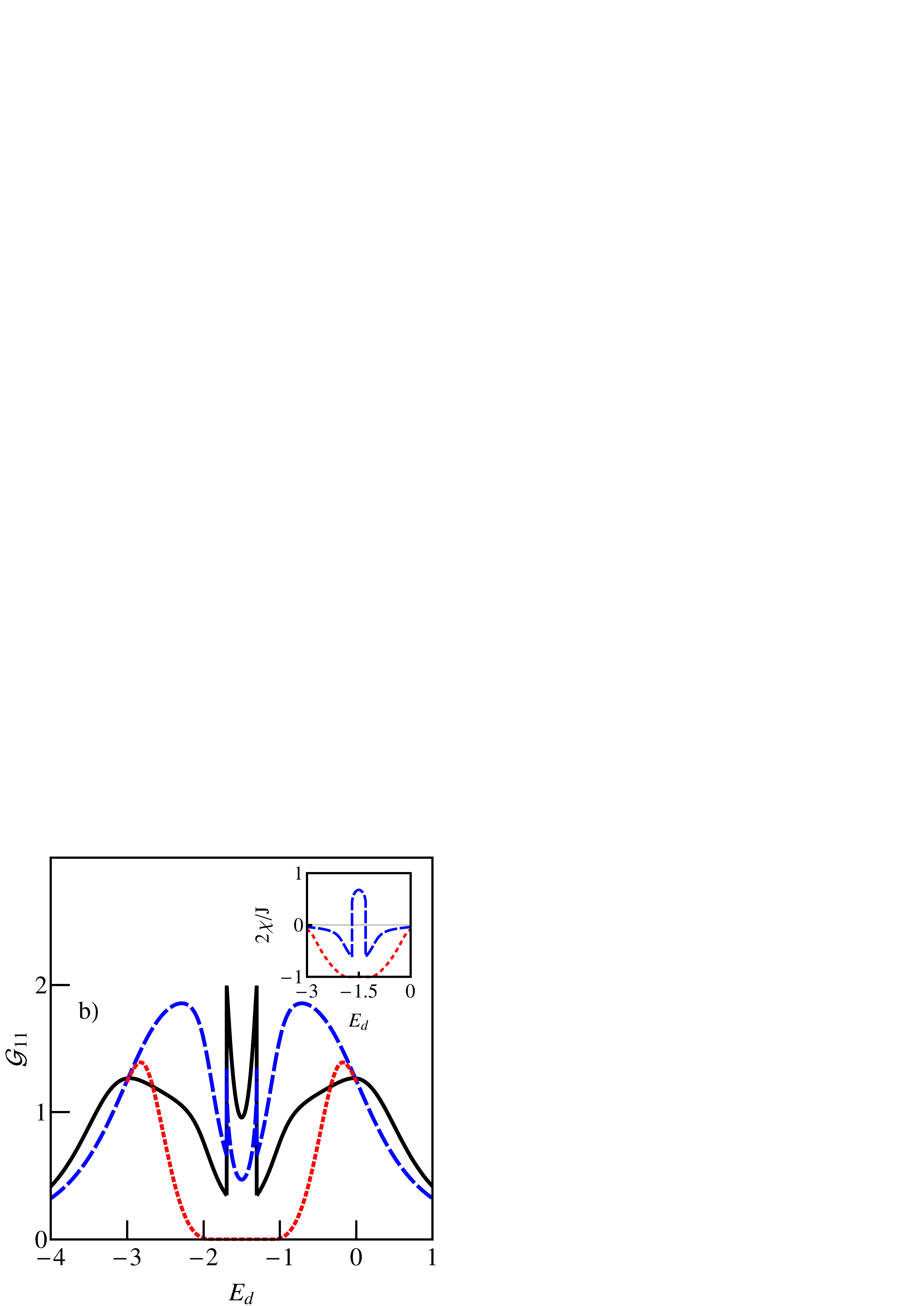}
\caption{\label{fig:epsart} (Color on line) Conductance of PDQD with interdot exchange for $\mathcal{U} =3$ and $\Gamma = 0.09$ and VB parameter in the insets (a) $t =0$ and several values of $\mathcal{J}$. (b) Conductance for $\mathcal{J} = 0.09$, $t = 0.006$ ( broken blue line),  $t =0.07$ ( solid black) and for $\mathcal{J}= 0.25$, $t =0.07$ (dotted red). Inset shows gate dependence of VB parameter for $t =0.07$ and $\mathcal{J} = 0.09$ and $0.25$. }
\end{figure}
This approximation together with SBMFA approach  renders the whole PDQD Hamilton into quadratic form in fermion operators and the MFA procedure is performed with additional minimizing parameter $\chi$. Let us first concentrate on the limit $\mathcal{U}\rightarrow\infty$ and discuss the  case $t = 0$.  The   gate voltage evolution of  conductance and VB order parameter is presented in  Figure 15a.  Upon making $E_{d}$  more negative i.e. decreasing Kondo temperature  and thus  increasing the ratio $\mathcal{J}/T_{K}$ at fixed $\mathcal{J}$, the transmission peaks split and narrow (Fig. 15b).  For still deeper dot energies or smaller hybridization to the leads peaks  abruptly transform at the  critical value $(\mathcal{J}/T_{K})_{c}\approx2.5$ into  Dirac $\delta$ functions  split by $2\mathcal{J}$ indicating the formation of magnetic states ($\chi=\pm\mathcal{J}/2$), for stronger hybridization  this transition smoothes and  magnetic states emerge before the Kondo state with unitary conductance is reached. When polarized electrodes are attached to the dots the spin degeneracy is removed, transmission in the intermediate range ($\chi<\mathcal{J}/2$) exhibits the multi peak structure (not presented), conductance becomes spin dependent and up and down  spin channels differently contribute to valence bond correlator. When $\chi$ reaches the limit $\mathcal{J}/2$ conductance is totally suppressed (Fig. 15c), for still deeper dot energies VB order is destroyed and finite spin dependent conductance is recovered. For $t\neq0$ bonding and antibonding entangled Kondo resonance are formed and the degeneracy is removed. In the coexistence region  of MV and magnetic states (unsaturated value of $\chi$ - inset of Fig. 15b) VB correlator can reverse sign at a critical value of interdot tunneling,  the relative position of bonding and antibonding energies are switched, or equivalently the triplet and singlet states interchange (Fig. 15d). The same effect can be achieved fixing $t$ and changing dot energies. A switch from negative to positive value of $\chi$ means that bonding resonance swaps position from below to above the Fermi level. This effect manifests also in the jumps of conductance (inset of Fig. 15d).  Fig. 16a  presents the competition of Kondo correlations and magnetic order for finite $\mathcal{U}$. Magnetic correlations are observed   in the region where in the absence of interdot exchange ($\mathcal{J} = 0$) Kondo correlations are the strongest i.e.  around $E_{d} = -\mathcal{U}/2$  ($N =1$). A complete singlet formation is hardly to achieve in this case ($\chi<\mathcal{J}/2$) and the limiting value of $\chi=\mathcal{J}/2$ is only reached for very large values of  $\mathcal{J}$ or $\mathcal{U}$. The swaps between entangled bonding and antibonding Kondo resonances for finite $\mathcal{U}$ are presented in Fig. 16b, where we show the  gate voltage dependencies of VB parameter (inset) and conductance for fixed interdot tunneling. The conductance peaks appear precisely at the points when $\chi$ changes sign.

\section{Concluding remarks}
In this paper, we compared transport properties of  pairs of strongly correlated  quantum dots of  SU(2) and SU(4) symmetries in parallel geometry.  The former case occurs for spin or orbital degeneracy and the latter when both these degeneracies occur simultaneously. Much less attention has been paid in literature to the last systems and they are of interest not only for the cognitive purposes  but also for applications  since the relevant temperature scale of many body effects can be much higher than for spin Kondo effect. For weakly coupled  dots Kondo effect with unitary limit of conductance occurs in the range of  single occupations of the dots for SU(2) symmetry, for SU(4)  this resonance is formed both for even  and  odd dot occupancies. In the latter case SU(4) Kondo peaks  are  shifted away from the Fermi level and are  characterized by relatively high Kondo temperatures. For half filling six states are engaged in cotunneling processes and the resulting resonance locates similarly to single level dots  close to $E_{F}$  and the total conductance is doubly enhanced in this case.  Different types of couplings between the dots  lower the symmetries and modify correlations and interference conditions. We examined  the impact  of interdot tunneling, interdot interaction, exchange coupling, mixing of electrode channels  and the effects of polarization  and singularities of the electronic structure of electrodes.  To study the  correlations we used the slave-boson mean-field approximation at finite U. Different regimes were analyzed in a wide range of parameter space. A crossover from a separate Kondo state on each of the dots (atomic-like) at half fillings to coherent bonding-antibonding superposition of many-body Kondo states of the dots (molecular-like) has been observed with the increase of interdot tunneling. For strong tunneling  the bonding orbitals around half filling are almost fully occupied and antibonding empty, Kondo effect is suppressed in this case, but the unitary conductance is observed  outside this range manifesting  the  single electron or single hole molecular Kondo effects. Depending on whether repulsive or attractive interaction  occurs between the dots the ranges of Kondo states extend or narrow and in the case of attractive interaction transitions into mixed valence state sharpen. For electrodes with divergent singularities in the density of states and consequently singular hybridizations as occurs for instance for carbon nanotubes discussed by us, dramatic changes of Kondo physics and interference condition are observed when Fermi level enters Van Hove singularity. This reflects in the dips of conductance, in close vicinity of VHs  the many-body resonances take the mixed valence character.  For dots coupled by exchange we have  discussed  a competition between Kondo correlations and magnetic ordering  showing that in the regions of unsaturated magnetic order it is possible to swap between the bonding and antibonding many-body resonances. The principal  motivation  of the present work  was to discuss how the  interplay of many-body correlation effects and interference might be exploited in spintronics and to highlight the potential of parallel quantum dots  in this field. The wide tunability of  PDQDs bodes well for future  applications. For illustration we have presented two examples: DQD spin filter device and  spin valve, both  gate controllable  and  we have shown how polarization of conductance can be transferred between tunnel coupled subsystems.

\begin{acknowledgments}
This project was supported by the Polish National Science Centre from the funds awarded through the decision No. DEC-2013/10/M/ST3/00488.
\end{acknowledgments}

\def\refname{References}

\end{document}